%% file: IntStrings.tex
\RequirePackage{ifpdf}
\documentclass[a4paper,12pt]{JHEP3}
\usepackage[latin1]{inputenc}
\pdfoutput=1
\usepackage{amsmath}
\usepackage{amsfonts}
\usepackage{amssymb}
\usepackage{graphicx}
\usepackage{subfig}
\usepackage{mathrsfs}
\usepackage{listings}
\usepackage{verbatim}
\usepackage{mathtools}
\usepackage{simplewick}
\usepackage{color}
\usepackage{subfig}

\usepackage[font=footnotesize]{caption}

\makeatletter
\def\@cite#1#2{$^{[\mbox{\scriptsize #1\if@tempswa , #2\fi}]}$}

\graphicspath{{Images/}}

\makeatother

\newcommand{\be}{\begin{equation}}
\newcommand{\ee}{\end{equation}}
\newcommand{\bq}{\begin{eqnarray}}
\newcommand{\eq}{\end{eqnarray}}

\newcommand{\one}{\hbox{\rm 1\kern-.27em I}}

\title{Delta-function Interactions for the Bosonic and Spinning Strings and the Generation of Abelian Gauge Theory}
\author{James P. Edwards and Paul Mansfield \\Centre for Particle
Theory, Department of Mathematical Sciences, \\
University of Durham, Durham DH1 3LE, UK \\ Email:
J.P.Edwards@durham.ac.uk,
P.R.W.Mansfield@durham.ac.uk}

\abstract{We construct contact interactions for bosonic and spinning strings. In the tensionless limit of the spinning string this reproduces the super-Wilson loop that couples spinor matter to Abelian gauge theory. Adding boundary terms that quantise the motion of charges results in a string model equivalent to spinor QED. The strings represent lines of electric flux connected to the charges. The purely bosonic model is spoilt by divergences that are excluded from the spinning model by world-sheet supersymmetry, indicating a preference for spinor matter.}


\preprint{DCPT-14/47}

\begin{document}

%\abstract{We construct contact interactions for bosonic and spinning strings. In the tensionless limit of the spinning string this reproduces the super-Wilson loop that couples spinor matter to Abelian gauge theory. Adding boundary terms that quantise the motion of charges results in a string model  equivalent to spinor QED. The strings represent lines of electric flux connected to the charges. The purely bosonic model is spoilt by divergences that are excluded from the spinning model by world-sheet supersymmetry.}

\section{Introduction}

In 1955 Dirac, attempting to eliminate infinities, proposed a 
new approach to QED in which the fermion operators create and
annihilate part of the electromagnetic field along with the
electrons and positrons themselves\cite{Dirac}. He made several
suggestions as to which part of the electromagnetic field to
include. In \cite{Dirac2} it was the Coulomb field -- in the
version we will consider here it is a single line of electric
flux created along a curve $C$ connecting equal positive and
negative charges
\be
{\bf E}({\bf x})={q\over\epsilon_0}\int_C \delta^3( {\bf x}-{\bf
y}) \,d{\bf y}
\,.\label{DiracString}
\ee
This satisfies Gauss' law
\begin{equation}
\nabla\cdot{\bf E}({\bf x})={q\over\epsilon_0}\int_C \nabla_{\bf
x}\delta^3( {\bf x}-{\bf y}) \cdot d{\bf y}
=-{q\over\epsilon_0}\int_C \nabla_{\bf y}\delta^3( {\bf x}-{\bf
y}) \cdot d{\bf y}={q\over\epsilon_0}\,\delta^3( {\bf x}-{\bf
a})-{q\over\epsilon_0}\,\delta^3( {\bf x}-{\bf b})\,,
\end{equation}
but $\nabla\times{\bf E}\neq 0$ and (\ref{DiracString}) clearly
differs from the dipole field that would arise classically. It
was hoped that the conventional electromagnetic field would
result from quantum averaging. Quantising this system of a pair
of equal and opposite charges and a single line of electric flux
running between them requires treating the curve $C$ amongst the
dynamical degrees of freedom. This hints at string theory and
the connection is reinforced if we consider what would be an
appropriate action to describe the dynamics of $C$. If the two
equal and opposite charges move along world-lines these will be
the boundary, $B$, of the world-sheet, $\Sigma$, swept out by
the curve $C$. The boundary could be a closed curve if for
example the charges were a virtual pair created from the vacuum
which later annihilate, or it could extend to infinity if the
charges are scattering particles. The generalisation of
(\ref{DiracString}) to this dynamical case is an expression for
the electromagnetic field strength
\be
F_{\mu\nu}(x)=
-q\int_\Sigma \delta^4(x-X)\, d\Sigma_{\mu\nu}(X)\,.
\label{Fsolnn}
\ee
where $d\Sigma_{\mu\nu}$ is an element of area on $\Sigma$. We
take $\Sigma$ to be given parametrically by $X^\mu=X^\mu(\xi)$
with the world-sheet co-ordinates $\xi^1$ and $\xi^2$ lying in a
parameter domain $D$ with boundary $\partial D$ on which $\left.X^\mu\right|_{\partial D}=w^\mu$, so
$2\,d\Sigma_{\mu\nu}=\epsilon^{ab}\partial_a
X^\mu\partial_bX^\nu$ with $\partial_a=\frac{\partial}{\partial\xi^a}$.
Again this satisfies Gauss' law
\be
\partial^\mu \, F_{\mu\nu}=J_\nu \,,
\ee
where the current density due to the charges on the boundary of
$\Sigma$ is
\be
J^\mu(x)=q\int_B \delta^4( x-w)\,dw^\mu.
\label{curr}
\ee
Nielsen and Olesen\cite{NO} have used such a field strength tensor to form a field theory describing the dual string from a basis of non-linear electrodynamics. It is also present in theories of electromagnetism with magnetic monopoles\cite{diracM} and can be used to derive an effective string theory describing the evolution of the Dirac string linking two such poles\cite{Nambu, Baker}. Using this field strength in the standard form of the action for
electromagnetism gives
$$
S_{EM}=-{1\over 4}\int d^4x\, F_{\mu\nu}F^{\mu\nu}={q^2\over
4\epsilon_0^2}\int_\Sigma
d\Sigma^{\mu\nu}(\xi)\delta^4\left(X(\xi)-X(\xi')\right)\,
d\Sigma_{\mu\nu}(\xi')\,.
$$
The argument of the delta-function vanishes when
$\xi^{\prime a}=\xi^a$, and also at points where the world-sheet
intersects itself. This gives two contributions,
\be
S_{EM}={q^2\over 4\epsilon_0^2}\delta^2(0) \,{\rm Area}(\Sigma)+{q^2\over 4\epsilon_0^2}\int_\Sigma\left.
d\Sigma^{\mu\nu}(\xi)\,\delta^4\left(X(\xi)-X(\xi')\right)\,
d\Sigma_{\mu\nu}(\xi')\right|_{\xi\neq\xi'}\, ,
\label{Sem}
\ee
the first is proportional to the Nambu-Goto action of string
theory, albeit with a divergent coefficient, the second is a
self-intersection interaction that we will study in detail
below. Such direct interactions have previously been discussed by Kalb and Ramond\cite{kalbR} and the one we propose here satisfies the consistency constraints they derive. This action has been applied before classically to the problem of confinement\cite{NRH, Olesen} but without the effects of self-intersections or quantisation that we shall consider here.

\bigskip
\noindent In \cite{Mansfield:2011eq} the connection to string theory was
taken further by showing that when (\ref{Fsolnn}) is averaged
over world-sheets with fixed boundary $B$ using Polyakov's
formulation of string theory the result is the classical Maxwell
field sourced by $J_\nu$. So, although the $F_{\mu\nu}$ in
(\ref{Fsolnn}) does not satisfy the remaining Maxwell's
equations, its average does. This follows from the result (which
we will re-derive in section \ref{GenExp})
\be
4\pi^2\langle \int_\Sigma \delta^4(x-X)\,
d\Sigma_{\mu\nu}(X)\rangle _{\Sigma}
=\partial_\mu \int_B {dw_\nu \over ||x-w||^2}-\partial_\nu
\int_B {dw_\mu \over ||x-w||^2}
\label{sol}
\ee
where the average over surfaces of any functional
$\Omega[\Sigma]$ in Polyakov's formulation is
\be
\langle \Omega\rangle_\Sigma={1\over Z}\int {\mathscr{D}}g\,{\mathscr{D}}_g {X}\,\Omega\,e^{-S[X,\,g]}\,;\quad
S[X,\,g]={
{1\over 4\pi\alpha'}\int_D g^{ab}\,{\partial{
X^\mu}\over\partial\xi^a}{\partial{
X^\mu}\over\partial\xi^b}\,\sqrt {g}\,d^2\xi}
\,,\label{avvS'}
\ee
\footnote{The normalisation constant $Z=\int {\mathscr{D}}g\,e^{-F}$
where $F$ is the sum of $S[X,g]$ minimised with respect to $X$,
and the logarithm of the functional determinants that give rise
to the Liouville theory\cite{poly}. The dependence of $Z$
on B is contained in the first term which vanishes in the
tensionless limit in which the size of $B$ tends to zero in
units of $\alpha'$. We denote the value of $Z$ in this
tensionless limit by $Z_0$.}This is computed in Euclidean space where the integrals are
better behaved, so that $||x-y||$ is the Euclidean distance
between $x$ and $y$. Minkowski space results are obtained by
Wick rotation. We can interpret (\ref{sol}) as a quantitative
realisation of Faraday's idea\cite{Faraday} that the dynamical degrees of
freedom of electromagnetism are the lines of force, rather than
the fields of Maxwell's theory, but with the addition that their
positions are to be averaged over with the natural geometric
weight used in string theory.

\bigskip
\noindent There are two unusual features of (\ref{sol}). The first is that
the result is independent of the string-scale $\alpha'$. If this
had not been the case then at length scales that are large in
terms of $\alpha'$ we might expect the average to be dominated
by the minimal surface spanning $B$, with the electromagnetic
field being close to zero away from this surface. This situation
would be more appropriate to a model of confinement rather than
electromagnetism and would have endangered the interpretation in
terms of fluctuating lines of force which pervade the whole of
space.

\bigskip
\noindent The second feature is that the presence of the delta-function in
(\ref{sol}) means that the average is `off-shell' in the usual
sense of string theory. Recall how the mass-shell condition
arises in Polyakov's approach. Computing the expectation value
of a product of re-parametrisation invariant vertices such as
$$
\kappa\int d^2\xi\,{\sqrt g}e^{ik\cdot X(\xi)}
$$
involves first the evaluation of the functional integral
\be
\int{\mathscr{D}}_g {X}\,e^{-S[X,\,g]}\ldots\kappa\int d^2\xi\,{\sqrt
g}e^{ik\cdot X(\xi)}\ldots\equiv\langle \ldots\kappa\int d^2\xi\,{\sqrt
g}e^{ik\cdot X(\xi)}\ldots\rangle_X
\label{abov}
\ee
where the dots stand for the other vertices. 
Using Wick's theorem the integral over $X$ leads to a factor
involving the Green function for the two-dimensional Laplacian
at coincident points, $G(\xi,\xi)$
\be
\kappa\int d^2\xi\,{\sqrt
g}e^{-\pi\alpha'k^2G(\xi,\,\xi)}\label{conv}
\ee
amongst other terms not of immediate relevance. $G(\xi,\xi)$
needs to be regulated which introduces a dependence on the scale
of the metric. With heat-kernel regularisation and choosing
$\xi^a$ to satisfy the conformal gauge
$g_{ab}=\delta_{ab}e^\varphi$ the leading behaviour at points
away from the boundary is $G(\xi,\xi)\sim (\varphi(\xi)-2\log
(\epsilon))/(4\pi)$ where $\epsilon$ is a short-distance
cut-off.
So (\ref{conv}) becomes
$$\kappa\epsilon^{\alpha'k^2/2}\int
d^2\xi\,e^\varphi\,e^{-\alpha'k^2\varphi/4}
\label{suu}
$$
and $\varphi$ decouples if $k$ satisfies the tachyon mass-shell
condition $k^2=4/\alpha'$; also the result is finite if $\kappa$
is renormalised to make $\kappa\epsilon^2$ finite.

\bigskip
\noindent The expectation value of the delta-function
also decouples from $\varphi$, but in such a different way that
it evades a mass-shell condition. If we decompose the
delta-function as a Fourier integral then 
\be
\delta^4(x-X)\,
d\Sigma^{\mu\nu}(X)=\int {d^4k\over (2\pi)^4}e^{ik\cdot x}
{1\over 2}V_{k}^{\mu\nu};~\quad V_{k}^{\mu\nu}(\xi)=\epsilon^{ab}\partial_a
X^\mu\partial_bX^\nu \,e^{-ik\cdot X(\xi)}
\ee
so that the computation
required to establish (\ref{sol}) involves the functional
integral akin to (\ref{abov})
\begin{align}
e^{ik\cdot x}\left< V_k^{\mu\nu}(\xi)\right>_X &=\int{\mathscr{D}}_g {X}\,e^{-S[X,\,g]}\,\epsilon^{ab}\partial_a
X^\mu\partial_bX^\nu \,e^{ik\cdot (x-X(\xi))} \nonumber\\
&\propto
\epsilon^{ab}\left( k^{[\mu}\partial_bX_c^{\nu ]} \,
2\pi\alpha'\partial_aG(\xi,\xi)\,
+\partial_a X_c^\mu\partial_b X_c^\nu \right)\,e^{ik\cdot
(x-X_c(\xi))}
e^{-\pi\alpha'k^2G(\xi,\xi)}
\label{su}
\end{align}
where $X_c$ is the value of $X$ that evolves from the boundary
value according to the classical equations of motion. Note that
we do not have the freedom to introduce a $k$-dependent factor
like $\kappa$ without ruining the Fourier decomposition of the delta function. For points away from the boundary
\be
e^{-k^2\pi\alpha'G(\xi,\,\xi)}\sim
\epsilon^{\alpha'k^2}e^{-\alpha'k^2\varphi/4}\label{suum}
\ee
is suppressed as $\epsilon\downarrow 0$, since in Euclidean
signature $k^2>0$. On the boundary itself $G(\xi,\xi)=0$ since
there $X$ coincides with the world-lines of the charges
requiring Dirichlet boundary conditions to be imposed on the
Laplacian. This differs from the conventional string setting
which employs Neumann conditions. So, precisely on the boundary
\be
e^{-k^2\pi\alpha'G(\xi,\,\xi)}=1
\ee
This means that (\ref{su}) is negligible except when $\xi$ is in
a thin strip close to the boundary, and the value precisely on
the boundary is independent of $\varphi$. The width of this
strip is determined by the length-scale introduced when we
regulate $G(\xi,\xi)$. When (\ref{su}) is integrated over the
parameter domain $D$ we only have to consider contributions
within this strip and so we can separate the various factors
into those like $G(\xi,\xi)$ and its derivative that vary
rapidly as $\xi$ moves from the boundary into the interior of
the world-sheet, and terms like $X_c$ and its derivatives that
vary slowly and can be approximated by their boundary values.
Arranging the parameters $\xi^a$ so that $\xi^2$ is constant on
the boundary and $\xi^1$ varies along it the integral over $\xi$
of the first term of (\ref{su}) contains
\begin{align}
&\int d^2\xi\, k^{[\mu}\partial_1X_c^{\nu ]} \,e^{ik\cdot
(x-X_c(\xi))}
2\pi\alpha'\partial_2G(\xi,\xi)\,e^{-\pi\alpha'k^2G(\xi,\xi)}
\nonumber\\
=&\int d\xi^1\,\left( k^{[\mu}\partial_1X_c^{\nu ]} \,e^{ik\cdot
(x-X_c(\xi))}
\int
d\xi^2\,2\pi\alpha'\partial_2G(\xi,\xi)\,e^{-\pi\alpha'k^2G(\xi,\xi)}\right)\nonumber \\
=&\int_B dw ^{[\mu}k^{\nu]}  e^{ik\cdot (x-w)}/k^2\label{reqres}
\end{align}
which is the Fourier transform of (\ref{sol}).
Note that this is independent of the cut-off scale, $\epsilon$, the
length-scale, $\alpha'$, and the scale of the metric, $\varphi$,
even though all of these entered the intermediate expressions.
The remaining terms in (\ref{su}) can be shown to vanish as the
cut-off is removed. They also vanish in the tensionless limit
for finite cut-off as we will see later. We will also need to consider functional integrals with mixed boundary conditions, i.e. where the boundary is divided into sections where Dirichlet conditions are imposed and sections where Neumann conditions are imposed. The result generalises so that the right hand side of (\ref{reqres}) receives contributions from just the Dirichlet sections of the boundary (Appendix A). 

%\subsection{Reformulation of QED}
\bigskip
\noindent Since the expectation value of the delta-function decouples
from the scale of the metric we have a chance of
being able to construct an interacting string theory based on
(\ref{Sem}). More specifically we will consider the system
consisting of a number, $N$, of surfaces $\{\Sigma_i\}$. These have boundary components including curves $w^\mu_i$. The curves can be either closed or open, in which case we impose Neumann boundary conditions on $X^\mu$ on the remaining boundary components of $\Sigma$.
The action is
$S_f=\sum_i S_i+\sum_{i,j} S_{ij}$ 
with 
$S_i=S[X_i,g_i]$
and
\be
S_{ij}=
{q^2\over 4\epsilon_0^2}\int_{\Sigma_i,\Sigma_j}
d\Sigma_i^{\mu\nu}(\xi)\,\delta^4\left(X_i(\xi)-X_j(\xi')\right)\,
d\Sigma_{j}^{\mu\nu}(\xi')
\ee

\bigskip
\noindent Our aim is to show that this interacting string theory reproduces the functional integral over the electromagnetic field, i.e. that
\be\int \prod_{i=1}^N {\mathscr{D}(X_i,g_i)\over Z_0}\, e^{-S_f}\label{maino}
\ee
is the same as
\be
\int {{\mathscr{D}}A\over N}\,e^{-S_{gf}}\,\prod_i
e^{-i\int dw_i\cdot A}=
\prod_{i,j}\, e^{-{q^2\over
4\epsilon_0^2}\int{dw_i^\mu \Delta_{\mu\nu} \,dw_j^\nu}}
\label{main}
\ee
where the Maxwell theory is gauge-fixed in the gauge $\partial \cdot A=0$ so that the propagator, $\Delta_{\mu\nu}$, has Fourier transform
$\delta_{\mu\nu}/k^2-k_\mu k_\nu/ (k^2)^2$. It will turn out that we are unable to completely establish this result for the case of the bosonic string due to the possible appearance of unwelcome divergences. However the world-sheet supersymmetry of the spinning string provides sufficient structure to eliminate these, allowing us to prove the supersymmetric generalisation. It is precisely this generalisation, in which the super-Wilson loop appears, that is needed for electric charges with spin. So it appears that this string model has a preference for the realistic case of spinor QED over that of scalar QED.

\bigskip
\noindent This can be made the basis of quantising the electromagnetic field using the world-line formulation of Strassler\cite{Strass1} elaborated by a number of authors\cite{Schu, McKeon, Bast} and which can be extended to include multiple loops\cite{Sch}. For simplicity begin with scalar electrodynamics with a single complex field $\phi$ coupled to electromagnetism with (Euclidean) action
\be
S_\phi=\int d^4x \,\bar\phi\left(-{\cal{D}}^2+m^2\right)\phi,\quad {\cal{D}}=\partial+iA
\ee
The generating functional for Green functions\footnote{Later we will include a source for the gauge field, but for the time being we suppress this to simplify our expressions}
\be
Z[\bar J, J]=\int {\mathscr{D}}(A,\bar \phi,\phi)\,e^{-S_\phi-S_{gf}+
\int d^4x\left(\bar J\phi+\bar\phi J\right)}
\ee
can be computed by first integrating over the scalar field leaving
\be
Z[\bar J, J]=\int {\mathscr{D}}A\,e^{-S_{gf}-{\rm log\,Det}\left(-{\cal{D}}^2+m^2\right)+
\int d^4x\bar J   \left(-{\cal{D}}^2+m^2\right)^{-1}      J}
\ee
which we expand as
\be
\sum_{r,s=1}^\infty{1\over r!s!}\int {\mathscr{D}}A\,e^{-S_{gf}}
\left(-{\rm log\,Det}\left(-{\cal{D}}^2+m^2\right)\right)^r\left(\int d^4x\bar J   \left(-{\cal{D}}^2+m^2\right)^{-1}      J\right)^s
\label{expanded}
\ee
The functional determinant and the inverse of the operator $\left(-{\cal{D}}^2+m^2\right)$ are then expressed as world-line functional integrals that have simple, exponential dependence on the gauge field. We now give reparametrisation invariant versions of the expressions used by Strassler. The Green function can be represented as 
\be
\left(-{\cal{D}}^2+m^2\right)^{-1}({ b},{a})=\int  {{\mathscr{D}} (h,{w})\over Z}\,e^{-S[w,h]}\,,\label{avnew}
\ee
with\cite{BdVH}
\be
S[w,h]=S_0[w,h]+i\int dw\cdot A;\quad S_0[w,\,h]=
\int_0^1 \left({1\over 2}{h^{-1}(\xi)}\,{d{ w(\xi)}\over 
d\xi}^2 +m^2\right)\,\sqrt{h(\xi)}\,d\xi 
\,,\label{bbdvs}
\ee
where ${ w(\xi)}$, $0\le \xi\le 1$ is a parametrisation of a world-line from ${ a}$ to ${b }$ depending on the arbitrary parameter $\xi$ and $h(\xi)>0$ is an intrinsic metric on the world-line. We take $h$ to have the dimensions of $[\rm length ]^4$ so that $S[w,h]$ is dimensionless. 
(\ref{bbdvs}) 
is invariant under diffeomorphisms $\xi\rightarrow\tilde\xi$ that preserve the parameter interval provided that $h(\xi)$ transforms as a metric $h(\xi)\,d\xi^2\rightarrow \tilde h(\tilde\xi)\,d\tilde\xi^2=h(\xi)\,d\xi^2$ and
${w(\xi)}\rightarrow \tilde{w}(\tilde\xi)
={w(\xi)}$. As we show in Appendix B, gauging fixing this parametrisation invariance\footnote{The gauge-fixing procedure follows from integrating with respect to $T$ equation  (2.12) of \cite{Mansfield:2011eq}} by choosing $h=T^2$, a constant, leads for open world-lines to ${\mathscr{D}}h\propto dT$ so that (\ref{avnew}) becomes
\be
\int_0^\infty dT
\left\{
\int {\mathscr{D}}{w}\,e^{-\int_0^T dt\,\dot{w}^2/2-m^2 T-i\int dw\cdot A}\right\}
\ee
The expression in braces is the functional integral representation of the heat-kernel for the operator $\left(-{\cal{D}}^2+m^2\right)$, i.e. $\exp \left(-T(-{\cal{D}}^2+m^2\right)) $, and integrating this with respect to $T$ gives the Green function as claimed. Similarly we can express the logarithm of the functional determinant as
\be
{\rm Tr\,log}\left(-{\cal{D}}^2+m^2\right)
=-
\int_0^\infty {dT\over T}\int d^4{ a}\left.
\left\{
\int {\cal D} { w}\,e^{-\int_0^T dt\,\dot{w}^2/2-m^2 T-i\int dw\cdot A}\right\}\right|_{{ a}={b} }
\nonumber
\ee
which has a reparametrisation form that is very similar to that of the right hand side of (\ref{avnew}), i.e.
\be
-\int {{\mathscr{D}} (h,{ w})\over Z}\,e^{-S[w,h]}\label{avnew2}
\ee
except that we have to integrate over closed curves which implements the functional trace and generates the inverse power of $T$. It is a curious and useful fact that the two components of (\ref{expanded}) that we need, i.e. the logarithm of the functional determinant and the Green function take on identical forms when expressed as reparametrisation invariant functional integrals, differing only by the topology of the world-lines to be integrated over. This is discussed in Appendix B. Using these representations we re-write (\ref{expanded}) as
\begin{align}
\sum_{r,s=1}^\infty{1\over r!s!}\int {\mathscr{D}}A\,e^{-S_{gf}}
&\prod_{j=1}^{r+s}{{\mathscr{D}} (h_j,{w}_j)\over Z}\,e^{-S_0[{w}_j,\,h_j]}\nonumber \\
\times &\!\!\!\prod_{k=r+1}^{r+s}
e^{-i \oint d{w}_{k}\cdot A} \prod_{l=1}^{s}
\int d^4b_{l}\,d^4a_{l}\bar J (b_{l})  e^{-i \int _{a_{l}}^{b_{l}}d{w}_{l}\cdot A}   J(a_{l})\label{intooo}
\end{align}
Were (\ref{maino}) equivalent to (\ref{main}) it could be used to represent the integral over the gauge-field as a set of integrals over surfaces
\be 
\sum_{r,s=1}^\infty{1\over r!s!}\int \prod_{l=1}^{s}
 d^4b_l\,d^4a_l\,\bar J (b_l)   J(a_l)
\prod_{i=1}^{r+s }{\mathscr{D}(X_i,g_i)\over Z_0}\, e^{-S_i}
\prod_{j=1}^{r+s}{{\mathscr{D}} (h_j,{w}_j)\over Z}\,e^{-S_0[{w}_j,\,h_j]}\prod_{i,j=1}^{r+s}e^{-S_{ij}}
\ee
Re-arranging this slightly we would have arrived at 
\be
Z[\bar J, J]=\sum_{r,s=1}^\infty{1\over r!s!}\int \prod_{l=1}^{s}
 d^4b_l\,d^4a_l\,\bar J (b_l)   J(a_l)
\prod_{i=1}^{r+s }{\mathscr{D}(X_i,g_i,h_i,{w}_i)\over ZZ_0}\, e^{-S_i-S_0}
\prod_{i,j=1}^{r+s}e^{-S_{ij}}\label{physo}
\ee
Although we will be unable to demonstrate the equivalence of (\ref{maino}) and (\ref{main}) in scalar QED we will demonstrate an exact relation for spinor QED. The physical interpretation of (\ref{physo}) would be that the field theory is equivalent to an indeterminate number of strings described by the usual free action, $S_i\equiv S[X_i,g_i]$, augmented with a boundary term $S_0[w,h]$ interacting with each other via the contact term
$S_{ij}$. There is some freedom in how we associate the curves $w_i$  to the world-sheet surfaces. For want of an obvious alternative we choose the simplest assignment by associating a zero genus surface to each each $y_i$. Thus the ${\rm log\,Det}\left(-{\cal{D}}^2+m^2\right)$ factors correspond to closed curves bordering world-sheets which together describe particle anti-particle pairs connected by lines of force. The Green function factors $\left(-{\cal{D}}^2+m^2\right)^{-1}$ correspond to world-sheets that have mixed boundary conditions, Dirichlet conditions for the curves $w_i$ which run from $a_i$ to $b_i$ and Neumann everywhere else, so these describe strings with the usual string theory Neumann conditions at one end and a charged particle (or anti-particle) at the other.

\bigskip
\noindent We return to the issue of including a source for the gauge field. Rather than using a general source we limit attention to one that generates scattering amplitudes via the LSZ procedure by shifting the gauge-fixed Maxwell action, $S_{gf}$, in (\ref{intooo})
\be
S_{gf}\rightarrow S_{gf}-{1\over q^2}\int d^4 x \,{\cal A}\cdot \partial^2 A
\ee
where the source is on-shell, i.e. $\partial^2{\cal A}=\partial\cdot{\cal A}=0$. (We revert to Lorentzian signature briefly to be able to invoke LSZ). 
\bigskip

\noindent We show in Appendix D that
\be
\int {{\mathscr{D}}A\over N}\,e^{-S_{gf}+{1\over q^2}\int d^4 x \,{\cal A}\cdot \partial^2 A-i\sum_j \int dw_j\cdot A}
=
\prod_{i,j}\, e^{-{q^2\over
2}\int{dw_i^\mu \Delta_{\mu\nu} \,dw_j^\nu}}\prod_i
e^{-i\int dw_i\cdot {\cal A}}
\label{main'}
\ee
so that the effect of including the source $\cal A$ is simply to add a term to the boundary part of the action:
\be
S_0[w,h]\rightarrow S_0[w,h]+i\int dw\cdot {\cal A}
\ee
We note in passing that if we were to consider the generating functional for scattering amplitudes of charged particles and anti-particles then we would replace the source terms $\bar J\phi$ and $\bar\phi J$ by
 $\bar J\{-\partial^2+m^2\}\phi$ and $(\{-\partial^2+m^2\}\bar\phi )J$ leading to insertions of $\{-\partial^2+m^2\}$ in (\ref{physo}). These insertions could be generated by functional derivatives with respect to $\sqrt h$ at the ends of the curves $w$.

\bigskip
\noindent As we have said we will not be able to fully achieve our aim of showing the equivalence of (\ref{maino}) and (\ref{main}) until we include world-sheet supersymmetry. In any case QED with spin-one-half matter is more interesting as a realistic theory, and we will see that it emerges naturally from the spinning string. The Euclidean action for the Dirac field coupled to electromagnetism is
\be
S_\Psi=\int d^4x \,\bar\Psi\left(\gamma\cdot{\cal{D}}+im\right)\Psi\,.
\ee
The generating functional for Green functions
\be
Z_D[\bar K, K]=\int {\mathscr{D}}(A,\bar \Psi,\Psi)\,e^{-S_\Psi-S_{gf}+
\int d^4x\,\left(\bar K\Psi+\bar\Psi K\right)}
\ee
can be computed by first integrating over the Dirac field leaving
\be
Z_D[\bar K, K]=\int {\mathscr{D}}A\,e^{-S_{gf}-{\rm log\,Det}\left(-{(\gamma\cdot\cal{D})}^2+m^2\right)+
\int d^4x\bar K   \left(\gamma\cdot{\cal{D}}+im\right)^{-1}      K}
\ee
which we expand as
\be
\sum_{r,s=1}^\infty{1\over r!s!}\int {\mathscr{D}}A\,e^{-S_{gf}}
\left(-{\rm log\,Det}\left(-{(\gamma\cdot\cal{D})}^2+m^2
\right)\right)^r\left(\int d^4x\bar K   \left(\gamma\cdot{\cal{D}}+im \right)^{-1}K\right)^s
\label{expandedspin}
\ee
As in the scalar case we can represent the two components of this expression, the functional determinant and the Green function, as functional integrals of the same form but for closed and open worldlines respectively.
\bigskip

\noindent The worldline formulation of spin-one-half matter goes back to Feynman, but we need reparametrisation invariant expressions so we use the action introduced by Brink, di Vecchia and Howe\cite{BdVH}. For simplicity we take the mass to be zero. They augment the scalar action $S_0[w,h]$, (\ref{bbdvs}), with a piece containing Grassmann numbers $\psi^\mu$ that will play the role of Dirac $\gamma$-matrices and $\chi$ that is a fermionic partner to $\sqrt h$. In Euclidean signature it is
\be
S_F=-{1\over 2}\int_0^1 \left(\psi\cdot {d \psi\over d\xi}+
{\chi \over\sqrt h} {d w\over d\xi}\cdot \psi\right)
\ee
This is reparametrisation invariant provided $\psi^\mu$ transforms as a world-line scalar (like $w^\mu$) and $\chi$ transform like $\sqrt h$:
$$
\delta_V \chi(\xi)=-\left\{V(\xi){d\over d\xi}+{dV(\xi)\over d\xi}\right\}\,\chi(\xi),
$$
and the complete action is also invariant under the local supersymmetric transformations:
\be
\delta_\alpha w=\delta\alpha\psi\,,\quad 
\delta_\alpha \psi={\delta\alpha\over\sqrt h}\left ({d w\over d\xi}
-{1\over 2}\chi\psi\right)\,,
\quad
\delta_\alpha \sqrt h=\delta\alpha\chi\,,\quad
\delta_\alpha\chi=2{d\, \delta\alpha\over d\xi}\,.
\label{wlsusy}
\ee
Supersymmetry is preserved when we include a coupling to a gauge-field provided we include an extra term in the action involving the field-strength
\be
S_A=i\int \left ({dw\over d\xi} \cdot A+{1\over 2}F_{\mu\nu}\psi^\mu\psi^\nu\sqrt h \right)\,d\xi
\ee
We will see later that this new coupling arises naturally if we take the world-sheet theory to be supersymmetric.
\bigskip

\noindent In Appendix C we show that for closed world-lines
\be\int {\mathscr{D}} (h,w,\chi,\psi)\,e^{-S_0-S_F-S_A}
=-{\rm ln\,Det}\left(\left(\gamma\cdot\cal D\right)^2\,\right)
\ee
whilst for open world-lines running from $w_i$ to $w_f$
\be
\int {\mathscr{D}} (h,w,\chi,\psi)\,e^{-S_0-S_F-S_A}\Big|_{ab}
=\langle w_f,a|\,\left(\gamma\cdot{\cal D}\right)^{-1}
|w_i,b\rangle
\ee
where the spinor indices $a,b$ correspond to boundary conditions on the $\psi$ integral. So now we can express the generating functional for the Dirac field, $Z_D$, as
\bq 
\sum_{r,s=1}^\infty&&{1\over r!s!}\int {\mathscr{D}}A\,e^{-S_{gf}}
\prod_{j=1}^{r+s}{{\mathscr{D}} (h_j,{w}_j,\chi_j,\psi_j)\over Z}\,e^{-S_0[{w}_j,\,h_j]-S_F[\psi_j,\,\chi_j]} \nonumber \\
&&\qquad\times\prod_{k=r+1}^{r+s}
e^{-i \left(\oint d{w}_{k}\cdot A+{1\over 2}\int F_{\mu\nu}\psi^\mu\psi^\nu\sqrt h d\xi\right)
}\nonumber\\
&&
\qquad\times\prod_{\hphantom{+}l=1\hphantom{r}}^{s}
\int d^4b_{l}\,d^4a_{l}\,\bar K (b_{l})\,  e^{-i\left( \int _{a_{l}}^{b_{l}}d{w}_{l}\cdot A +{1\over 2}\int F_{\mu\nu}\psi^\mu\psi^\nu\sqrt h d\xi\right) }   K(a_{l})\label{intooD}
\eq
which contains the expectation value of supersymmetric exponentiated line integrals generalising the bosonic case. We will show that the these expectation values can be calculated by introducing fermionic degrees of freedom onto the worldsheets spanned by the open and closed curves representing the Green functions and determinants in the field theory. That is,
\begin{equation}
	\prod_{i = 1}^{n} \frac{\mathscr{D}\left(X_{i},\psi_{i},g_{i}\right)}{Z_{0}} e^{-S_{s}} = \int \frac{\mathscr{D}A}{N}e^{-S^{\prime}_{gf}}\prod_{i}e^{-S_{A}}
	\label{mainf}
\end{equation}
where $S^{\prime}_{gf}$ is the equivalent gauged fixed action for the fermionic quantum theory and $S_{s}$ is the action for the spinning string augmented by a supersymmetric generalisation of the contact interaction discussed above. We shall give explicit expressions for these objects when dealing with the spinning string in section \ref{susy}. This equivalence can then be used to rewrite the integral over the gauge field in equation (\ref{intooD}) in terms of open and closed spinning strings with contact interactions.

\bigskip
\noindent The main burden of this paper is to investigate the relationship between (\ref{maino}) and (\ref{main}) for the bosonic theory and then establish the supersymmetric version (\ref{mainf}), showing that spinor QED is equivalent to tensionless spinning strings\footnote{The tensionless limit of bosonic string theory has a degenerate worldsheet metric\cite{Karl} and as such can be reformulated on the level of the action by introducing a vector density to replace the metric in Polyakov's formulation\cite{Amor, Lind1}. The equation of motion of this auxiliary field imposes the null-metric condition and the formulation extends to the spinning string\cite{Lind2}. In the spinning case however the metric is no longer degenerate. Here we prefer to keep the tension arbitrary throughout the calculation to demonstrate how the tensionless limit suppresses unwanted quantities.} with a contact interaction. We use the perturbative expansion in powers of $q^2$, building on the
result (\ref{sol}). We begin with the purely bosonic theory. In section \ref{GenExp} we describe some basic tools including the regulator, and apply these to the derivation of (\ref{reqres}). In section \ref{calc0} we consider the first order in perturbation theory, studying potential divergences in some detail as a warm-up for higher order calculations, and also discuss how the split in the action (\ref{Sem}) between the free string action and the contact term is affected by the regulator. Higher orders in perturbation theory for the bosonic case are discussed in section \ref{Hoc} which includes a discussion of potential problems associated with divergences that might be generated when the interaction terms approach each other close to the world-sheet boundary. Concluding that our bosonic string model is incomplete we turn to the more realistic case of spinor matter and show that this is naturally described by the spinning string in section \ref{susy}. We discuss the gauge-fixed action and regulator and the 
residual supersymmetry, and then use this to 
restrict the divergences that can occur in the perturbative expansion enabling us to establish the connection between the spinning string model and spinor QED.

\section{General expectation values}
\label{GenExp}
Before proceeding to the evaluation of the partition function we describe our general approach to the computation of such functional integrals which will be essentially standard.
These functional integrals are computed conventionally by first integrating over the $X_i$ with
source terms to generate the insertions of
$\partial_{a}X^{\mu}\left(\xi\right)$ and the exponents
resulting from the Fourier decomposition of the
$\delta$-functions. So we consider
\begin{equation}
\mathcal{Z}\left(j,k\right) = \int\!\!\mathscr{D}_{g}X~
\exp{\left(-S[X,g]+\int d^2\xi \,J^{\mu} X^{\mu}\right)}
	\end{equation} 
	where
\begin{equation}
J^{\mu}\left(\xi\right) = -\partial_{a}j^{\mu a}\left(\xi\right)
+ i \sum_jk_j^{\mu} \left(\delta\left(\xi - \xi_j\right) \right)
	\label{currentJ}
\end{equation}
We write the field itself as the sum of three terms $X^{\mu} =
X_{c}^{\mu} + \tilde{X}^{\mu}+\bar X^{\mu}$ where $\bar X^{\mu}$ is the
quantum fluctuation to be functionally integrated over and
$X_{c}^{\mu}$ and $\tilde{X}^{\mu}$ satisfy Euler-Lagrange
equations. $X_{c}^{\mu}$ absorbs the boundary values of the
original $X$. Denoting the two-dimensional Laplacian as
$\Delta$:

\begin{equation}
-\frac{1}{\sqrt{g}}\partial_{a}\left(\sqrt{g}g^{ab}\partial_{b}X^{\mu}_{c}(\xi)\right)
\equiv \Delta X_c^{\mu}\left(\xi\right)=0;
\quad 
{X_c^{\mu}}|_{\partial D}=w^{\mu}, 
\end{equation}
and $\tilde X$ absorbs the sources we have just introduced
 
\begin{equation}
-\Delta \tilde{X}^{\mu}(\xi)=
2\pi\alpha^{\prime}\left(i\sum_jk_j^{\mu} \left(\delta\left(\xi
- \xi_j\right) \right) + \partial_{a}j^{\mu
a}(\xi)\right)
\end{equation} 
and is required to vanish on $\partial D$. $X_c$ and $\tilde X$ can both be found in terms of the Green function for the Laplacian with Dirichlet boundary conditions (which satisfies $\Delta G(\xi,
\xi') = \delta^{2}\left(\xi - \xi'\right)/\sqrt{{g}}$, $G(\xi,
\xi') =0$ for $\xi$ or $\xi'\in \partial D$):

\begin{equation}
X^{\mu}_{c}\left(\xi\right) = \oint_{\partial
D}\!\!d\tilde{\xi}^{c}~\epsilon_{ac}\sqrt{\tilde{g}}\tilde{g}^{ab}\tilde{\partial}_{b}G\left(\xi,
\tilde{\xi}\right) w^{\mu}(\tilde{\xi}),~\quad 
\tilde X\left(\xi\right)=-2\pi\alpha^{\prime}\int d^2\tilde{\xi}~G(\xi,
\tilde\xi)\,J^\mu(\tilde\xi)
\end{equation}
\bigskip
\noindent Integrating out the
quantum fluctuation generates a determinant so
\begin{align}
\mathcal{Z}\left(j, k\right) =&
\exp{\left.\bigg( \!-\!\pi\alpha^{\prime}\sum_{rs}k_r\cdot
k_s\,G(\xi_{r}, \xi_{s})\right.} +S[X_c,g]-2\log ({\rm Det}
\Delta)
\nonumber\\&
+ 2\pi\alpha^{\prime}i\left.\int d^{2}\xi~ j^{\mu a}
\sum_rk_r^{\mu} \partial_{a} G\left(\xi, \xi_r\right) \right.
%\nonumber \\&
\left.
+ ~2\pi\alpha^{\prime}\int\!\!\int d^{2}\xi d^{2}\xi'
~j^{\mu a} j^{\mu b} \partial _{a} \partial_{b} G\left(\xi,
\xi'\right) \right.\nonumber \\
&\left. + \int d^{2}\xi~ j^{\mu a}
\partial_{a}X_{c}^{\mu}\left(\xi\right) + i\sum_rk_r\cdot
X_{c}\left(\xi_r\right) \right)
		\label{Z0} 
\end{align}
$\log ({\rm Det} \Delta))$depends only on the scale of the
metric, and not on the sources, so will cancel out of our
computations because of the decoupling of the expectation value of the delta function discussed in the introduction. An alternative approach would be to assume the existence of further internal degrees of freedom to cancel the dependence on the Liouville mode.

\bigskip
\noindent The Green function is divergent at coincident points and so we replace it with a regulated version constructed from the heat-kernel
\begin{equation}
G_{\epsilon}\left(\xi, \xi'\right) = \int_{\epsilon}^{\infty}
\!d\tau~\mathcal{G}\left(\xi, \xi; \tau\right)\,,
\quad
{\partial\over\partial \tau}\mathcal{G}=-\Delta \mathcal{G}\,,
\quad
\mathcal{G}\left(\xi, \xi'; 0^{+}\right) = \delta^{2}\left(\xi
- \xi'\right)/\sqrt g
\end{equation}
This cut-off
procedure is reparameterisation invariant since the definition
of the kernel and Green function do not require a choice of
coordinates. It is not, however, Weyl invariant since it
introduces a distance cut off. 
The effect of
$\epsilon$ is to modify for high modes the spectral decomposition of the
Green function in terms of the eigenfunctions of $\Delta$, $u_{n}$, belonging to eigenvalues $\lambda_n$
to\cite{poly}
\begin{equation}
G_{\epsilon}\left(\xi, \xi'\right) = \sum_{n}
u_{n}\left(\xi\right)u_{n}\left(\xi'\right)\frac{e^{-\epsilon
\lambda_n}}{\lambda_n}.
\end{equation}
The short-distance divergence of the Green function is
associated with the short-time behaviour of the heat
kernel. Information can be extracted by expansion about a flat
metric because in a short time the heat from the 
delta-function source cannot diffuse too far meaning that the
heat kernel is sensitive to variations in the metric only over a distance of size on the order of $\sqrt{\epsilon}$. 

\bigskip

\noindent The general form of the heat kernel for small times can be determined
using the Seeley-DeWitt expansion\cite{SdW} which can be modified to take into account the presence of the boundary\cite{mcA1, mcA2}. If $\sigma$ denotes the square of the distance of the shortest path from $\xi$ to itself via a reflection from the boundary then the dominant small $\epsilon$ behaviour of the coincident Green function can be expressed as
\begin{align}
G_\epsilon(\xi,\xi)\equiv \psi\left(\xi \right) &\sim \int_{\epsilon}^{\infty}
\frac{d\tau}{4\pi\tau} \left(1 -
\exp{\left(-\frac{\sigma}{4\tau}\right)}\right)\left(1 +
\frac{1}{6}R\left(\xi\right)\tau\right) \label{psi1} \\
&= \begin{cases}\frac{\sigma}{16\pi\epsilon} - \frac{\sigma
\ln{\epsilon} R}{96\pi} & \sigma \ll \epsilon \\
\frac{1}{4\pi}\ln{\frac{\sigma}{4\epsilon}} - \frac{\epsilon
R}{24 \pi} & \sigma \gg \epsilon \end{cases} \label{psi2}
\end{align}
This reveals that as $\xi$ varies from being on the boundary to moving into the bulk, $\psi$ varies from $0$ to order $\log \epsilon$
over a distance $\epsilon^{\frac{1}{2}}$.

\bigskip

\noindent We will need the form of the
above functions in conformally flat gauge. We may choose complex
coordinates $z= x+iy$ with $ds^{2} = e^{\phi}dzd\bar{z}$. 
With this
choice $R\left(z\right) = e^{-\phi} \partial
\bar{\partial} \phi = 0$. For much of our work it will be sufficient to take $\phi$ to be constant and work on the
half-plane $y\ge 0$ whereby
\begin{equation}
	\sigma\left(z, z'\right) = e^{\phi} \left|z - z'\right|^{2}
\end{equation}
and for the coincident Green function the distance of the
shortest path from $z$ reflected from the boundary is $\sigma =
4e^{\phi} y^{2}$. So
\begin{align}
\mathcal{G}\left(z, z'; \tau\right) &=
\frac{1}{4\pi\tau}\left(\exp{\left(-\frac{e^{\phi}\left|z -
z'\right|^{2}}{4\tau}\right)} -
\exp{\left(-\frac{e^{\phi}\left|z -
\bar{z'}\right|^{2}}{\tau}\right)}\right) \\
\psi\left(\xi\right) &\sim \begin{cases} \frac{e^{\phi}y^{2}}{4
\pi \epsilon}& \sigma \ll \epsilon \\ \frac{1}{4\pi}
\ln{\frac{e^{\phi}y^{2}}{\epsilon}} & \sigma \gg \epsilon
\end{cases}
	\label{psiCases}
\end{align}
This provides a useful method to track the appearance of $\phi$
through the calculation. However in the more general case
$\sigma$ picks up non-trivial $\phi$-dependent corrections and
the heat-kernel picks up curvature corrections according to
(\ref{psi1}). These do not contribute to our calculation at leading order in $\epsilon$ so it will be sufficient to 
specialise to $\phi=0$ and  introduce a function $f$ by:

\begin{align}
G_\epsilon\left(z, z'\right) &= \int_{\epsilon}^{\infty}
\frac{d\tau}{4\pi\tau}\left(\exp{\left(-\frac{\left|z -
z'\right|^{2}}{4\tau}\right)} - \exp{\left(-\frac{\left|z -
\bar{z'}\right|^{2}}{4\tau}\right)}\right)\nonumber\\
&= \int_{\epsilon}^{\infty}
\frac{d\tau}{4\pi\tau}\left[\left(\exp{\left(-\frac{\left|z -
z'\right|^{2}}{4\tau}\right)} - 1\right) -
\left(\exp{\left(-\frac{\left|z -
\bar{z'}\right|^{2}}{4\tau}\right)}-1\right)\right] \nonumber \\
&= -f\left(\frac{\left|z - z'\right|}{2\sqrt{\epsilon}}\right) +
f\left(\frac{\left|z - \bar{z'}\right|}{2\sqrt{\epsilon}}\right)
\label{Greg}
\end{align}
where 
\be
f\left(s\right) = \int_{1}^{\infty} \frac{d\tau}{4\pi\tau}
\left(1-\exp{\left(-\frac{s^{2}}{\tau}\right)} \right),
\label{fdef}
\ee
so that $\psi(\xi)=f(y/{\sqrt \epsilon})$.
$f$ is monotonically increasing and can be approximated for small ($s<a$) and large ($s>b$) values of $s$ by
\be
f(s)\approx\begin{cases} \frac{s^{2}}{4\pi} & s<a \ll 1 \\
\frac{1}{4\pi}\ln{s^{2}} & s>b\gg 1 \end{cases}
	\label{fBound}
\ee

\bigskip

\noindent We illustrate this by revisiting (\ref{su}) to show that the unwanted terms on the right hand side vanish as the cut-off is removed and are also (independently) suppressed in the tensionless limit. In terms of the function $f$ the right hand side of (\ref{su}) becomes

\begin{align}
\epsilon^{ab} &\left(k^{[\mu}\partial_bX_c^{\nu ]} \,
2\pi\alpha'\partial_aG(\xi,\xi)\,
+\partial_a X_c^\mu\partial_bX_c^\nu\right) \,e^{ik\cdot
(x-X_c(\xi))}
e^{-\pi\alpha'k^2G(\xi,\xi)}\nonumber\\
=
&\left(-k^{[\mu}\partial_xX_c^{\nu ]} \,
2\pi\alpha'\partial_yf(y/{\sqrt \epsilon})\,+\epsilon^{ab}\partial_a X_c^\mu\partial_bX_c^\nu \right)\,e^{ik\cdot
(x-X_c(\xi))}
e^{-\pi\alpha'k^2f(y/{\sqrt \epsilon})}
\label{suk}
\end{align}
which is to be integrated over $y>0$ and over $k$. The first term leads to the required result (\ref{reqres}). Integrating the second over $k$ gives
\be
{\epsilon^{ab}\partial_a X_c^\mu\partial_bX_c^\nu \,e^{
-(x-X_c(\xi))^2/(4\pi\alpha'f(y/{\sqrt \epsilon}))}
\over
(\alpha'f(y/{\sqrt \epsilon}))^2}\label{ter}
\ee
As described in section 1 the integral over $y$ is suppressed outside a thin strip of width $\Lambda$, say, bordering the boundary. Taking $\Lambda>b\sqrt\epsilon$ shows that outside this strip $f(y/{\sqrt \epsilon})>{1\over 2\pi}\log (y/\sqrt\epsilon)$ which becomes large as $\epsilon\downarrow 0$ and so damps (\ref{ter}) provided $\Lambda/\sqrt\epsilon$ also becomes large (which can be arranged whilst still taking $\Lambda$ to zero). When we integrate (\ref{ter}) over the strip we can treat $X_c$ as a slowly varying quantity, independent of $y$ to leading order, leaving just the following integral to be computed, which we separate into three pieces
\begin{align}
\int_0^\Lambda dy\,{e^{
-(x-X_c(\xi))^2/(4\pi\alpha'f(y/{\sqrt \epsilon}))}
\over
(\alpha'f(y/{\sqrt \epsilon}))^2}&=\nonumber\\
\quad{\sqrt{\epsilon}\over\alpha'^2}\left( 
\int_0^a dy\,{(4\pi)^2e^{
-(x-X_c(\xi))^2/(\alpha' y^2)}
\over y^4}
\right.&+\left.\int_a^b dy\,{e^{
-(x-X_c(\xi))^2/(4\pi\alpha'f(y))}\over (f(y))^2}\right. \nonumber \\
&+\left.\int_b^{\Lambda/\sqrt\epsilon} dy\,{(4\pi)^2e^{
-(x-X_c(\xi))^2/(\alpha' \log y^2)}
\over (\log y^2)^4}\right)\label{into}
\end{align}
The first two integrals inside the brackets are independent of $\epsilon$ so the overall factor of $\sqrt\epsilon$ outside the brackets damps these terms as $\epsilon\downarrow 0$. The last term can be bounded:
\be
\left|\int_b^{\Lambda/\sqrt\epsilon} dy\,{(4\pi)^2e^{
-(x-X_c(\xi))^2/(\alpha' \log y^2)}
\over (\log y^2)^4}\right|<(\Lambda/\sqrt\epsilon-b)(4\pi)^2/(\log b^2)^4\,.
\ee
Combining this with the overall factor of $\sqrt\epsilon$ causes this to go to zero with $\epsilon$ because $\Lambda$ does. Consequently (\ref{into}) goes to zero as the cut-off is removed.

\bigskip
\noindent We note that these integrals also vanish independently in the tensionless limit. When $\alpha'$ is large in comparison to the length scale of the boundary the exponents in the last two integrals can be ignored and the first integral simplifies on scaling $y$ so that (\ref{into}) becomes
\be{\sqrt{\epsilon}\over\alpha'^2}\left( 
\alpha'^{3/2}\int_0^\infty dy\,{(4\pi)^2e^{
-(x-X_c(\xi))^2/y^2}
\over y^4}
+\int_a^b {dy\over (f(y))^2}
+\int_b^{\Lambda/\sqrt\epsilon} dy\,{(4\pi)^2
\over (\log y^2)^4}\right)\label{intoo}
\ee
which is suppressed in the tensionless limit, $\alpha^{\prime} \rightarrow \infty$, with the leading term coming from the first integral incorporating the small-$y$ behaviour. We shall now turn to apply similar techniques to calculating the effect of the interaction term in on the string theory partition function.

\section{The first order interaction of the bosonic theory}
\label{calc0}
\noindent In this section we will carry out the calculation to
first order in the expansion of the interaction term which is proportional to
$$
\sum_{j,k} \int \left(\prod_i {{\mathscr{D}}(X_i,g_i)\over Z_0}\,
e^{- S_i}\right)\int_{\Sigma_j,\Sigma_k}
d\Sigma_j^{\mu\nu}(\xi)\,\delta^4\left(X_j(\xi)-X_k(\xi')\right)\,
d\Sigma_{k}^{\mu\nu}(\xi').
$$
We shall show that the form of the coincident Green function suppresses the integrand for a general configuration of the points $\xi$ and $\xi^{\prime}$ except for two cases. The result we seek will arise when both points are separately close to the boundary where we have seen that the Green function is of order 1. Secondly, divergences appear when the points become close in the bulk of the worldsheet but we shall discuss how these can be interpreted in terms of a renormalisation of the free action and are consistent with the original splitting of the action in (\ref{Sem}). The question of the two points meeting one another close to the boundary will be discussed and this could provide corrections to the equality we are trying to prove. In the spinning string neither divergences nor unwanted boundary contributions will arise, as will be demonstrated in section \ref{susy}.

\bigskip
\noindent There are two kinds of term in this sum. The first is when
$j\neq k$, in which case we can organise the integrals to reduce
the computation to our previous result (\ref{sol}):
\begin{align}
&\sum_{j\neq  k} \int  {{\mathscr{D}}(X_j,g_j)\over Z_0}\, e^{- S_j}
 {{\mathscr{D}}(X_k,g_k)\over Z_0}\, e^{- S_k}
\int_{\Sigma_j,\Sigma_k}
d\Sigma_j^{\mu\nu}(\xi)\,\delta^4\left(X_j(\xi)-X_k(\xi')\right)\,
d\Sigma_{k}^{\mu\nu}(\xi') \nonumber \\
=&\sum_{j\neq k} \int  {{\mathscr{D}}(X_j,g_j)\over Z_0}\, e^{- S_j}
\int_{\Sigma_j} d\Sigma_j^{\mu\nu}(\xi)\,\left<\int_{\Sigma_k}
\delta^4\left(X_j(\xi)-X_k(\xi')\right)\,
d\Sigma_{k}^{\mu\nu}(\xi')\right>_{\Sigma_k}\nonumber \\
=&\sum_{j\neq k} \int  {{\mathscr{D}}(X_j,g_j)\over Z_0}\, e^{- S_j}
\int_{\Sigma_j} d\Sigma_j^{\mu\nu}(\xi)\,{1\over
4\pi^2}\left(\partial_\mu \int_{B_k} {dw_{k,\nu} \over
||x_j-w_k||^2}-\partial_\nu \int_{B_k} {dw_{k,\mu} \over
||x_j-w_k||^2}\right)
\end{align}
Applying Stokes' theorem and observing that the boundary $B_j$
is held fixed during the functional integration over $\Sigma_j$
reduces this to
\be
\sum_{j\neq k} {1\over 2\pi^2}\int_{B_j,B_k}{dw_j\cdot
dw_k\over||w_j-w_k||^2}
\ee

\bigskip
\noindent The second type of term that occurs in the sum has $j=k$, in
which case we have to consider
$$
\sum_{j} \int  {{\mathscr{D}}(X_j,g_j)\over Z_0}\, e^{- S_j}
\int_{\Sigma_j}
d\Sigma_j^{\mu\nu}(\xi)\,\delta^4\left(X_j(\xi)-X_j(\xi')\right)\,
d\Sigma_{j}^{\mu\nu}(\xi')
$$
where both integrals are over the same worldsheet. If, as before, we make a Fourier decomposition of the $\delta$-function 
\be
\int_{\Sigma}
d\Sigma^{\mu\nu}(\xi)\,\delta^4\left(X(\xi)-X(\xi')\right)\,
d\Sigma^{\mu\nu}(\xi')=\int {d^4k\over 64\pi^4}
\int d^2\xi \,d^2\xi'\,V_{-k}(\xi)\,V_{k}(\xi')
\ee
this requires the computation of $\left< V_{-k}(\xi)\,V_{k}(\xi')\right>_X$ which involves two insertions of $V$ on the world-sheet in contrast to the single insertion of (\ref{su}). To evaluate this we shall use Wick's
theorem, based on (\ref{Z0}), to write the expectation of  products of fields as an
expansion in terms of all possible contractions of $X$. The simplest is
\begin{equation}
\bcontraction{X^{\mu} X^{\nu} \,\cong \,\,:\!X^{\mu}X^{\nu}\!: + } {X}
{^{\mu}} {X}
	X^{\mu} X^{\nu} \,\cong \,\,:\!X^{\mu}X^{\nu}\!: + X^{\mu}X^{\nu},
\end{equation}
where by the normal ordering colons we mean that all
contractions have been carried out between the fields contained
within and the basic contraction is
\begin{equation}
\bcontraction{} {\tilde{X}} {^{\mu}\left(\xi\right)} {\tilde{X}}
%\bcontraction{\tilde{X}^{\mu}\left(\xi\right)\tilde{X}^{\nu}\left(\xi'\right)
%= \delta^{\mu\nu}G\left(\xi, \xi'\right) 
%\qquad \textrm{and}
%\qquad} {e} {^{ik\cdot \left(\tilde{X}\left(\xi\right) -
%\tilde{X} \right.}} {^{\left.\left(\xi'\right) \right)}}
%	%
\tilde{X}^{\mu}\left(\xi\right)\tilde{X}^{\nu}\left(\xi'\right)
= \alpha^{\prime}\delta^{\mu\nu}G\left(\xi, \xi'\right)
%\qquad e^{ik\cdot \left(\tilde{X}\left(\xi\right) -
%\tilde{X}\left(\xi'\right)\right)} =
%e^{-\pi\alpha^{\prime}k^{2}\Psi\left(\xi,\xi'\right)}.
\label{wicks}
\end{equation}
We use the $\cong$ sign to denote that the equality is meant to hold inside the functional integral $\langle \,\,\,\,\,\,\rangle_X$.
Because (\ref{Z0}) was obtained by expanding about a classical field $X_c$ that contains the information about the boundary value the expectation of the normal ordered part of the product contains $X_c$, thus
\begin{equation}
	\label{bound}	
	\left<:\!X^{\mu}X^{\nu}\!:\right>/\left<1\right> = X^{\mu}_{c}X^{\nu}_{c}.
\end{equation}
with similar expressions holding for greater numbers of
operators in the product. ($\left<1\right>$ is included because it contains boundary data $S[X_c,g]$ as well as functional determinants.)  The exponential
\begin{align}
e^{i k\cdot\left(X(\xi) - X(\xi')\right)}& \,\cong\,\, :e^{i k \cdot
\left(X(\xi) - X(\xi')\right)}: e^{-\pi\alpha^{\prime}k^{2}\Psi}
\label{exp}
\end{align}
with
\begin{equation}
\Psi\left(\xi, \xi'\right) = \psi\left(\xi\right) +
\psi\left(\xi'\right) - 2 G\left(\xi, \xi'\right).
\end{equation}
will be crucial in what follows. In the parametrisation of $D$ of the previous section this is 
\begin{align}
\Psi &= -f\left(0\right) +
f\left(\frac{y}{\sqrt{\epsilon}}\right) - f\left(0\right) +
f\left(\frac{y^{\prime}}{\sqrt{\epsilon}}\right) +
2\left(f\left(\frac{\left|z - z'\right|}{\sqrt{\epsilon}}\right)
- f\left(\frac{\left|z -
\bar{z'}\right|}{\sqrt{\epsilon}}\right)\right)\\
&= f\left(\frac{y}{\sqrt{\epsilon}}\right) +
f\left(\frac{y'}{\sqrt{\epsilon}}\right) +
2\left(f\left(\frac{\left|z - z'\right|}{2\sqrt{\epsilon}}\right)
- f\left(\frac{\left|z -
\bar{z'}\right|}{2\sqrt{\epsilon}}\right)\right)
\label{Psif}
\end{align}

\bigskip

\noindent Applying Wick's theorem to $\left< V_{-k}(\xi)\,V_{k}(\xi')\right>_X$ and carrying out the functional integration over $X$ gives (with $X(\xi)$ and $X(\xi')$ renamed as $X^{1}$ and $X^{2}$ for brevity)
\bigskip

\begin{equation}
\langle \,\!\epsilon^{ab}\epsilon^{cd} \partial_{a}^{1}
X^{1\left[\mu\right.} \partial_{b}^{1}X^{1\left.\nu\right]} e^{i
k \cdot \left(X^{1} -\, X^{2}\right)}
\partial_{c}^{2}X^{2\left[\mu\right.}\partial_{d}^{2}X^{2\left.\nu\right]}\,\rangle_X/\langle \,1\,\rangle_X
=\label{termslots}
\end{equation}
\begin{align}
\epsilon^{ab}\epsilon^{cd}~e^{i k \cdot \left(X^{1}_c -
\,X^{2}_c\right)} e^{-\pi\alpha^{\prime}k^{2}\Psi\left(\xi^{1},
\xi^{2} \right)}\bigg(&\partial_{a}^{1}
X^{1\left[\mu\right.}_{c}
\partial_{b}^{1}X^{1\left.\nu\right]}_{c}
\partial_{c}^{2}X^{2\left[\mu\right.}_{c}\partial_{d}^{2}X^{2\left.\nu\right]}_{c}
\label{I}\tag{I}\\
+&2.4\pi\alpha^{\prime}~ik^{\left[\mu\right.}\partial_{a}^{1}
\Psi \cdot \partial_{b}^{1} X^{1\left.\nu\right]}_{c}
\partial_{c}^{2}
X^{2\left[\mu\right.}_{c}\partial_{d}^{2}X^{2\left.\nu\right]}_{c}\label{II}\tag{II}\\
+&\left(4\pi\alpha^{\prime}\right)^{2}ik^{\left[\mu\right.}
\partial_{a}^{1} \Psi \cdot 
ik^{\left[\mu\right.}\partial_{c}^{2} \Psi \cdot
\partial_{b}^{1}X^{1\left.\nu\right]}_{c}\partial_{d}^{2}X^{2\left.\nu\right]}_{c}
\label{III}\tag{III} \\
+&8.3. 4\pi\alpha^{\prime}
~\partial_{a}^{1}\partial_{c}^{2} G \cdot
\partial_{b}^{1}X_{c}^{1\nu}\partial_{d}^{2}X^{2\nu}_{c}
\label{IV}\tag{IV} \\
+&4.3 \left(4\pi\alpha^{\prime}\right)^{
2}~\partial_{a}^{1}\partial_{c}^{2} G \cdot  i
k^{\nu} \partial_{b}^{1} \Psi \partial_{d}^{2}X^{2\nu}_{c}
\label{V}\tag{V}\\
+&2.3 \left(4\pi\alpha^{\prime}\right)^{3}
~\partial_{b}^{1}\partial_{d}^{2} G \cdot ik^{\nu}
\partial_{a}^{1} \Psi \cdot  ik^{\nu}
\partial_{c}^{2} \Psi \label{VI}\tag{VI} \\
+&4.4.3 \left(4\pi\alpha^{\prime}\right)^{2}~\partial_{a}^{1}\partial_{c}^{2}
G\cdot\partial_{b}^{1}\partial_{d}^{2} G\bigg) 
\label{VII}\tag{VII}
\end{align}
where $G = G\left(\xi^{1}, \xi^{2}\right)$ and we have made use of the results of Appendix E.
\bigskip 

\noindent The exponential factor
$e^{-\pi\alpha^{\prime} k^{2} \Psi}$ depends on the configuration of the two points, as depicted in Figure 1,
and will be important. For generic values of $\xi^{1}$ and $\xi^{2}$  in $D$, neither close to the
boundary nor close to one another, $\Psi$ is of order
$\ln{\epsilon}$ so that its exponential damps the integrand. As
one of these points, say $\xi^{1}$, approaches the boundary
$\psi\left(\xi^{1}\right)$ becomes of order unity, but with
$\xi^{2}$ still in the bulk the factor of
$\psi\left(\xi^{2}\right)$ keeps $\Psi$ of order
$\ln{\epsilon}$. So the only values of $\xi^{1}$ and $\xi^{2}$ that lead to non-zero contributions as the cut-off is removed are those for which both points are close to the boundary or close to each other in the interior of $D$. We will describe these two cases separately in the next two sub-sections.

\begin{figure}[h]
\centering
\subfloat[][An arbitrary configuration with both points in the
bulk. $\Psi$ is of order $\ln{\epsilon}$.]
	{
		\def\svgwidth{0.4 \columnwidth}
		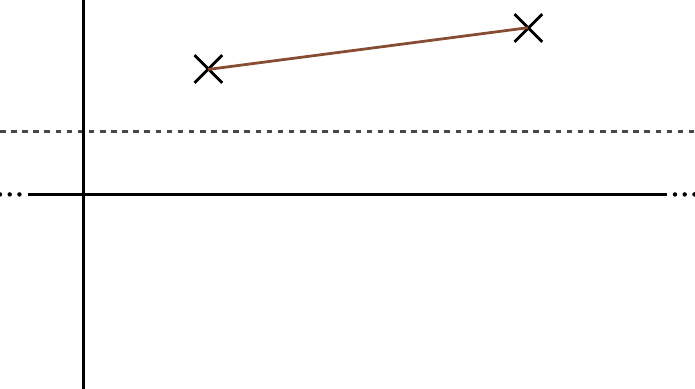
	}
	\hspace{20mm}
\subfloat[][The case that one of the points approaches the
boundary. The second point in the bulk holds $\Psi$ at order
$\ln{\epsilon}$.]
	{
		\def\svgwidth{0.4 \columnwidth}
		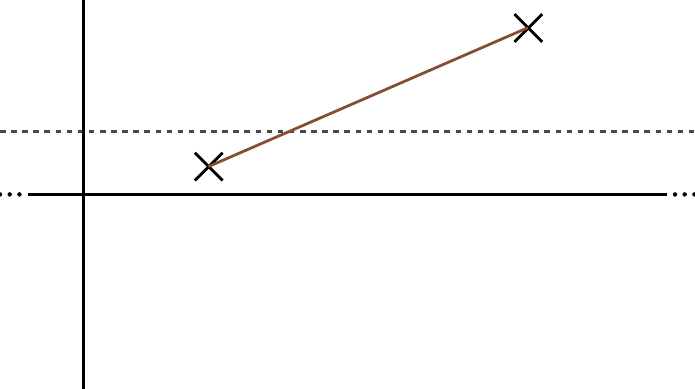
	}
	
\subfloat[][The case that both points are with a distance of
$\Lambda$ of the boundary. As each point is integrated through
this strip into the bulk $\Psi$ varies from order $1$ to order
$\ln{\epsilon}$.]
	{
		\def\svgwidth{0.4 \columnwidth}
		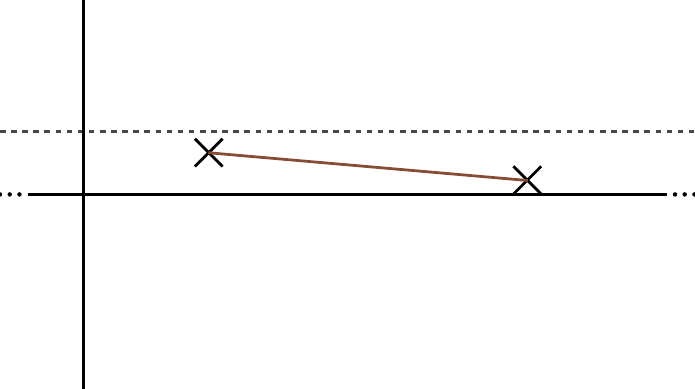
	}
	\hspace{20mm}
\subfloat[][The case that the two points are within a distance
$\Lambda$ of one another in the bulk. As one of the points is
integrated about this region $\Psi$ varies from order unity to
order $\ln{\epsilon}$.]
	{
		\def\svgwidth{0.4 \columnwidth}
		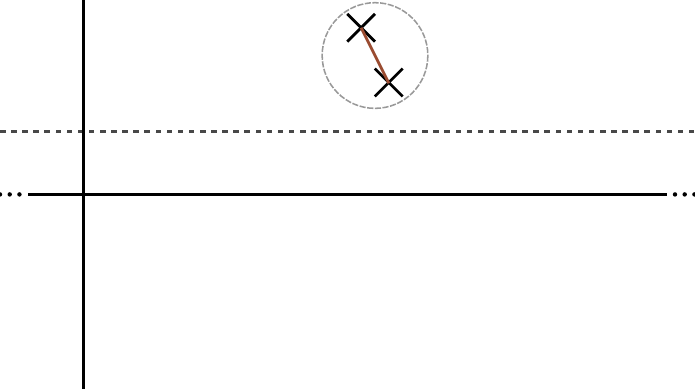
	}
\caption{The possible configurations of the two points within
the integration domain. The cases illustrated in the top line
lead to heavy suppression of the integrand and it is the bottom
two cases which will make finite contributions. The line joining
the points is to represent that they are linked by the
$\delta^{D}\left(X\left(\xi^{1}\right) -
X\left(\xi^{2}\right)\right)$ which will prove more useful at
higher order.}
	\label{figFirst}
\end{figure}

\bigskip

\noindent 
\subsection{Boundary contribution}
\label{bound1}
The first case to consider is $\left< V_{-k}(\xi)\,V_{k}(\xi')\right>_X$ with both points $\xi$ and
$\xi'$ close to the boundary. It is this case which will lead to our result. We need only integrate across a
small suitably chosen strip, say of size $\Lambda$, since the
integrand is suppressed moving into the bulk. In the parametrisation of the previous section $0<y<\Lambda$, $0<y'<\Lambda$, and we will also consider only the generic case of $|x-x'|>\Lambda$. For this configuration the rapidly varying functions in (\ref{termslots}) are just $\psi(\xi)$, $\psi(\xi')$ and their derivatives. In contrast, the fields, their
derivatives and the Green function between the two points all
vary smoothly and slowly. We shall consequently carry out this
part of the integration by replacing slowly varying fields with
their values on the boundary and then integrating the rapidly
varying fields into the bulk. These arise from contractions between the component pieces within each $V$ separately and not between them. We can anticipate the result by applying Wick's theorem to $V_k$ by itself as 

\bq
V_k\,\cong\,\,
\epsilon^{ab} \left(\colon k^{[\mu}\partial_bX_c^{\nu ]} \, e^{-ik\cdot
X}\colon
2\pi\alpha'\partial_a\psi
+\colon \partial_a X^\mu\partial_bX^\nu \,e^{-ik\cdot
X}\colon\right)
e^{-\pi\alpha'k^2\psi}
\label{sul}
\eq
(This is similar in form to (\ref{su}) because that equation is obtained from the expectation value of this). 
So, as $y$ and $y'$ are integrated over the strip, approximating the slowly varying functions in $\left< V_{-k}(\xi)\,V_{k}(\xi')\right>_X$ as constant means that we can approximate 
\begin{align}
\int_0^\Lambda dy~ V_k\,&\cong\,\,
\epsilon^{ab} \colon k^{[\mu}\partial_bX^{\nu ]} \, e^{-ik\cdot
X}\colon
2\pi\alpha'\int_0^\Lambda dy\,\partial_a\psi\,e^{-\pi\alpha'k^2\psi}\nonumber \\
&+\epsilon^{ab}\colon \partial_a X^\mu\partial_bX^\nu \,e^{-ik\cdot
X}\colon\int_0^\Lambda dy\,
e^{-\pi\alpha'k^2\psi}\nonumber
\end{align}
which parallels the derivation of (\ref{reqres}) so that by a similar argument we can neglect the second integral and compute the first to obtain (as the cut-off is removed)
\be
\int_0^\Lambda dy\, V_k\,\cong\,\,
-2\colon k^{[\mu}\partial_xX^{\nu ]} \, e^{-ik\cdot
X}\colon
/k^2,
\ee
so that the boundary contribution from the product of two vertex operators is
\be
\left< \int_{|y|<\Lambda} \!\!\!\!d^2\xi \,V_{-k}(\xi)\,\int_{|y'|<\Lambda} \!\!\!\!d^2\xi'\, V_{k}(\xi')\right>_X
={4\over (k^2)^2}\int_B\left<\colon k^{[\mu}dX^{\nu ]} \, e^{-ik\cdot
X}\colon \colon
k^{[\mu}dX^{\nu ]} \, e^{ik\cdot
X}\colon\right>_X
\ee
Contractions between the two normal ordered expressions involve the Green function that vanishes on the boundary, so in evaluating this expression we simply have to replace $X$ by its classical value $X_c$ which reduces to the boundary value $w$ on $B$. This gives (up to a factor of $\langle 1\rangle_X$)
\be
\int_{B}{dw\cdot dw' ~{e^{ik\cdot(w-w')}\over k^2}}
\ee
\bigskip
which is the required result.

\bigskip
\noindent We will now give a more careful treatment of the same calculation, based on the explicit expression (\ref{termslots}), to show that the less rapidly varying parts of (\ref{termslots}) do not change the result. Beginning with term (\ref{I}) we consider 
\begin{align}
\int\!\int dx dx^{\prime}\int_{0}^{\Lambda} dy\int_{0}^{\Lambda}
dy^{\prime}~
\epsilon^{ab}\epsilon^{rs}~\partial_{a}^{1}
X^{1\left[\mu\right.}_c
\partial_{b}^{1}X^{1\left.\nu\right]}_c 
\partial_{r}^{2}X^{2\left[\mu\right.}_c\partial_{s}^{2}X^{2\left.\nu\right]}_c \nonumber \\
\times e^{i k \cdot
\left(X^{1}_c - \,X^{2}_c\right)} e^{-\pi\alpha^{\prime}k^{2}\Psi\left(x,
x^{\prime}; y, y^{\prime} \right)}
\end{align}
The rapidly varying part of this integral is contained in
$\Psi\left(x, x^{\prime}; y,y^{\prime}\right)$ and for $\linebreak \left|x - x^{\prime}\right| > \Lambda$ the last two terms of (\ref{Psif}) are slowly varying and sum to zero on the boundary. Their subleading pieces are higher order in $\epsilon$ and, since we will find no divergences for this boundary case, will not be important. We are consequently left with the integral
\begin{equation}
\int\!\int dx
dx^{\prime}~\epsilon^{ab}\epsilon^{rs}~\partial_{a}
X_c^{1\left[\mu\right.}
\partial_{b}X_c^{1\left.\nu\right]}
e^{i k \cdot \left(w - \,w^{\prime}\right)}
\partial_{r}^{\prime}X_c^{2
\left[\mu\right.}\partial_{s}^{\prime}X_c^{2\left.\nu\right]}\int_{0}^{\Lambda} dy\int_{0}^{\Lambda} dy^{\prime}~
e^{-\pi\alpha^{\prime}k^{2}\Psi\left(x,x^{\prime}; y, y^{\prime}
\right)}
\end{equation}
Using (\ref{Psif}) the integrals over $y$ and $y^{\prime}$
factorise:
\begin{equation}
\int_{0}^{\Lambda} dy~ e^{-\pi\alpha^{\prime}k^{2}
f\left(\frac{y}{\sqrt{\epsilon}} \right)}\int_{0}^{\Lambda}
dy^{\prime}
~e^{-\pi\alpha^{\prime}k^{2}f\left(\frac{y^{\prime}}{\sqrt{\epsilon}}
\right)}
	\label{yInts}
\end{equation}
\begin{figure}
	\centering
	\def\svgwidth{0.35 \columnwidth}
	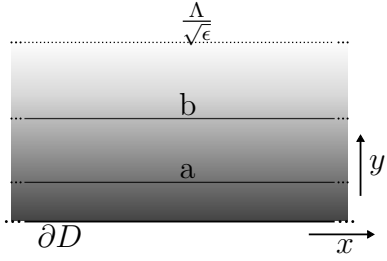
	\caption{The regions of interest for the integral into the bulk. The lower line represents the boundary $\partial D$. $a, b$ and $\Lambda$ are chosen to allow the application of the approximate forms of $\Psi$. From the boundary up to $a$ the lower approximation holds and from $b$ to $\frac{\Lambda}{\sqrt{\epsilon}}$ the upper approximation holds. In between there is no explicit form of $\Psi$ but integrals in this region are independent of $\epsilon$.}
	\label{figStrip}
\end{figure}

\bigskip
\noindent We make a change of
variables to scale out $\epsilon$, $\frac{y^{2}}{\epsilon}
\rightarrow y^{2}$, and then split the integral into the three
parts described in the previous section. This is illustrated in Fig. \ref{figStrip} and allows the use of (\ref{fBound}) in the first and
third regions:
\begin{align}
\int_{0}^{\Lambda} dy ~e^{-\pi\alpha^{\prime}k^{2} f
\left(\frac{y}{\sqrt{\epsilon}} \right)} &=
\epsilon^{\frac{1}{2}}
\int_{0}^{\frac{\Lambda}{\sqrt{\epsilon}}}dy~
e^{-\pi\alpha^{\prime}k^{2} \psi \left(y^{2} \right)} \nonumber
\\
&= \epsilon^{\frac{1}{2}}\left(\int_{0}^{a} dy
~e^{-\pi\alpha^{\prime}k^{2} \cdot \frac{1}{4\pi} y^{2}} +
\int_{a}^{b} dy
~e^{-\pi\alpha^{\prime}k^{2} f \left(y^{2} \right)} \right. \nonumber \\
&+\left.
\int_{b}^{\frac{\Lambda}{\sqrt{\epsilon}}}
dy~e^{-\pi\alpha^{\prime}k^{2} \cdot \frac{1}{4\pi}
\ln{\left(y^{2}\right)}}\right).
\end{align}
The first two terms on the bottom line have no divergences in their
integrands and so evaluate to some (k-dependent) constant
multiplied into $\epsilon^{\frac{1}{2}}$. The last term is
\begin{align}
\epsilon^{\frac{1}{2}}
\int_{b}^{\frac{\Lambda}{\sqrt{\epsilon}}}
dy~e^{-\frac{1}{4}\alpha^{\prime}k^{2} \ln{\left(y^{2}\right)}}
&= {\Lambda\over 1-
\frac{1}{2}\alpha^{\prime}k^{2}}\left({\frac{\Lambda}{\sqrt{\epsilon}}}\right)^{-
\frac{1}{2}\alpha^{\prime}k^{2}}-{\sqrt\epsilon}{b^{1-
\frac{1}{2}\alpha^{\prime}k^{2}}\over 1-
\frac{1}{2}\alpha^{\prime}k^{2}}
\end{align}
Both terms vanish as the cut-off is removed since $\Lambda$ also goes to zero in this limit and because $k^2\ge 0$. These terms multiply 
a corresponding contribution from $y^{\prime}$ 
with
identical $\epsilon$-dependence, so that overall the product goes to zero as the cut-off is removed.\footnote{We have worked with the Fourier transform, implying that we should integrate our final expressions over $k$ so there is a question as to whether this integral converges. To explore this we can in fact carry out
the $k$-integral first (as at the end of section 2) which we now do for the strip close to
the boundary, bearing in mind that our expression has the usual Fourier exponential $e^{ik\cdot l}$ where $l^\mu=w^\mu-w^{\prime\mu}$:
\begin{align}
\epsilon^{\frac{1}{2}}\int_{0}^{a} dy \int d^{D}k
~e^{-\pi\alpha^{\prime}k^{2} \cdot \frac{1}{4\pi}
y^{2}}e^{ik\cdot l} &\sim
\epsilon^{\frac{1}{2}} \int_{0}^{a} dy ~y^{-D}
e^{-\frac{2l^{2}}{\alpha^{\prime}y^{2}}} 
%\nonumber \\
%&\sim \epsilon^{\frac{1}{2}} \int_{a^{-1}}^{\infty} du~ u^{D-2}
%e^{-\frac{2l^{2}}{\alpha^{\prime}}
%u^{2}}
\end{align}
The final integral is well-defined for any
value of $D$ so this contribution vanishes as $\epsilon \rightarrow 0$. Furthermore we can consider the
same situation in the upper region of integration where we have
\be
\epsilon^{\frac{1}{2}}
\int_{b}^{\frac{\Lambda}{\sqrt{\epsilon}}} dy~\int d^{D}k~
e^{-\pi\alpha^{\prime}k^{2} \cdot \frac{1}{4\pi} \ln{y^{2}}}
e^{ik \cdot l} \sim
\epsilon^{\frac{1}{2}}
\int_{b}^{\frac{\Lambda}{\sqrt{\epsilon}}} dy
~\left(\ln{y}\right)^{-\frac{D}{2}}
e^{-\frac{2}{\alpha^{\prime}\ln{y}}l^{2}} 
\ee
which is bounded by $\Lambda$ multiplied by the greatest value of the integrand, which is in turn smaller than $(\ln b)^{-D/2}$ and so vanishes with the cut-off as required.}

\bigskip
\noindent For the remaining terms (II-VII) we shall
determine their $\epsilon$-dependence by picking out
the rapidly varying bits of each expression and evaluating the
integrals. The derivative structure of the above terms 
determines the $\epsilon$-dependence, since a derivative normal to the boundary cancels the factor of $\epsilon^{\frac{1}{2}}$ which arises under the scaling of $y$. So the terms which we
may expect to contribute to the expectation value will have two
derivatives of the rapidly varying $\Psi$; one with respect to
$y$ and one with respect to $y^{\prime}$. Since the Green function is slowly and smoothly
varying when the two points are not close together terms (\ref{IV}) and (\ref{VII}) actually have the
same rapid variation as term (\ref{I}) above, though they are
multiplied by different powers of $\alpha^{\prime}$. The $\epsilon$-dependence of terms (\ref{II}) and
(\ref{V}) is the same, whilst terms (\ref{III}) and (\ref{VI})
can be expected to share the same dependence.

\bigskip
\noindent We consider the $y$-integral of term (\ref{II}):
\begin{equation}
\alpha^{\prime}ik^{\mu} \int_{0}^{\Lambda}
dy~\epsilon^{ab}\partial_{a} \Psi \cdot e^{-\pi\alpha^{\prime}
k^{2} \Psi}
\end{equation}
Only the $f\left(\frac{y}{\sqrt{\epsilon}}\right)$ part of
$\Psi$ varies rapidly with $y$. Since this is a function of $y$
only, the non-zero contribution arises when $a = 2$ and the presence of this derivative makes the integrand
invariant to scaling:
\begin{align}
\alpha^{\prime}ik^{\mu}
\int_{0}^{\frac{\Lambda}{\sqrt{\epsilon}}} \partial_{y}
f\left(y\right) \cdot
e^{-\pi\alpha^{\prime}k^{2}f\left(y\right)} &= \frac{-2
ik^{\mu}}{\pi k^{2}} \int_{0}^{\frac{\Lambda}{\sqrt{\epsilon}}}
\partial_{y}\left(e^{-\pi\alpha^{\prime}k^{2}f\left(y\right)}\right) \nonumber \\
&=\frac{2i k^{\mu}}{\pi k^{2}} \left(1 -
\left(\frac{\Lambda}{\sqrt{\epsilon}}\right)^{-\frac{1}{2}\alpha^{\prime}k^{2}}\right).
\end{align}
The second term here vanishes as the regulator is removed because $k^{2} \ge 0$. We must also
combine the above answer with the $y^{\prime}$ integral which is
of the same form as that evaluated for term (\ref{I}). We find that their product vanishes as $\epsilon \rightarrow 0$, as will the contribution from term (\ref{V}).

\bigskip

\begin{figure}[h]
	\centering
	\def\svgwidth{0.35 \columnwidth}
	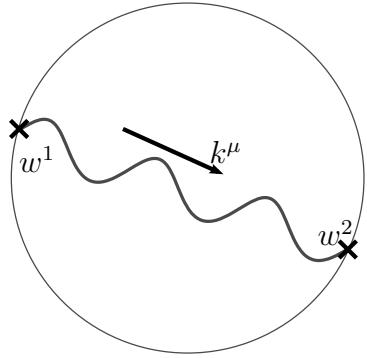
	\caption{At first order the correction is given by a massless vector propagator between the boundary points $x^{1}$ and $x^{2}$ to be integrated around the boundary with respect to both points.}
	\label{figO1}
\end{figure}

\noindent Both terms (\ref{III}) and (\ref{VI}) have two
derivatives acting on $\Psi$ so we expect to get two copies of the
form of the $y$-integral evaluated above. The $y$ and $y^{\prime}$ dependent part of Term (\ref{III}) take the
form
\begin{align}
\frac{1}{4} \alpha^{\prime 2}k^{\mu} k^{\mu} \partial_{y} \Psi
\partial^{\prime}_{y} \Psi e^{-\pi\alpha^{\prime}k^{2} \Psi}.
\end{align}
Reinstating the remaining boundary factors and the antisymmetry on
the worldsheet indices and scaling $\epsilon$ out of
the integrand gives
\begin{align}
&\frac{1}{4}\alpha^{\prime
2}k^{\left[\mu\right.}k^{\left[\mu\right.}\int\!\int_{\partial
D} dx dx^{\prime}~
\partial_{x}X_{c}^{1\left.\nu\right]}
\partial^{\prime}_{x} X_{c}^{2
\left.\nu\right]} e^{i k \cdot l}\int_{0}^{\frac{\Lambda}{\sqrt{\epsilon}}}
dy~ \partial_{y}f\left(y\right)
e^{-\pi\alpha^{\prime}k^{2}f\left(y\right)}~ \times \nonumber \\
&\phantom{\frac{1}{4}\alpha^{\prime
2}k^{\left[\mu\right.}k^{\left[\mu\right.}\int\!\int_{\partial
D} dx dx^{\prime}~
\partial_{x}X_{c}^{1\left.\nu\right]}
\partial^{\prime}_{x} X_{c}^{2
\left.\nu\right]} e^{i k \cdot l}}\int_{0}^{\frac{\Lambda}{\sqrt{\epsilon}}}
dy^{\prime } ~\partial_{y}f\left(y^{\prime}\right)
e^{-\pi\alpha^{\prime}k^{2}f\left(y^{\prime}\right)} \nonumber
\\
=&\frac{1}{4}\alpha^{\prime
2}k^{\left[\mu\right.}k^{\left[\mu\right.} \int\!\int_{\partial
D} dx dx^{\prime}~
\partial_{x}X_{c}^{1\left.\nu\right]}
\partial^{\prime}_{x} X_{c}^{2
\left.\nu\right]} e^{i k \cdot l} \int_{0}^{\infty} df
e^{-\pi\alpha^{\prime}k^{2}f} \int_{0}^{\infty} df'
e^{-\pi\alpha^{\prime}k^{2}f'} \nonumber \\
=&\int\!\int_{C} 
~\frac{d w^{\left[\nu\right.} d
w^{\prime\left[\nu\right.}k^{\left.\mu\right]}k^{\left.\mu\right]}}{\pi^{2} \left(k^{2}\right)^{2}}
e^{i k \cdot l}.
	\label{intProp}
\end{align}
On the second line we have made we removed the regulator taking $\epsilon \rightarrow 0$. The final step is to integrate over all
values of $k$ and
to apply the contraction of the target
space indices so that the full expression reads
\begin{equation}
\int\!\int_{B} d w \cdot d w^{\prime} \int d^{D}
k~\frac{2\left(D - 1\right)}{\pi^{2}k^{2}} e^{i k \cdot \left(w
-\, w^{\prime}\right)}
	\label{prop}
\end{equation}
We see here
the Fourier representation for a massless vector propagator integrated
around the boundary which is depicted in Fig \ref{figO1}. This result is independent of the metric on the worldsheet and thus on its scale and our integral over $k$ was not on-shell. We discuss
this further in the next section but first turn to the
calculation of the final term (\ref{VI}) and demonstrate that it
is in fact vanishing by our choice of coordinates.

\bigskip

\noindent The only contribution to term (\ref{VI}) arises when the derivatives of $\Psi$ are with respect to $y$ and $y^{\prime}$. This then contains
\begin{equation}
\alpha^{\prime 3}\delta^{\nu \nu}
\partial_{x}\partial^{\prime}_{x} G \frac{1}{4} k^{\mu} k^{\mu}
\partial_{y} \Psi \partial^{\prime}_{y} \Psi
e^{-\pi\alpha^{\prime} k^{2}\Psi},
	\label{fullC2}
\end{equation}
The smoothly varying field $G$ has a Taylor expansion based at
the boundary where its value is identically zero. The partial
derivatives $\partial_{x}\partial^{\prime}_{x}G$ are then along
the boundary so vanish identically. All other contributions are
slowly varying and are subleading in $\epsilon$ so vanish as $\epsilon \rightarrow 0$. We have thus demonstrated
that there is only one contribution to the correlation functions
from close to the boundary -- (\ref{prop}). We postpone further
discussion of this result until we have considered the
contribution from the two points coming close together in the
bulk.

\subsection{Bulk divergences}
\label{bulk}

\noindent We now study what happens to $V_{-k}(\xi)\,V_{k}(\xi')$ as the two points $\xi$ and $\xi'$ approach each other but remain far from the boundary. Corresponding to the split between the free action and the interaction term in (\ref{Sem}) we will show that this leads to a renormalisation of the free action. This computation is also useful in considering the more general case that occurs at higher order of several $V_k$ approaching each other in the bulk.

\bigskip
\noindent We consider (\ref{termslots}) for $\xi$ close to $\xi'$ but far from the boundary, so that $\Psi$ can be separated into rapidly and slowly varying parts:
\bq
\Psi =
-2f\left(\frac{\left|z - z'\right|}{2\sqrt{\epsilon}}\right)
+\left(2f\left(\frac{\left|z -
\bar{z'}\right|}{2\sqrt{\epsilon}}\right)
-f\left(\frac{y}{\sqrt{\epsilon}}\right) -
f\left(\frac{y'}{\sqrt{\epsilon}}\right)\right)
\label{Psiff}
\eq
using the large distance behaviour (\ref{fBound}) this is
 \bq
\Psi =
-2f\left(\frac{\left|z - z'\right|}{2\sqrt{\epsilon}}\right)
+{1\over 2\pi}\log \left({(x-x')^2+(y+y')^2\over 4y y'}\right)
\label{Psifff}
\eq
We will integrate firstly over $\xi'$, keeping $\xi$ fixed. Then the first term in (\ref{Psifff}) varies rapidly over a disk with centre $\xi$ of size $\Lambda$, from $0$ at the centre to order $\log{(\Lambda/\sqrt\epsilon)}$ on the edge. $\Psi$ acts as a damping factor for $\xi'$ outside this disk if $\Lambda/\sqrt\epsilon$ is taken large as $\epsilon$ is taken to zero. The second is slowly varying and vanishes when the two points are coincident. The first subleading term is quadratic in $\left(x - x^{\prime}\right)$ so under the scaling we will carry out is of order $\epsilon$. The exponent $\exp{\left(i k \cdot \left(X - X^{\prime}\right)\right)}$ is unity at zeroth order in $\epsilon$ and its first correction is of order $\sqrt{\epsilon}$. We shall see that it is only for terms (\ref{V}), (\ref{VI}) and (\ref{VII}) that these corrections are relevant due to divergences which we will encounter for these terms. Our general strategy will be to concentrate on the rapidly varying parts of the integrands we 
need and to replace the slowly 
varying fields by their values at the point $\xi$.

%
%\begin{figure}
%	\centering
%	\def\svgwidth{0.4 \columnwidth}
%	\input{./bulkBar.pdf_tex}
%\caption{Fixing the gauge to complex coordinates in the upper
%half plane we demonstrate the integral of $z^{\prime}$ about
%$z$. The image point $\bar{z^{\prime}}$ is shown in the lower
%half plane. Expressions that do not depend on the separation $z
%- z^{\prime}$ are slowly varying will be replaced by their value
%at centre of the region where $z^{\prime} = z$. This includes
%dependence on $z - \bar{z^{\prime}}$, the distance between the
%point $z$ and the image of $z^{\prime}$: the distance $\left|z -
%\bar{z^{\prime}}\right|$ is approximately constant. We can thus
%also make the replacement $\bar{z^{\prime}} \rightarrow
%\bar{z}$; that is replace the image point of $z^{\prime}$ by
%that of $z$ at the centre of the region.}
%\label{figIm}
%\end{figure}

\begin{figure}
	\centering
	\def\svgwidth{0.3 \columnwidth}
	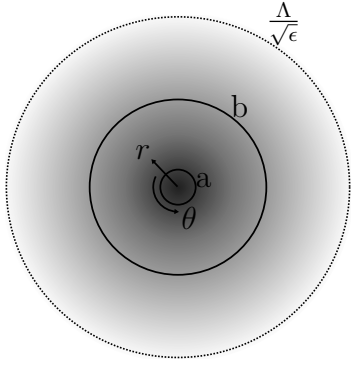
\caption{The regions of interest for the integral into the bulk.
The centre represents the coincidence of the two points in the
bulk. $a,~ b$ and $\Lambda$ are chosen to allow the application
of the approximate forms of $\Psi$. From the centre out to $a$
the lower approximation holds and from $b$ to
$\frac{\Lambda}{\epsilon}$ the upper approximation holds. In
between (the shaded part) there is no explicit form of $\Psi$
but integrals in this region are independent of $\epsilon$.}
	\label{figBulk}
\end{figure}

\bigskip

\noindent First we consider (\ref{I}). Instead of the two integrals with respect to $y$ and $y'$ that we had to consider for points close to the boundary in (\ref{bound1}) we now have to integrate over the disk. The rapidly varying parts of the integrand are
\begin{align}
\int dx dy \int_{ |z-z'|\leqslant
\Lambda}dx^{\prime}dy^{\prime}~e^{-\pi\alpha^{\prime}k^{2}
f\left(\frac{r}{\sqrt{\epsilon}}\right)} &= 2\pi\epsilon \int
dxdy\int_{0}^{\frac{\Lambda}{\sqrt{\epsilon}}} dr~r
e^{-\pi\alpha^{\prime}k^{2}f\left(r\right)}
	\label{rInt}
\end{align}
where we have used polars and scaled by $\sqrt\epsilon$. We split the integration region
into the three parts demonstrated in Fig. \ref{figBulk} enabling us to use the short and large-distance approximations for $f$. This provides
\begin{align}
2\pi \epsilon \int_{0}^{\frac{\Lambda}{\sqrt{\epsilon}}} dr~r
e^{-\pi\alpha^{\prime}k^{2}f\left(r\right)} &= 2\pi\epsilon\left(\int_{0}^{a} dr~r
e^{-\alpha^{\prime}k^{2}\frac{r^{2}}{4}} +
\int_{a}^{b} dr~r
e^{-\pi\alpha^{\prime}k^{2}f\left(r\right)}\right. \nonumber\\
&+\left. \int_{b}^{\frac{\Lambda}{\sqrt{\epsilon}}}
dr~r^{1 - \frac{1}{2}\alpha^{\prime}k^{2}}\right)
\end{align}
The explicit factors of $\epsilon$ multiplying the first two integrals cause these terms to vanish as the cut-off is removed. The final term evaluates to
\be
{\pi\over{1-\alpha'k^2/4}} \left(\Lambda\left({\Lambda\over\sqrt\epsilon}\right)^{-\alpha'k^2/2}-\epsilon b^{2-\alpha'k^2/2}\right)
\ee
which goes to zero as the cut-off is removed because $k^2\ge 0$ and $\Lambda$ goes to zero. 

\bigskip
\noindent Turning to (\ref{II}), the rapidly varying part that we have to integrate over the disk is 
\begin{equation}
\alpha^{\prime}\int^{r \leqslant \Lambda} dx^{\prime}dy^{\prime}
~\partial^{\prime}_{c} f\left(\frac{r}{\sqrt{\epsilon}}\right)
e^{-\pi\alpha^{\prime}k^{2}f\left(\frac{r}{\sqrt{\epsilon}}\right)}
	\label{bulkOneD}
\end{equation}
which vanishes by rotational symmetry. There are corrections to this arising from terms subleading in $\epsilon$ and also from the derivative acting on the slowly varying part of $\Psi$ but these have the same $\epsilon$ dependence as Term (\ref{I}) because they have the same rapidly varying content.

\bigskip 

\noindent 
We now find the terms which lead to renormalisation of the string action. Using $\partial_{a}g\left(r\right) = - \partial^{\prime}_{a}g\left(r\right)$ for any function of $r$, the rapidly varying part of (\ref{III}) can be written
\be
\alpha^{\prime 2}k^{\mu}k^{\alpha}\int dx dy \int^{r
\leqslant \Lambda} dx^{\prime}dy^{\prime} ~
\partial^{\prime}_{a}f\left(\frac{r}{\sqrt{\epsilon}}\right)
\partial^{\prime}_{c}f\left(\frac{r}{\sqrt{\epsilon}}\right)
e^{-\pi\alpha^{\prime}k^{2}f\left(\frac{r}{\sqrt{\epsilon}}\right)}
\ee
The integral over the primed variables must be proportional to
$\delta_{ac}$ by symmetry and we can extract the constant of
proportionality by contracting these indices; splitting up the
integration region once again gives
\begin{align}
\int^{r \leqslant \frac{\Lambda}{\sqrt{\epsilon}}}
dx^{\prime}dy^{\prime} ~ \partial^{\prime}_{a}f\left(r\right)
\partial^{\prime}_{a}f\left(r\right)
e^{-\pi\alpha^{\prime}k^{2}f\left(r\right)} &= 8\pi \int_{0}^{a}
dr~ r^{3} e^{-\alpha^{\prime}k^{2}\frac{r^{2}}{4}} \nonumber \\
 &+ 2\pi\int_{a}^{b} dr ~ r \partial^{\prime}_{a}f\left(r\right)
\partial^{\prime}_{a}f\left(r\right)
e^{-\pi\alpha^{\prime}k^{2}f\left(r\right)} \nonumber \\
&+ 8\pi\int_{b}^{\frac{\Lambda}{\sqrt{\epsilon}}} dr~ r^{-1 -
\frac{1}{2}\alpha^{\prime}k^{2}} 
\end{align}
which remains finite as the regulator is removed. There are again further contributions from the slowly varying fields but these vanish as we take $\epsilon \rightarrow 0$\footnote{Again we have still to integrate over $k$ and that leads to ultra-violet divergences which we can regulate by dimensional regularisation. Large $k$ corresponds to small $r$ and causes the first integral to diverge at the origin, but we keep the spacetime dimension $D$ arbitrary (in a range where the integral exists) and compute
\be
\int_{0}^{a}dr\int d^{D}k~ k^{\mu}k^{\alpha} r^{3}
e^{-\frac{1}{4}\alpha^{\prime}r^{2}k^{2}} 
=\delta^{\mu\alpha} {2^{D+1}a^{2-D}\over 2-D}\left({\pi\over\alpha'}\right)^{D/2+1}
\ee
which continues to the physical value of $D=4$.}. Putting this together with the slowly varying parts of (\ref{III}) gives a term proportional to

\begin{equation}
\alpha^{\prime -1}\int d^{2}\xi^{1}~ \delta^{ab}\partial^{1}_{a}
X^{\mu}_c \partial^{1}_{b} X^{\mu}_c
\end{equation}
which is simply a renormalisation of the free string theory action in the conformal gauge we have chosen. Note that this is suppressed in the tensionless limit.

\bigskip
\noindent The remaining terms involve derivatives of the
Green function and these are rapidly varying fields. However
it is possible to simplify matters by noting that 
\begin{equation}
\partial^{1}_{a}\partial^{2}_{c} G = -\frac{1}{2}
\partial^{1}_{a}\partial^{2}_{c}\Psi.
	\label{GBulk}
\end{equation}

\noindent 
Turning to term (\ref{IV}), the rapidly
varying piece is
\begin{equation}
\alpha^{\prime}\int^{r \leqslant \Lambda}
dx^{\prime}dy^{\prime}~\partial^{\prime}_{a}\partial^{\prime}_{c}f\left(\frac{r}{\sqrt{\epsilon}}\right)e^{-\pi\alpha^{\prime}k^{2}f\left(\frac{r}{\sqrt{\epsilon}}\right)}
	\label{IVDel}
\end{equation}
which by symmetry must be proportional to $ \delta_{ac}$, so it is sufficient to consider its trace. However, the defining equation of the
heat kernel implies that the function $f$ obeys
$\partial^{\prime}_{a}\partial^{\prime}_{a}f\left(r\right) =
\frac{1}{\pi}e^{-r^{2}}$ so that we may immediately calculate this as 
\begin{align}
\frac{1}{2\pi}\int_{0}^{\Lambda/\sqrt\epsilon}dr~
re^{-r^{2}}e^{-\pi\alpha^{\prime}k^{2}f\left(r\right)}
\end{align}
which is also finite as the cut-off is removed. As with the previous term the $X-$ dependence of the slowly varying contributions leads to a renormalisation of the free action which is also suppressed in the tensionless limit.

\bigskip

\noindent The analysis of the (\ref{V}) is more involved. Naively the calculation of the rapidly varying piece follows that of term (\ref{II}) because it vanishes by rotational invariance:
\begin{equation}
\alpha^{\prime 2}\int dx dy ~ik\cdot \partial_{d}X\int^{r \leqslant \Lambda}
dx^{\prime}dy^{\prime}~
\partial^{\prime}_{a}\partial^{\prime}_{c}f\left(\frac{r}{\sqrt{\epsilon}}\right)\partial^{\prime}_{b}f\left(\frac{r}{\sqrt{\epsilon}}\right)
e^{-\pi\alpha^{\prime}k^{2}f\left(\frac{r}{\sqrt{\epsilon}}\right)}=0.
	\label{rot3}
\end{equation}
However by scaling $r$ by $\sqrt{\epsilon}$ the three derivatives imply an overall factor of $1/\sqrt{\epsilon}$ so that we must expand the slowly varying fields beyond leading order to find contributions that could remain finite as the regulator is removed. This can be found from the expansion
\begin{align}
\exp\left[ik\cdot \left(X_{c} - X^{\prime}_{c}\right)\right)]
&=1-i \left[\left(x - x{\prime}\right)^{\alpha}\partial_{\alpha} X_{c}+\ldots \right]\cdot k \nonumber \\
&-{1\over 2}\left[\left(\left(x - x^{\prime}\right)^{\alpha}\partial_{\alpha} X_{c}+\ldots\right)\cdot k\right]^2+\ldots
\label{expand}
\end{align}
where after scaling the first subleading term is $-i\sqrt{\epsilon} \left(x - x{\prime}\right)^{\alpha}\partial_{\alpha} X_{c} \cdot k$. This offers a correction
\begin{equation}
	 \epsilon^{ab}\epsilon^{cd}\alpha^{\prime 2}\int dx dy ~ik\cdot \partial_{d} X~ i k\cdot \partial_{\alpha} X_{c} \int^{r \leqslant \frac{\Lambda}{\sqrt{\epsilon}}}
dx^{\prime}dy^{\prime}~ \left(x - x{\prime}\right)^{\alpha}
\partial^{\prime}_{a}\partial^{\prime}_{c}f\left(r\right)\partial^{\prime}_{b}f\left(r\right)
e^{-\pi\alpha^{\prime}k^{2}f\left(r\right)}.
\end{equation}
This can be integrated by parts to reduce it to the same form as ({III}). In particular the procedure contracts the indices $d$ and $a$ and the integral over $k$ contributes only its trace so that again we find a renormalisation of the free action which is suppressed in the tensionless limit. 

\bigskip
\noindent 
This leaves only terms (\ref{VI}) and
(\ref{VII}) to analyse. In fact, in the bulk the rapidly varying parts of (\ref{VI}) and (\ref{VII}) are related by integration by parts:
\begin{align}
&k^{2}~\alpha^{\prime 3}\epsilon^{ab}\epsilon^{cd}\int dxdy\int
dx^{\prime}dy^{\prime}
e^{ik\cdot\left(X_{c} - X^{\prime}_{c}\right)}
\partial_{b}^{\prime}\partial^{\prime}_{d} f
\partial^{\prime}_{a}f \partial^{\prime}_{c} f
e^{-\pi\alpha^{\prime}k^{2}f} \propto \nonumber \\
%\propto&-\alpha^{\prime 2}\epsilon^{ab}\epsilon^{cd}\int
%dxdy\int dx^{\prime}dy^{\prime} 
%e^{ik\cdot\left(X_{c} - X^{\prime}_{c}\right)}
%\partial_{b}^{\prime}\partial^{\prime}_{d} f
%\partial^{\prime}_{a}f \partial^{\prime}_{c} \left(
%e^{-\pi\alpha^{\prime}k^{2}f}\right) \nonumber \\
%=&~~~~\alpha^{\prime 2}\epsilon^{ab}\epsilon^{cd}\int dxdy\int
%dx^{\prime}dy^{\prime} 
%\partial^{\prime}_{c}\left(e^{ik\cdot\left(X_{c} -
%X^{\prime}_{c}\right)}\partial_{b}^{\prime}\partial^{\prime}_{d}
%f \partial^{\prime}_{a}f \right) e^{-\pi\alpha^{\prime}k^{2}f} 
%\nonumber \\
&~~~~\alpha^{\prime 2}\epsilon^{ab}\epsilon^{cd}\int dxdy \int
dx^{\prime}dy^{\prime}e^{ik\cdot\left(X_{c}
- X^{\prime}_{c}\right)}
\partial^{\prime}_{b}\partial^{\prime}_{d}f \left(
\partial^{\prime}_{c}\partial^{\prime}_{a}f -  ik\cdot \partial^{\prime}_{c}X^{\prime}_{c}\partial^{\prime}_{a}f\right)
e^{-\pi\alpha^{\prime}k^{2}f}
%&-\alpha^{\prime 2}\epsilon^{ab}\epsilon^{cd}\int dxdy\int
%dx^{\prime}dy^{\prime} 
%ik\cdot \partial^{\prime}_{c}X^{\prime}_{c} e^{ik\cdot\left(X_{c} -
%X^{\prime}_{c}\right)}\partial_{b}^{\prime}\partial^{\prime}_{d}
%f \partial^{\prime}_{a}f  e^{-\pi\alpha^{\prime}k^{2}f}
	\label{T67}
\end{align}
where the boundary contribution is exponentially suppressed as $\epsilon \rightarrow 0$. The second term in brackets has the same rapidly varying structure as term (\ref{V}). The presence of four derivatives of $f$ in the first term implies that when we scale $r$ by $\sqrt\epsilon$ there will be an overall $1/\epsilon$ multiplying the integral. This time we will use (\ref{expand}) and must also expand the slowly varying part of $\Psi$:
\begin{equation}
	{1\over 2\pi}\log \left({(x-x')^2+(y+y')^2\over 4y y'}\right) = \frac{\left(x - x^{\prime}\right)^{2} + \left(y - y^{\prime}\right)^{2}}{8\pi y^{2}} + \ldots
	\label{expandpsi}
\end{equation}
which under the scaling we apply is of order $\epsilon$ but is independent of $X$. 

\bigskip
\noindent To begin with consider just the first term contributing to (\ref{VII}):
\begin{align}
\alpha^{\prime 2}\epsilon^{ab}\epsilon^{cd} &\int^{r \leqslant
\Lambda} dx^{\prime}dy^{\prime}~
\partial^{\prime}_{b}\partial^{\prime}_{d}f\left(\frac{r}{\sqrt{\epsilon}}\right)
\partial^{\prime}_{c}\partial^{\prime}_{a}f\left(\frac{r}{\sqrt{\epsilon}}\right)
e^{-\pi\alpha^{\prime}k^{2}f\left(\frac{r}{\sqrt{\epsilon}}\right)}
= \nonumber \\
\frac{\alpha^{\prime 2}}{\epsilon}&\int^{r \leqslant
\frac{\Lambda}{\sqrt{\epsilon}}} dx^{\prime}dy^{\prime}~\left(
\partial^{\prime}_{x}\partial^{\prime}_{x}f\left(r\right)\partial^{\prime}_{y}\partial^{\prime}_{y}f\left(r\right)
- \left( \partial^{\prime}_{x}\partial^{\prime}_{y}
f\left(r\right) \right)^{2} \right)
e^{-\pi\alpha^{\prime}k^{2}f\left(r\right)}.
\end{align}
For the first region of integration ($0 \leqslant r \leqslant
a$) the second term in brackets is zero and the first is simply
equal to four. For the outer region of integration -- where $b
\leqslant r \leqslant \frac{\Lambda}{\sqrt{\epsilon}}$ -- both
terms contribute and we find the bracketed terms evaluate to
$-4r^{-4}$ so that we must determine
\begin{align}
\frac{4\alpha^{\prime 2}}{\epsilon}\int_{0}^{a} dr~r
e^{-\alpha^{\prime}k^{2}\frac{r^{2}}{4}} &+
\frac{\alpha^{\prime 2}}{\epsilon}\int_{a}^{b} dr~r\left(
\partial^{\prime}_{x}\partial^{\prime}_{x}f\left(r\right)\partial^{\prime}_{y}\partial^{\prime}_{y}f\left(r\right)
- \left( \partial^{\prime}_{x}\partial^{\prime}_{y}
f\left(r\right) \right)^{2} \right)
e^{-\pi\alpha^{\prime}k^{2}f\left(r\right)} \nonumber \\
&-\frac{4\alpha^{\prime 2}}{\epsilon}
\int_{b}^{\frac{\Lambda}{\sqrt{\epsilon}}} dr ~r^{-3 -
\frac{1}{2}\alpha^{\prime}k^{2}} \nonumber \\
\end{align}
showing a $\frac{1}{\epsilon}$ divergence\footnote{This too has to be integrated over $k$ leading to a divergent integral that we again regulate by working in $D$ spacetime dimensions. The first integral captures the
$\frac{1}{\epsilon}$ divergence so we consider the inner-most
region of integration and do the $k$-integration first:
\begin{align}
\frac{4\alpha^{\prime 2}}{\epsilon}\int_{0}^{a}dr\int
d^{D}k~re^{-\alpha^{\prime}k^{2}\frac{r^{2}}{4}}e^{ik\cdot
\left(X_{c} - X^{\prime}_{c}\right)} &= \frac{4^{1+D/2}\alpha^{\prime
(2-D/2)}}{\epsilon}\int_{0}^{a} dr~ r^{1-D} e^{-\frac{\left(X_{c} -
X^{\prime}_{c}\right)^{2}}{\alpha^{\prime}r^{2}}} 
\end{align}
The integral with respect to $r$ could be defined for $D < 2$ to
avoid the logarithmic divergence there and be analytically
continued to physical values of $D$.\label{footK}}.

\bigskip
\noindent The $1/\epsilon$ divergence here is independent of $X$ and corresponds to an infinite renormalisation of the cosmological term $\int d^{2}\xi\sqrt{g}$ which is implicit when considering the quantisation of the string. There is also a finite renormalisation of this term arising out of the subleading term in (\ref{expandpsi}). These renormalisations are not suppressed in the tensionless limit $\alpha^{\prime} \rightarrow \infty$. 
 
\bigskip
\noindent We consider also the exponential factor that remains and use ({\ref{expand}). It is more economic to carry out the integral over $k$ first as in footnote \ref{footK}. Then we expand
\begin{align}
\left(X_{c}- X^{\prime}_{c}\right)^{2} &= \left[\left(x -
x^{\prime}\right)^{\alpha}\partial_{\alpha} X_{c}+\cdots
\right]^{2} \nonumber \\
&\sim \epsilon r^2 \partial_{\alpha}
X_{c} \cdot \partial_{\alpha} X_{c}
\end{align}
where the second line follows because it is to be inserted into an integral over a rotationally symmetric domain. After exponentiation we obtain a term independent of $\epsilon$ that renormalises the string action; again this renormalisation is not suppressed in the tensionless limit. 

\bigskip 
\noindent The renormalisations we have found in this section are just as we expected to find given the original derivation of (\ref{Sem}). No further non-renormalisable divergences appear which justifies this consistency of the contact interaction we have introduced. An appropriate way to split the action into a free and interacting piece is to take the latter to explicitly exclude the coincidence of the two-points 
$\xi$ and $\xi'$, which really requires that in the presence of the regulator which smears out the $\delta$-function we take
$|\xi-\xi'|>\Lambda$ in the interaction.

\bigskip

\noindent Returning to our aim of showing that the conformal scale of the worldsheet metric decouples from the expectation value we also note this is the only time it is necessary to consider higher order terms corresponding to variations about constant $\phi$. We have worked with a constant worldsheet metric and absorbed the conformal scale into the cutoff. Had we explicitly tracked it through the calculation it would appear in this expression as $1 / \epsilon e^{-\phi}$ and the Green function would pick up further dependence on $\phi$ which is subleading in $\epsilon$. We can expand $\phi\left(x'\right)$ about the point $x$ -- the linear terms vanish when averaging in a disk about the point $x$ so that the leading correction to our calculations is of order $\epsilon\nabla^{2}\phi$. This combines with the $\frac{1}{\epsilon}$ divergence found above to produce a finite term dependent on $\phi$. It is proportional, however to $e^{\phi}R$, where $R$ is the curvature on the worldsheet so integrating this 
term with respect to $x$ provides simply
\begin{equation}
	\int \sqrt{g}R
\end{equation}
which is a topological invariant, independent of $\phi$. The higher order terms in the expansion vanish with the cut-off. This completes our discussion of the first order correction of the contact interaction term we propose. Up to renormalisations of the free string action and cosmological term we have found the result we sought and have shown that the conformal scale $\phi$ decouples from the calculation. 

\section{Higher order corrections}
\label{Hoc}
We now proceed to give a general analysis of the higher order interactions present in the theory. We follow the same procedure of extracting the rapidly varying parts of the integrands. We consider the order $N$ expansion of the interaction term with $2N$ vertex operator insertions (corresponding to $2N$ points $\xi_{i}$ placed around the worldsheet) and consider
\begin{equation}
	\left<V_{-k_{1}}^{\alpha\beta}\left(\xi_{1}\right)V_{k_{1}}^{\gamma \delta}\left(\xi_{2}\right)\cdots V_{k_{i}}^{\mu\nu}\left(\xi_{i}\right)\cdots V_{-k_{N}}^{\rho \sigma}\left(\xi_{2N-1}\right)V_{k_{N}}^{\tau\chi}\left(\xi_{2N}\right)\right>.
	\label{VN}
\end{equation} 
which must be integrated with respect to each point $\xi_{i}$ about the worldsheet as well as with respect to each of the momenta. Applying Wick's theorem to this product will produce a factor common to all terms
\begin{equation}
	\exp{\left(-\pi\alpha^{\prime} \sum_{ij}k_{i}\cdot k_{j} G\left(\xi_{i}, \xi_{j}\right)\right)}e^{i\sum_{i} k_{i}\cdot X_{c}\left(\xi_{i}\right)}
	\label{GN}
\end{equation}
which will determine the damping of the integrand. The contractions which generate terms that are rapidly varying depend upon the placement of the 2N points in the bulk. The first exponent in (\ref{GN}), however, can be split into parts containing the coincident Green function for each point $\psi_{i} \equiv \psi\left(\xi_{i}\right)$ and those involving the Green function between two points $G_{ij}\equiv G\left(\xi_{i}, \xi_{j}\right)$:
\begin{equation}
	\sum_{ij}k_{i}\cdot k_{j} G\left(\xi_{i}, \xi_{j}\right) = \sum_{i}k_{i}^{2}\psi_{i} + \sum_{i\neq j}k_{i}\cdot k_{j} G_{ij}.
	\label{Gpsi}
\end{equation}
For a general placement of the 2N points the sum involving the $\psi_{i}$ will ensure that the integrand is damped by a factor of order \begin{equation}
\sqrt{\epsilon}^{\frac{\alpha}{4}^{\prime}\sum_{i}k_{i}^{2}},
\label{damp}
\end{equation}
but we must consider what happens when the points approach the boundary or when points approach one another in the bulk since here the effects of the $G_{ij}$ become important.

\subsection{Points close to the boundary}
\label{boundn}
The first case to consider is when we locate each of the points within a small strip close to the boundary. We continue to work on the upper half plane with coordinates $x_{i}$ and $y_{i}$ for each point. The $2N$ points will then be integrated a distance $\Lambda$ into the bulk and in this section we continue to consider only the generic case where $\left|x_{i} - x_{j}\right| > \Lambda$ for all $i$ and $j$ -- see Fig. \ref{figNbound}. Consequently in this region the $G_{ij}$ are slowly varying fields, whilst the $\psi_{i}$ vary rapidly with $y_{i}$. We can again replace the slowly varying fields with their values at the boundary; in particular $G_{ij} = 0$ whenever either argument is on the boundary. To leading order in $\epsilon$ equation (\ref{GN}) therefore factorises as
\begin{equation}
	\prod_{i}\exp{ \left(-\pi\alpha^{\prime} k_{i}^{2} f\left(\frac{y_{i}}{\sqrt{\epsilon}} \right)\right)}e^{i k_{i} \cdot w_{i}}
\end{equation}
where we have also replaced the field $X_{c}\left(\xi_{i}\right)$ by its boundary value $w_{i}$.

\begin{figure}
	\centering
	\def\svgwidth{0.4 \columnwidth}
	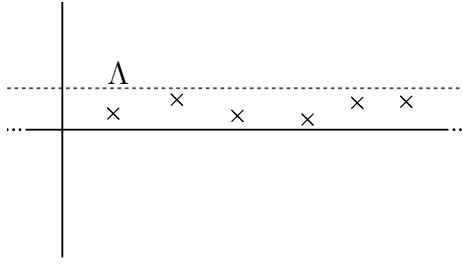
\caption{We consider $N$ points located within a distance $\Lambda$ of the boundary of the worldsheet but not within a distance $\Lambda$ of one another.}
\label{figNbound}
\end{figure}

\bigskip

\noindent The contractions in (\ref{VN}) which will lead to the appearance of rapidly varying terms are those which will produce dependence on $\psi_{i}$. This occurs when we consider contractions amongst the component pieces in each $V_{k_{i}}$ alone rather than those between different vertex operators. Contractions arising out of the pieces of $V_{k_{i}}^{\mu\nu}\left(\xi_{i}\right)$ provide 
\begin{equation}
	2\pi i\alpha^{\prime}\epsilon^{ab} k_{i}^{\left[\mu\right.} : \partial_{b}X_{i}^{\left.\nu\right]}e^{-ik_{i}\cdot X_{i}}: \partial^{i}_{a}\psi_{i}e^{-\pi\alpha^{\prime}k_{i}^{2}\psi_{i}} + \epsilon^{ab}:\partial^{i}_{a}X_{i}^{\mu}\partial^{i}_{b}X_{i}^{\nu}e^{-i k_{i}\cdot X_{i}}:e^{-\pi\alpha^{\prime}k_{i}^{2}\psi_{i}}.
\end{equation}
No further contractions are possible because of the antisymmetry of the worldsheet indices. Since $\psi_{i}$ is a function of the distance into the bulk only we may limit consideration to derivatives with respect to $y_{i}$. We thus consider the general case where the integrand, (\ref{VN}), contains $r$ contractions of the form $\partial_{y}^{i}\psi_{i}$. The remaining factors in the integrand can be replaced by their boundary values so the rapidly varying parts of integrals into the bulk can be expressed
\begin{alignat}{2}
	 &\prod_{j=1}^{r}\alpha^{\prime}k_{j}^{\mu_{j}}\int_{0}^{\Lambda} ~dy_{j}~\partial^{j}_{y}f\left(\frac{y_{j}} {\sqrt{\epsilon}}\right)\exp{\left(-\pi \alpha^{\prime}k_{j}^{2}f\left(\frac{y_{j}}{\sqrt{\epsilon}}\right)\right)}\nonumber \\  &\hphantom{\prod_{j=1}^{r}\alpha^{\prime}k_{j}^{\mu_{j}}\int_{0}^{\Lambda} ~dy_{j}~\partial^{j}_{y}f\left(\frac{y_{j}} {\sqrt{\epsilon}}\right)}\times\prod_{i=r+1}^{2N}\int_{0}^{\Lambda} ~dy_{i}\exp{\left(-\pi \alpha^{\prime}k_{i}^{2}f\left(\frac{y_{i}}{\sqrt{\epsilon}}\right)\right)} \nonumber \\
	=\sqrt{\epsilon}^{2N-r}  ~&\prod_{j=1}^{r}\alpha^{\prime}k_{j}^{\mu_{j}}\int_{0}^{\frac{\Lambda}{\sqrt{\epsilon}}} dy_{j}~\partial^{j}_{y}f\left(y_{j}\right)\exp{\left(-\pi \alpha^{\prime} k_{j}^{2} f\left(y_{j}^{2}\right)\right)}\nonumber \\
	&\hphantom{\prod_{j=1}^{r}\alpha^{\prime}k_{j}^{\mu_{j}}\int_{0}^{\Lambda} ~dy_{j}~\partial^{j}_{y}f\left(\frac{y_{j}} {\sqrt{\epsilon}}\right)}\times \prod_{i=r+1}^{2N}\int_{0}^{\frac{\Lambda}{\sqrt{\epsilon}}} dy_{i}\exp{\left(-\pi \alpha^{\prime} k_{i}^{2} f\left(y_{i}\right)\right)} \nonumber \\
	=\sqrt{\epsilon}^{2N-r} ~&\prod_{j=1}^{r}\frac{2 k_{j}^{\mu_{j}}}{\pi k_{j}^{2}}\left(1 - \left(\frac{\Lambda}{\sqrt{\epsilon}}\right)^{-\frac{1}{2}\alpha^{\prime}k^{2}_{j}}\right)\prod_{i=r+1}^{2N}\int_{0}^{\frac{\Lambda}{\sqrt{\epsilon}}} dy_{i}\exp{\left(-\pi \alpha^{\prime} k_{i}^{2} f\left(y_{i}\right)\right)}
	\label{epsBound}
\end{alignat}
where the second equality follows after a scaling $\frac{y_{i}^{2}}{\epsilon} \rightarrow y_{i}^{2}$. This determines the $\epsilon$-dependence of a term with $r$-contractions. Since $k_{j}^{2} \ge 0$ in Euclidean signature and $2N - r \ge 0$ we see that the second term in rounded brackets will always vanish as the regulator is removed. This allows us to focus on the final product of integrals which can be bounded:
\begin{align}
	\sqrt{\epsilon}^{2N-r} \left|\prod_{i=r+1}^{2N}\int_{0}^{\frac{\Lambda}{\sqrt{\epsilon}}} dy_{i}\exp{\left(-\pi \alpha^{\prime} k_{i}^{2} f\left(y_{i}\right)\right)}\right| &\le \sqrt{\epsilon}^{2N - r} \prod_{i=r+1}^{2N} \left(\frac{\Lambda}{\sqrt{\epsilon}}\right) \exp{\left(-\pi\alpha^{\prime}k_{i}^{2}f\left(0\right)\right)} \nonumber \\
	&= \sqrt{\epsilon}^{2N-r} \left(\frac{\Lambda}{\sqrt{\epsilon}}\right)^{2N - r} \nonumber \\
	&= \Lambda^{2N - r}
\end{align}
The maximum value of $r$ is at $r = 2N$ because each vertex operator can only supply one rapidly varying contribution; then since $\Lambda \rightarrow 0$ with $\epsilon$ this is the only case that will provide a non-vanishing contribution when the regulator is removed. The integral into the bulk in this case takes the form (removing the regulator)
\begin{equation}
	\prod_{j=1}^{2N}\alpha^{\prime}k_{j}^{\mu_{j}}\int_{0}^{\infty} dy_{j}~\partial^{j}_{y}f\left(y_{j}^{2}\right)\exp{\left(\pi \alpha^{\prime}k_{j}^{2}f\left(y_{j}\right)\right)},  
	\label{boundfin}
\end{equation}
providing
\begin{equation}
	\prod_{j=1}^{2N}\frac{2k^{\mu_{j}}}{\pi k_{j}^{2}}.
\end{equation}
This must now be combined with the remainder of the slowly varying fields and the integrals about the boundary. This involves some number of second derivatives $\partial_{a}^{i}\partial_{b}^{j}G_{ij}$ and the remaining field derivatives $\partial_{a}^{i}X_{c}\left(\xi_{i}\right)$.  The only arrangement of derivatives which provides a non-vanishing contribution as the regulator is removed involves 2N derivatives $\partial^{i}_{y}\psi_{i}$ meaning that the only derivatives remaining are with respect to each $x_{i}$. Since at leading order the Green function is to be evaluated on the boundary, where it is identically zero, all derivatives $\partial^{i}_{x}\partial^{j}_{x}G_{ij}$ vanish. We are consequently free to consider the 
case of having $2N$ of the fields $X_{c}\left(\xi_{i}\right)$ uncontracted which gives the only non vanishing contribution close to the boundary as
\begin{align}
	&\prod_{j = 1}^{N} \frac{4}{\left(k_{j}^{2}\right)^{2}} \int_{B} k_{j}^{\left[\mu\right.} \partial_{x} X_{c}^{\left.\nu\right]} k_{j}^{\left[\mu\right.}\partial^{\prime}_{x}X_{c}^{\prime \left.\nu\right]} e^{i k_{j}\cdot \left(X - X^{\prime}\right)} \\
	=4^{N}&\prod_{j = 1}^{N} \int_{B} dw_{j} \cdot dw^{\prime}_{j}\frac{e^{i k_{j}\cdot \left(w_{j}- w^{\prime}_{j}\right)}}{k_{j}^{2}}	
	\label{Nbndy}
\end{align}
where we have left the result in its Fourier representation and the points $w_{j}$ and $w^{\prime}_{j}$ have opposite momenta. For the above expression we have also reinstated the antisymmetry on worldsheet and target space indices and have contracted the indices of the fields corresponding to vertex operators with opposite momenta.
\begin{figure}
	\centering
	\def\svgwidth{0.35 \columnwidth}
	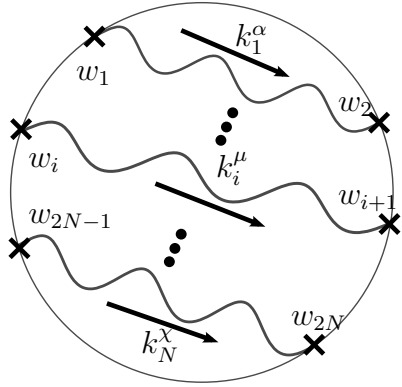
\caption{Using the unit disk representation of the worldsheet domain we demonstrate the physical meaning behind the result. There are N propagators with momenta $k_i$ joining 2N points restricted to the boundary. This mirrors the calculation in Maxwell field theory for the expectation value of a Wilson loop given by the worldline of a pair of quarks, if this worldline is taken to be the fixed boundary of the string in our theory.}
\label{figNBdy}
\end{figure}

\bigskip

\noindent We reiterate that (\ref{Nbndy}) is the only contribution from this regime that does not vanish as the regulator is removed and also point out that it is independent of $\alpha^{\prime}$. The physical interpretation is of $N$ massless propagators pairing off the $2N$ points on the boundary, as depicted in Fig \ref{figNBdy}. The pairs of points joined together are those from vertex operators with equal and opposite momenta $\pm k_{i}$. These momenta are to be integrated over but (at least so far) a dependence on the scale of the worldsheet metric has not arisen so there are no mass shell conditions to be imposed. It remains to consider the other cases where the damping of (\ref{damp}) is not present to investigate whether any dependence on this scale arises to ensure that these expectation values do indeed evade mass shell conditions.

\subsection{Points clustered in the bulk}
\label{bulkn}
When we consider pairs of points meeting in the bulk the Green function between nearby points $\xi_{i}$ and $\xi_{j}$ becomes of the same order as $\psi_{i}$ and $\psi_{j}$ when their distance is less than $\sqrt{\epsilon}$. Furthermore $G_{ij}$ is then rapidly varying as the two points are moved apart. In this subsection we again work at order $N$ but consider the effect of having $n$ of these points clustered in the bulk about a reference point, $\xi_{n+1}$, as is illustrated in Fig \ref{figNbulk}. We shall calculate the contribution of this configuration to the expectation value (\ref{VN}) by integrating the $n$ points about that reference point. The reference point $\xi_{n+ 1}$ would remain to be integrated about the entire worldsheet.

\bigskip

\noindent We proceed by considering the form of the integrand due to Wick contractions between the $n + 1$ vertex operators $V_{k_{1}}\cdots V_{k_{n + 1}}$ because carrying out these contractions is sufficient to extract the leading order behaviour when these $n + 1$ points become close. In the following we shall extract the $\epsilon$ and $\alpha^{\prime}$ dependence arising from the integral of the $n$ points about the reference point before discussing the effect of the remaining points.
\begin{figure}
	\centering
	\def\svgwidth{0.4 \columnwidth}
	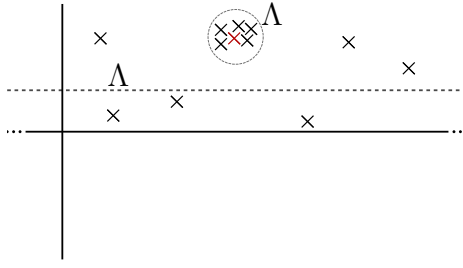
\caption{We imagine that $n$ points are clustered within a distance $\Lambda$ of a point $\xi_{n+1}$ (in red) in the bulk of the worldsheet. The remaining points are elsewhere on the worldsheet and will be discussed at the end of this section. We repeat that points of equal and opposite momenta $\pm k_{i}$ are excluded from meeting in the bulk so that the $n + 1$ points have distinct momenta.}
\label{figNbulk}
\end{figure}
\bigskip

\noindent A string of $n + 1$ vertex operators of the form $V_{k_{1}}^{\mu_{1}\nu_{1}}\cdots V^{\mu_{n+1}\nu_{n+1}}_{k_{n + 1}}$ corresponds to a product of fields
\begin{equation}
	e^{i k_{1}\cdot X_{1}}\partial_{a_{1}}X^{\mu_{1}}_{1}\partial_{b_{1}}X^{\mu_{1}}_{1}\cdots \partial_{a_{n+1}}X^{\mu_{n+1}}_{n+1}\partial_{b_{n + 1}}X^{\mu_{n+1}}_{n+1}e^{i k_{n+1}\cdot X_{n+1}}.
	\label{vBulk}
\end{equation}
In carrying out the analysis of this section we once again consider functions which vary rapidly within the region of integration and those which vary slowly. In this case, slowly varying expressions will be replaced by their values at the reference point $\xi_{n+1}$. In particular the Green function is given by
\begin{equation}
	G_{ij} = f\left(\frac{\left|\xi_{i} - \xi_{j}\right|}{2\sqrt{\epsilon}}\right) - f\left(\frac{\left|\xi_{i} - \xi_{j}^{\ast}\right|}{2\sqrt{\epsilon}}\right).
\end{equation}
Since the second term varies slowly as the $n$ points are integrated about the reference point $\xi_{n+1}$, it will be replaced at first order in $\epsilon$ by
\begin{equation}
f\left(\frac{\left|\xi_{n+1} - \xi_{n+1}^{\ast}\right|}{2\sqrt{\epsilon}}\right)
\label{fBulk}
\end{equation}
which is approximately $\frac{1}{2\pi}\ln{\frac{y_{n+1}}{\sqrt{\epsilon}}}$. The slow variation also implies that a derivative with respect to the relative displacement $\xi_{i} - \xi_{j}$ acting on the first term in $G_{ij}$ produces a factor of $\frac{1}{\sqrt{\epsilon}}$ so is enhanced in comparison to derivatives acting on the second term. As a further consequence, the coincident Green functions $\psi_{i}$ contain only (\ref{fBulk}) so are slowly varying and to first order in $\epsilon$ the following replacement can be made:
\begin{equation}
	\psi_{i} \approx \frac{1}{2\pi} \ln{\frac{y_{n+1}}{\sqrt{\epsilon}}}.
\end{equation}
The derivatives of these functions are therefore also subleading in $\epsilon$ and independent of $X$.

\bigskip

\noindent These properties allow us to consider a general term arising out of the expectation of (\ref{vBulk}) as follows. Wick contractions could generate $q$ terms of the form $\partial_{a_{i}}\partial_{a_{j}}G_{ij}$ and $r$ of the form $k^{\mu_{j}}_{j}\partial_{a_{i}}G_{ij}$ and uncontracted fields will offer $p$ terms of the form $\partial_{a_{i}}X\left(\xi_{i}\right)$. These numbers are constrained by the necessity $2q + r + p = 2\left(n + 1\right)$ and in forming the product of $r$ first derivatives the antisymmetry of the indices must be considered; we return to this later. With (\ref{Gpsi}) we are thus led to consider
\begin{align}
	\alpha^{\prime q+r}\int \prod_{i = 1}^{n}d^{2}\xi_{i} ~\overbrace{\partial_{a_{j}}\partial_{a_{k}}f\left(\frac{\left|\xi_{j} - \xi_{k}\right|}{2\sqrt{\epsilon}}\right)\cdots}^{q\textrm{ terms}}\overbrace{k^{\mu_{l}}_{l}\partial_{b_{m}}f\left(\frac{\left|\xi_{l} - \xi_{m}\right|}{2\sqrt{\epsilon}}\right)\cdots}^{r\textrm{ terms}}\overbrace{\vphantom{f\left(\frac{\left|\xi_{l} - \xi_{m}\right|}{2\sqrt{\epsilon}}\right)}\partial_{a_{w}}X^{\mu_{w}}\left(\xi_{w}\right)\cdots}^{p\textrm{ terms}} \times \nonumber \\ 
	\exp{\left(-\pi\alpha^{\prime} \sum_{i}k_{i}^{2}\psi_{i} - \pi\alpha^{\prime}\sum_{i\neq j}k_{i}\cdot k_{j} G_{ij}\right)} e^{i k_{i} \cdot X_{i}}.
\end{align}

\bigskip

\noindent The contribution from the first sum in the exponent and the corresponding factors of (\ref{fBulk}) from the second sum allows us to rewrite the exponential as
\begin{equation}
	\exp{\left(\frac{ -\alpha^{\prime}}{2}\sum_{i, j}k_{i}\cdot k_{j} \ln{\frac{y_{n+1}}{\sqrt{\epsilon}}}-  \pi\alpha^{\prime}\sum_{i \neq j}k_{i}\cdot k_{j}f\left(\frac{\left|\xi_{i} - \xi_{j}\right|}{\sqrt{\epsilon}}\right)\right)}
	\label{Gdelta}
\end{equation}
Since we are interested in eventually removing the regulator we may think of $\epsilon$ as a small quantity. Recalling that the expectation value of the vertex operators (\ref{VN}) is to be integrated with respect to each of the momenta we consider the effect of the first term in (\ref{Gdelta}) on such an integral. In the limit as $\epsilon \rightarrow 0$ Laplace's approximation shows that this term behaves effectively as
\begin{equation}
	\frac{\delta\left(\sum_{i}k_{i}\right)}{\left(\frac{\alpha^{\prime}}{2\pi}\ln{\frac{y_{n+1}}{\sqrt{\epsilon}}}\right)^{\frac{D}{2}}}
\end{equation}
which we shall use as a means of tracking the $\epsilon$ and $\alpha^{\prime}$ dependence it carries. 

\bigskip

\noindent The integrals with respect to the $\xi_{i}$ can be carried out by using the upper half plane geometry $z_{i} = x_{i} + iy_{i}$.
% and making a change of variables to a polar system for each of the $n$ points centred at $\xi_{n+1}$:
%\begin{align}
%	\mathbf{r}_{i} = \begin{pmatrix}x_{i}-x_{n+1}\\y_{i}-y_{n+1}\end{pmatrix}
%\end{align}
%which in particular gives $\left|z_{i} - z_{j}\right| = \left|\mathbf{r}_{i} - \mathbf{r}_{j}\right|$. 
We now consider integrating each of these points $\xi_{i}$ about a circular region of size $\Lambda$, centred on $\xi_{n+1}$:
%\begin{align}
%\frac{\delta\left(\sum_{i}k_{i}\right)}{\left(\frac{\alpha^{\prime}}{2\pi}\ln{\frac{y_{n+1}}{\sqrt{\epsilon}}}\right)^{\frac{D}{2}}}\int_{0}^{2\pi}\!\int_{0}^{\Lambda} \prod_{i = 1}^{n}r_{i}dr_{i}d\theta_{i} ~\overbrace{\partial_{a_{j}}\partial_{a_{k}}f\left(\frac{\left|\mathbf{r}_{j} - \mathbf{r}_{k}\right|}{2\sqrt{\epsilon}}\right)\cdots}^{q\textrm{ terms}}\overbrace{k^{\mu_{l}}_{l}\partial_{b_{m}}f\left(\frac{\left|\mathbf{r}_{l} - \mathbf{r}_{m}\right|}{2\sqrt{\epsilon}}\right)\cdots}^{r\textrm{ terms}}\overbrace{\vphantom{f\left(\frac{\left|\mathbf{r}_{l} - \mathbf{r}_{m}\right|}{2\sqrt{\epsilon}}\right)}\partial_{a_{w}}X^{\mu_{w}}\left(\mathbf{r}_{w}\right)\cdots}^{p\textrm{ terms}} \times \nonumber \\ 
%	\exp{\left( - \pi\alpha^{\prime}\sum_{i\neq j}k_{i}\cdot k_{j} f\left(\frac{\left|\mathbf{r}_{i} - \mathbf{r}_{j}\right|}{\sqrt{\epsilon}}\right)\right)} e^{i k_{i} \cdot X_{i}}.
%\end{align}
\begin{align}
	\frac{\alpha^{\prime q+r}\delta\left(\sum_{i}k_{i}\right)}{\left(\frac{\alpha^{\prime}}{2\pi}\ln{\frac{y_{n+1}}{\sqrt{\epsilon}}}\right)^{\frac{D}{2}}}\int_{\left|\xi_{i} - \xi_{n+1}\right| < \Lambda} \prod_{i = 1}^{n}d^{2}\xi_{i} ~\overbrace{\partial_{a_{j}}\partial_{a_{k}}f\left(\frac{\left|\xi_{j} - \xi_{k}\right|}{2\sqrt{\epsilon}}\right)\cdots}^{q\textrm{ terms}}\overbrace{k^{\mu_{l}}_{l}\partial_{b_{m}}f\left(\frac{\left|\xi_{l} - \xi_{m}\right|}{2\sqrt{\epsilon}}\right)\cdots}^{r\textrm{ terms}}\times \nonumber \\
	\overbrace{\vphantom{f\left(\frac{\left|\xi_{l} - \xi_{m}\right|}{2\sqrt{\epsilon}}\right)}\partial_{a_{w}}X^{\mu_{w}}\left(\xi_{w}\right)\cdots}^{p\textrm{ terms}}  
	\exp{\left(- \pi\alpha^{\prime}\sum_{i\neq j}k_{i}\cdot k_{j} G_{ij}\right)} e^{i k_{i} \cdot X_{i}}.
\end{align}
At this point we split the integration region into three sections corresponding to the regions where we may employ the approximate forms of $f$ for very large or very small argument. We have seen, however, that the divergences we stand to encounter manifest themselves when considering the short distance behaviour so we will concentrate here on the innermost region, where $0 \leq \left|\xi_{i} - \xi_{j}\right| \leq \sqrt{\epsilon}a$. Furthermore we anticipate taking the tensionless limit whereby the exponential factor $\exp{\left( - \pi\alpha^{\prime}\sum_{i\neq j}k_{i}\cdot k_{j} f\left(\frac{\left|\xi_{i} - \xi_{j}\right|}{\sqrt{\epsilon}}\right)\right)} 
$ damps the integrand for large $\alpha^{\prime}$ except when $G_{ij}$ is small -- precisely in the innermost region where we shall focus. This produces the contribution that is leading order in $\alpha^{\prime}$. In this region the function $f$ is approximated by a quadratic expression
\begin{equation}
f\left(s\right) \approx \frac{s^{2}}{4\pi}
\end{equation}
which implies that the exponent above takes on a Gaussian form. As previously we shall scale each of the $n$ displacement variables $\frac{\xi_{i} - \xi_{n+1}}{\sqrt{\epsilon}} \rightarrow \xi_{i} - \xi_{n+1}$ so as to remove the $\epsilon$-dependence from the integrand. We can finally replace any derivatives with respect to $x_{n+1}$ or $y_{n+1}$ acting on a function of $\left|\mathbf{\xi}_{i} - \mathbf{\xi}_{n+1}\right|$ by derivatives with respect to $x_{i}$ or $y_{i}$. The expression becomes
\begin{align}
\epsilon^{n - q - \frac{r}{2}}&\alpha^{\prime q+r-\frac{D}{2}}\frac{\delta\left(\sum_{i}k_{i}\right)}{\left(4\ln{\frac{y_{n+1}}{\sqrt{\epsilon}}}\right)^{\frac{D}{2}}}\int_{0}^{2\pi}\!\int_{\left|\xi_{i} - \xi_{n+1}\right| < a} \prod_{i = 1}^{n}d^{2}\xi_{i} ~\overbrace{\partial_{a_{j}}\partial_{a_{k}}f\left(\frac{\left|\xi_{j} - \xi_{k}\right|}{2\sqrt{\epsilon}}\right)\cdots}^{q\textrm{ terms}}\times \nonumber \\ 
&\overbrace{k^{\mu_{l}}_{l}\partial_{b_{m}}f\left(\frac{\left|\xi_{l} - \xi_{m}\right|}{2\sqrt{\epsilon}}\right)\cdots}^{r\textrm{ terms}}
\overbrace{\vphantom{f\left(\frac{\left|\xi_{l} - \xi_{m}\right|}{2\sqrt{\epsilon}}\right)}\partial_{a_{w}}X^{\mu_{w}}\left(\xi_{w}\right)\cdots}^{p\textrm{ terms}} 
	\exp{\left(- \pi\alpha^{\prime}\sum_{i\neq j}k_{i}\cdot k_{j} G_{ij}\right)} e^{i k_{i} \cdot X_{i}}.
\label{epNpts}
\end{align}
For large $\alpha^{\prime}$ the exponential factor damps the integrand outside of the region where $\left|\xi_{i} - \xi_{j}\right|^{2} << \alpha^{\prime}k_{i}\cdot k_{j}$ which is by construction inside the innermost region we are concerned with here. In this limit the integral can be safely approximated by taking the upper bound of the integration over relative displacements to infinity. Also in this region the $q$ second order derivatives are independent of the $\xi_{i}$ whilst the $r$ first order derivatives are linear in the differences $\xi_{l} - \xi_{m}$; these therefore lead to moments of a Gaussian integral. The exponent can be written $\mathbf{\xi}^{T} N \mathbf{\xi}$ where the vector $\mathbf{\xi}$ has $i^\textrm{th}$ component $\xi_{i} - \xi_{n+1}$ and the matrix $N$ has components $N_{ij} \equiv \delta_{ij}\left(\sum_{l} k_{l}\right)^{2} - k_{i}\cdot k_{j}$.
%$N_{ij} \equiv \delta_{ij}\left(\sum_{l} k_{l}\right)^{2} - k_{i}\cdot k_{j}$. Also in this region the $q$ second order derivatives are independent of the $\mathbf{r}_{i}$ whilst the $r$ first order derivatives are linear in the differences $\mathbf{r}_{l} - \mathbf{r}_{m}$; these therefore lead to moments of a Gaussian integral. We shall not calculate explicitly each term, instead focussing on the $\epsilon$-dependence to pick out those contributions which would not vanish with $\epsilon$.

\bigskip

\noindent It is clear that the smallest power of $\epsilon$ arises by maximising $q + \frac{r}{2}$. With the constraint $2q + r + p = 2\left(n + 1\right)$ this is done by setting $p = 0$ which automatically leads to a term of order $\frac{1}{\epsilon}$, mirroring the worst behaviour found in the previous section. Following the procedure used for term (\ref{VI}) of the first order calculation the $r$ first order derivatives can be removed via an integration by parts which leads to an integral with respect to the relative displacements of the form
\begin{align}
	\frac{\alpha^{\prime-\frac{D}{2}+ n + 1}}{\epsilon \left(\ln{\epsilon}\right)^{\frac{D}{2}}}  \left( \ln{y_{n+1}} \right)^ {-\frac{D}{2}} \delta\left(\sum_{i}k_{i}\right)\int\prod_{i=1}^{n} d^{2}\xi_{i} \exp{\left(-\pi\alpha^{\prime}\mathbf{\xi}^{T}N\mathbf{\xi} + ik_{i}\cdot X_{i}\right)}
\end{align}
which is
\begin{equation}
 \frac{1}{\epsilon \left(\ln{\epsilon}\right)^{\frac{D}{2}}} \frac{\left(\ln{y_{n+1}}\right)^{-\frac{D}{2}}}{\alpha^{\prime\frac{D}{2} - 1}}  \delta\left(i\sum_{i}k_{i}\right)\frac{1}{\det{N}} e^{\sum_{i}k_{i}X_{n + 1}}.
\end{equation}
This pole in $\epsilon$ can be suppressed by taking the tensionless limit of the string theory $\alpha^{\prime} k_{i}^{2} \rightarrow \infty$ for all momenta due to the overall factor
\begin{equation}
	\frac{1}{\alpha^{\prime \frac{D}{2} - 1}}.
\end{equation}
Note also that in this case there are no finite corrections arising from an expansion of the slowly varying fields due to the suppression caused by the denominator $\left(\ln{\epsilon}\right)^{D/2}$. The integration by parts leads to a complicated index structure but it is constrained. Contract (\ref{VN}) with one of the momenta, say $k_{i\mu}$, and integrate the point $\xi_{i}$ throughout the domain. The effect of the contraction $k_{i\mu}V^{\mu\nu}_{k_{i}}\left(\xi_{i}\right)$ can be written
\begin{align}
	\int_{D} d^{2}\xi_{i} ~\epsilon^{ab}k_{i\mu} \partial_{a}X_{i}^{\mu}\partial_{b}X_{i}^{\nu}e^{i k_{i}\cdot X_{i}} &= \int_{D} d^{2}\xi_{i}~ \epsilon^{ab}\partial_{a}\left(\partial_{b}X_{i}^{\nu}e^{i k_{i}\cdot X_{i}}\right) \\
	&= \oint_{\partial D} d\xi_{i}~ \partial X_{ci}^{\nu}e^{i k_{i}\cdot X_{ci}}
	\label{genGauss}
\end{align}
providing only a boundary contribution. This leaves a total of $n-1$ points to be integrated about the point $\xi_{n+1}$ but the divergences that arose out of contractions involving the $X\left(\xi_{i}\right)$ can no longer appear. The structure of the divergence which appears because of the presence of the vertex operator $V^{\mu\nu}_{k_{i}}$ must therefore be such that it vanishes when contracted with $k_{i}$. Integrating by parts to remove the $r$ first order derivatives is responsible for the formation of this index structure\footnote{This can be illustrated by considering two such operators and taking the leading order $\frac{1}{\epsilon}$ piece of
\begin{equation}
	\int d^{2}\xi d^{2}\xi^{\prime} ~\epsilon^{ab}\epsilon^{cd}\left<\partial_{a}X^{\mu}\partial_{b}X^{\nu}e^{ik\cdot X}e^{ij\cdot X^{\prime}}\partial^{\prime}_{c}X^{\prime \alpha}\partial^{\prime}_{d}X^{\prime\beta}\right> \equiv \frac{1}{\epsilon}H^{\mu\nu\alpha\beta} + \cdots
\end{equation}
where the $\cdots$ represent terms which are regular in $\epsilon$ (which should be familiar from section \ref{bulk} where we had $j^{\mu} = -k^{\mu}$ at first order). $H^{\mu\nu\alpha\beta}$ holds the tensor structure and is a function of the momenta. In our work we are concerned only with the piece antisymmetric in $\mu$ and $\nu$ and also in $\alpha$ and $\beta$ and as a consequence linear in momenta so that
\begin{align}
	H^{\mu\nu\alpha\beta} \propto A\left[\delta^{\mu\alpha}\delta^{\nu\beta} - \delta^{\mu\beta}\delta^{\nu\alpha}\right] &+ B\left[k^{\mu}j^{\alpha}\delta^{\nu\beta} - k^{\mu}j^{\beta}\delta^{\nu\alpha} - k^{\nu}j^{\alpha}\delta^{\mu\beta} + k^{\nu}j^{\beta}\delta^{\mu\alpha}\right] \nonumber \\
	& + C\left[j^{\mu}k^{\alpha}\delta^{\nu\beta} - j^{\mu}k^{\beta}\delta^{\nu\alpha} - j^{\nu}k^{\alpha}\delta^{\mu\beta} + j^{\nu}k^{\beta}\delta^{\mu\alpha}\right]
\end{align}
The requirement $k_{\mu}H^{\mu\nu\alpha\beta} = 0$ implies that $B = 0$ and $A = -k\cdot jC$. Taking $n = 1$ in (\ref{epNpts}) the term proportional to $A$  arises out of $q = 2$ second order derivatives (so $r = p = 0$) and the term proportional to $C$ comes from $r = 2$ first order derivatives and $q = 1$ second order derivatives ($p = 0$ again), with the dependence on $k$ and $j$ appearing after integrating by parts to remove the first order derivatives. To compare to the first order calculation presented in section \ref{bulk} it is necessary to set the momenta equal and opposite and to contract indices $\mu$ with $\alpha$ and $\nu$ with $\beta$.}.
\bigskip

\noindent Continuing with the general case of $n + 1$ points clustered in the bulk the next singular behaviour which may appear comes from $q = n$ and $r = 1$, which gives a term of order $\epsilon^{-\frac{1}{2}}$. This corresponds to exchanging one second derivative of the function $f$ for a single derivative which leaves an uncontracted derivative of a field $X$ ($p = 1$) leaving $2q$ second derivatives and 1 first derivative acting on Green functions. There is no rotationally invariant tensor with odd rank so the integral of this term vanishes when integrated about the point $\xi_{n + 1}$. All further terms are of order $1$ or a positive power of $\epsilon$ multiplied by the $\left(\ln{\epsilon}\right)^{-\frac{D}{2}}$ common to all terms. For this reason they vanish as the regulator is removed.
%The first order derivative leads to the first moment of the Gaussian integral and dimensionally provides an extra factor of $\alpha^{\prime -\frac{1}{2}}$. The expression can be evaluated and contributes a term of order
%\begin{equation}
%	\frac{1}{\epsilon^{\frac{1}{2}}\left(\ln{\epsilon}\right)^{\frac{D}{2}}}\frac{1}{\alpha^{\prime \frac{D}{2} - \frac{1}{2}}}
%\end{equation}
%which, whilst suffering from a pole in $\epsilon$, is also suppressed in the tensionless limit. 

\bigskip 

\noindent We have thus argued that in the tensionless limit the contribution from $n+1$ points meeting in the bulk is vanishing. For an arbitrary placement of the remaining points on the worldsheet the coincident Green functions of individual points, $\psi_{i}$, damp the integrand. A collection of points in the region of the boundary offers a finite contribution as these points are integrated into the bulk but the problem factorises into this and the cluster of points in the bulk. In general the integrand would be sensitive to the scale of the metric when considering the Green function of points which are located in the bulk. That this contribution vanishes as the regulator is 
removed completes the argument that the result of the previous subsection evades a mass shell condition on the momenta so it generalises to all orders. The scale of the worldsheet metric has decoupled from the calculation and integrating this degree of freedom simply cancels the same contribution arising in the normalisation of the amplitude. This is a significant result for the interacting string theory presented here because as well as evading a mass shell condition we also find no constraint on the dimensionality of target space. We are free to specify $D = 4$ where the result of the calculation at order $N$ reads:
\begin{equation}
	4^{N}\prod_{j = 1}^{N} \int_{B} dw \cdot dw^{\prime}~\frac{e^{i k_{j}\cdot \left(w- w^{\prime}\right)}}{k_{j}^{2}}	
	\label{Nbnd}
\end{equation}
Integrating the $N$ momenta gives the position space representation of the product of Wilson loops for the curves fixing the boundary of the string worldsheet.

\subsection{Discussion}
\label{disc}
We have presented an argument for the evaluation of a generic product of vertices by focussing on the configuration of points where the Green function does not damp the integrand. Close to the boundary we also dealt only with the generic case that the points could not come within a distance $\Lambda$ of one another and this deserves attention. Recall that there the coincident Green functions $\psi\left(\xi_{i}\right)$ vary from $0$ on the boundary to order $\ln{\epsilon}$ moving into the bulk so that close to the boundary they do not damp the integrand. 

\bigskip 

\noindent Considering two such vertex operators we have derived the form of the integrand for $\left|x - x^{\prime}\right| >> \Lambda$ in section \ref{bound1}. We can also consider the configuration where both points are close enough to apply the quadratic approximation to the terms making up the regulated Green function to derive
\begin{align}
	\Psi\left(z, z^{\prime}\right) &= -f\left(\frac{y}{\sqrt{\epsilon}}\right) - f\left(\frac{y^{\prime}}{\sqrt{\epsilon}}\right) - 2\left[f\left(\frac{\left|z - z^{\prime}\right|}{2\sqrt{\epsilon}}\right)-f\left(\frac{\left|z - \bar{z^{\prime}}\right|}{2\sqrt{\epsilon}}\right)\right] \nonumber \\ 
	&\approx  \frac{y^{2}}{4\pi\epsilon} + \frac{y^{\prime 2}}{4\pi\epsilon} - 2\left[\frac{\left(x - x^{\prime}\right)^{2}+\left(y - y^{\prime}\right)^{2}}{16\pi\epsilon}-\frac{\left(x - x^{\prime}\right)^{2}+\left(y + y^{\prime}\right)^{2}}{16\pi\epsilon}\right] \nonumber \\
	&= \frac{\left(y - y^{\prime}\right)^{2}}{4\pi\epsilon}
\end{align}
which is independent of the separation $x - x^{\prime}$. In this region it is feasible to carry out the integrals of the various terms which arise. However we must look ahead to the intermediate region where $a\sqrt{\epsilon} <\left|x - x^{\prime}\right| < b\sqrt{\epsilon}$ . As the transverse separation between the points increases the relative separation $x - x^{\prime}$ appears in $\Psi$ in a non-trivial manner and with both points close to the boundary this dependence is no longer subleading. The form of the answer is still constrained by the generalised Gauss' law (\ref{genGauss}) but a finite contribution is not suppressed by $\alpha^{\prime}$ and could be present. 

\bigskip

\noindent We are therefore been unable to complete our programme for the bosonic string due to the possible appearance of divergences when points meet in the vicinity of the boundary. In the next section we turn to spinor QED since this is a more realistic model. We discussed in the introduction that the one dimensional quantum theory on the worldlines used to describe spinor matter has a local supersymmetry and this motivates us to consider including supersymmetry in our interacting string theory. The next section introduces the necessary preliminaries and the equivalent calculations for the spinning string. The extra symmetry gained will be shown to lead to a cancelling of the $\frac{1}{\epsilon}$ divergences which arose in the purely bosonic case (both in the bulk and on the boundary) and the finite contribution shall be shown to provide precisely the expectation value of the supersymmetric Wilson loops which appear in the worldline formalism of spinor QED.

\section{Spinor QED}
\label{susy}
The worldline formalism of spinor QED enjoys a local supersymmetry which suggests a generalisation of our interacting string theory to include spinning degrees of freedom on the worldsheets. The gauge field $A$ appears in the supersymmetric Wilson loop
\begin{equation}
	W_{A} = \exp{\left(i \int \frac{d \omega}{d \xi} \cdot A + \frac{1}{2}F_{\mu\nu}\psi^{\mu}\psi^{\nu}\sqrt{h} ~d\xi \right)} 
\end{equation}
where our notation follows that of the introduction -- $\psi^{\mu}$ is the super-partner to the coordinate $w^{\mu}$. Our aim in this section is to replace the integral over $A$ of a product of these objects by a functional integral over spinning worldsheets supplemented by a contact interaction which generalises the bosonic theory presented so far. This interacting string theory will have worldsheet supersymmetry. A perturbative expansion of the contact interaction implies we must calculate the expectation value of products of supersymmetric vertices inserted at different points in the worldsheet. 

\bigskip
\noindent We shall demonstrate that the result we seek arises in a similar way to the bosonic calculation in that the contribution comes from vertices located close to the worldsheet boundary. The divergences encountered when the vertices cluster in the bulk will not be present for the spinning string because they are forbidden by the residual supersymmetry which we shall preserve throughout regularisation. There can also be no correction to the result arising when the points are close to one another and to the boundary; this time both supersymmetry and the generalisation of Gauss' law (\ref{genGauss}) prevent such a contribution from arising.

\subsection{The spinning string}
\label{spst}
Rather than dealing with the locally supersymmetric and reparameterisation invariant spinning string we shall use the gauge-fixed action\cite{Fried} 
\begin{equation}
	S_{\textrm{spin}} = \frac{1}{4\pi\alpha^{\prime}}\left(\int_{H} d^{2}z d^{2}\theta~\bar{D}\mathbf{X}\cdot D\mathbf{X} - \int_{y=0} dx~ \bar{\Psi} \cdot \Psi\right)	.
	\label{Sspin}
\end{equation}
The parameter domain is taken to be the upper-half complex plane $z = x + iy$ extended by the anticommuting variables $\theta$ and $\bar{\theta}$ which together make up the derivatives
\begin{equation}
	D \equiv \frac{\partial}{\partial \theta} + \theta \frac{\partial}{\partial z};~~~~\bar{D} \equiv \frac{\partial}{\partial \bar{\theta}} + \bar{\theta} \frac{\partial}{\partial \bar{z}}.
\end{equation}
We have introduced the superfield
\begin{equation}
	\mathbf{X} \equiv X + \theta\Psi + \bar{\theta}\bar{\Psi} + \theta\bar{\theta}B
\end{equation}
where $X$ is the bosonic coordinate and $\Psi$ and $\bar{\Psi}$ make up its fermionic super-partner; these fields have dimension of length. $B$ is an auxiliary field required for the supersymmetry which can be disregarded for the purposes of our calculation. This gauge-fixed form of the action has a residual global supersymmetry
%\footnote{This residual symmetry can be thought of as arising from the super-diffeomorphism invariance which preserves the domain: $z \rightarrow z + \theta \eta, ~\bar{z} \rightarrow \bar{z} + \bar{\theta}\eta; ~\theta \rightarrow \theta + \eta, ~ \bar{\theta} \rightarrow \bar{\theta} + \eta$. The transformation induces the changes in the fields $\delta X = \eta\left(\Psi + \bar{\Psi}\right);~\delta \Psi = \eta \partial X,~ \delta \bar{\Psi} = \eta\bar{\partial}X$.}
\begin{equation}
	\delta{\bf X}=\eta\left({\partial\over\partial\theta}-\theta{\partial\over\partial z}
+{\partial\over\partial\bar\theta}-\bar\theta{\partial\over\partial \bar z}
\right){\bf X},
\label{deltasusy}
\end{equation}
corresponding to the domain preserving transformation on the co-ordinates 
\be
z \rightarrow z + \theta \eta; ~~~\bar{z} \rightarrow \bar{z} + \bar{\theta}\eta; ~~~\theta \rightarrow \theta + \eta; ~~~ \bar{\theta} \rightarrow \bar{\theta} + \eta\,.
\label{susyz} 
\ee
The boundary term in (\ref{Sspin}) would not be present in the conventional string theory since it would vanish under the usual Ramond or Neveu-Schwarz boundary conditions. We have introduced it because in order to relate the worldsheet variables to those on the worldline we will have to enforce the Dirchlet boundary conditions
\begin{equation}
	\left.X\right|_{y=0}=w,\quad~ \left.\left(\Psi+\bar\Psi\right)\right|_{y=0}=h^{1/4}\,\psi\,.
	\label{bconSusy}
\end{equation}
Since $\psi$ is a world-line scalar the factor of $h^{\frac{1}{4}}$ is natural and will also be seen to lead to the correct formation of the supersymmetric Wilson loop when we consider the effect of the contact interaction to be introduced below. The relation between the local supersymmetry on the worldlines and the global supersymmetry of the worldsheet is understood by noting that under (\ref{deltasusy}) the boundary conditions (\ref{bconSusy}) are preserved if a simultaneous transformation of the worldline variables is made with the local supersymmetry parameter $\alpha$ in (\ref{wlsusy}) related to the global parameter $\eta$ by $\alpha=h^{1/4}\eta$.

\bigskip
\noindent We generalise the contact interaction of the bosonic string by writing in gauge-fixed form
\begin{align}
S_{\textrm{int}}\left[\mathbf{X}_{i}, \mathbf{X}_{j}\right]=q^2\int d^2\theta_{i}\left(\int d^2z_i ~\bar D_{i} {\bf X}_{i}^{[\mu} D_{i}{\bf X}_{i}^{\nu]}-\int_{y_{i}=0}\!\! dx_{i} ~\theta_{i}\bar{\theta}_{i}{\bar\Psi}_{i}^{[\mu} {\Psi}_{i}^{\nu]}\right) 
\delta^{d}\left({\bf X}_{i}-{\bf X}_{j}\right) \times& \nonumber \\
d^2\theta_{j} \left(\int d^2z_{j}\bar D_{j} {\bf X}_{j}^{[\mu} D_{j}{\bf X}_{j}^{\nu]}-\int_{y_{j} = 0}\!\!dx_{j}~\theta_{j}\bar{\theta}_{j}{\bar\Psi}_{j}^{[\mu} {\Psi}_{j}^{\nu]}\right)&
\label{susyInt}
\end{align}
where we again use the shorthand $X_{i} \equiv X\left(z_{i}\right).$ The inclusion of the boundary terms ensure that this contact interaction is also invariant under the residual supersymmetry and we use it to form a theory of a set of spinning strings spanning fixed boundaries 
\begin{equation}
	S_s=\sum_i S_{\textrm{spin}}[{\bf X}_j]+\sum_{ij}S_{\textrm{int}}[{\bf X}_i,{\bf X}_{j}]
\end{equation}
which is the generalisation of the bosonic theory considered in previous sections. We shall calculate the partition function for this interacting string theory by perturbative expansion of the interaction term in order to establish the equality (\ref{mainf})
\begin{equation}
	\prod_{i = 1}^{n} \frac{\mathscr{D}\left(X_{i},\psi_{i},g_{i}\right)}{Z_{0}} e^{-S_{s}} = \int \frac{\mathscr{D}A}{N}e^{-S^{\prime}_{gf}}\prod_{i}e^{-S_{A}}
\end{equation}
which replaces the functional integral over the gauge field of a product of supersymmetric Wilson loops by an integration over the string worldsheets whose boundaries are those curves. The delta-function in the interaction term can be Fourier decomposed to reduce the problem to the expectation value of insertions of vertices
\begin{equation}
	\bar D {\bf X}^{[\mu} D{\bf X}^{\nu]}
\delta^{d}\left({\bf X}-{\bf x}\right)=\int \frac{d^{d}k}{\left(2\pi\right)^{d}}e^{-i k\cdot \mathbf{x}}\frac{1}{2}V^{\mu\nu}\left(k\right),\quad V^{\mu\nu}\left(k\right)=\bar D {\bf X}^{[\mu} D{\bf X}^{\nu]}\,e^{ik\cdot {\bf X}}\,.
\end{equation}

\bigskip
\noindent As in the bosonic case we shall show that the expectation value of this delta function decouples from the super-conformal scale. This too is unusual because upon quantisation $V^{\mu\nu}\left(k\right)$ acquires an anomalous dimension which would impose a mass shell condition ($k^{2} = 0$). But again the Dirichlet boundary conditions and the self-contraction of the exponential which give rise to the anomalous dimension will ensure that the insertion is suppressed for all points $z$ that are not close to the boundary. Points close to the boundary, as measured with respect to the short distance cut-off we shall use to regulate the Green function, will provide the finite, scale independent contribution which makes up (\ref{mainf}).

\bigskip
\noindent To demonstrate the decoupling and establish the regularisation we shall use we begin with the zeroth-order calculation
\begin{equation}
	\int\mathscr{D}{\bf X}\, e^{-S_{\textrm{spin}}}\int d^{2}\theta \left(\int d^2z~\bar D {\bf X}^{[\mu} D{\bf X}^{\nu]}-\int_{y=0}\!\! dx~\theta\bar{\theta}{\bar\Psi}^{[\mu} {\Psi}^{\nu]}\right)e^{ik\cdot {\bf X}}
\,.
\end{equation}
The super-field can be split as $\mathbf{X} = \mathbf{X}_{c} + \tilde{\mathbf{X}} + \bar{\mathbf{X}}$ which is a classical piece -- $\bar{D}D\mathbf{X}_{c} = 0$ -- which also satisfies the boundary conditions, another solution $\tilde{\mathbf{X}}$ which absorbs the sources produced by the insertion and a quantum fluctuation $\bar{\mathbf{X}}$. Functionally integrating over $\bar{\mathbf{X}}$ gives
\begin{align}
	e^{-S_{\textrm{spin}}[{\bf X}_c]-S_L}\left(\int d^2z d^2\theta\,e^{ik\cdot {\bf X}_c-\pi\alpha' k^2 G_0}
\right.&\left.\left(\bar D {\bf X}_c^{[\mu} D{\bf X}_c^{\nu]}-2\pi\alpha'\left(\bar D {\bf X}_c^{[\mu} (DG)_0ik^{\nu]} \right.\right. \right.\nonumber \\
+&\left.\left.\left.(\bar D G)_0ik^{[\mu} D{\bf X}_c^{\nu]}
\right)\right) -\int_{y=0}\!\!dx~{\bar\Psi}^{[\mu} {\Psi}^{\nu]}e^{i k \cdot X_{c}}\right)
\label{intosusy}
\end{align}
where $S_{L}$ contains the functional determinants which give rise to the super-Liouville action\cite{polyF} and $G_{0}$ is the Green function evaluated at coincident points. The defining equation of the Green function is 
\begin{equation}
	-\bar{D}DG\left(z_{1}, \theta_{1}; z_{2}, \theta_{2}\right) = \delta^{2}\left(\theta_{1} - \theta_{2}\right)\delta^{2}\left(z_{1} - z_{2}\right)
\end{equation}
subject to the boundary conditions $G = 0$ if $y_{i} = 0$ and $\theta_{i} = \bar{\theta_{i}}$ ($i = 1$ or $2$) which has solution that generalises the bosonic case
\begin{equation}
	G\left(z_{1}, \theta_{1}; z_{2}, \theta_{2}\right) = \log{\left(\vphantom{z_{12}^{R}}z_{12}\bar{z}_{12}\right)} - \log{\left(z^{R}_{12}\bar{z}^{R}_{12}\right)}
\end{equation}
where
\begin{equation}
	z_{12}=z_1-z_2-\theta_1\theta_2,\,\,\,\bar z_{12}=\bar z_1-\bar z_2-\bar\theta_1\bar\theta_2,\,\,\,
z_{12}^R=z_1-\bar z_2-\theta_1\bar \theta_2,\,\,\,\bar z_{12}^R=\bar z_1-z_2-\bar\theta_1\theta_2\,.
\end{equation}
Evaluated at coincident points the Green function is singular and we regulate it via heat kernel regularisation with the obvious generalisation of (\ref{Greg}):
\begin{equation}
	G^\epsilon=-f\left(\frac{\sqrt{\vphantom{^{R}}z_{12}\bar z_{12}}}{\sqrt\epsilon}\right)+f\left(\frac{\sqrt{z_{12}^R\bar z_{12}^R}}{\sqrt\epsilon}\right)
\end{equation}
with $\epsilon$ again a short distance cut-off and $f$ defined as in (\ref{fdef}). This function satisfies the boundary conditions and to verify this is a regularisation of the Green function it is easy to determine
\begin{equation}
	-\bar D DG^\epsilon=(\theta_1-\theta_2)(\bar\theta_1-\bar\theta_2){e^{-\frac{\vphantom{z^{R}}z_{12}\bar z_{12}}{\epsilon}}\over 4\pi\epsilon}
-(\theta_1-\bar\theta_2)(\bar\theta_1-\theta_2){e^{-z_{12}^R\bar z_{12}^R\over \epsilon}\over 4\pi\epsilon}.
\end{equation}
Upon taking the limit $\epsilon \rightarrow 0$ we recover Green's equation. Furthermore this regulated Green function is invariant under the residual supersymmetry 
because $z_{12}$, $z_{12}^R$, $\bar z_{12}$ and $\bar z_{12}^R$ are all separately invariant under (\ref{susyz}). This will be crucial in allowing us to constrain the form of the integrals we will calculate. Using this regulator we can determine the coincident limits as an expansion in $\theta$ and $\bar{\theta}$:
\begin{equation}
	G^\epsilon_0=\left(1 + \frac{i}{2}\theta\bar{\theta}\frac{\partial}{\partial y}\right)f\left(\frac{2y}{\sqrt{\epsilon}}\right);\quad
(DG^\epsilon)_0=(\bar D G^\epsilon)_0={i\over 2}(\theta-\bar\theta){\partial \over\partial y}f\left(\frac{2y}{\sqrt\epsilon}\right)
\end{equation}
and we can also expand the common exponential term in (\ref{intosusy}) as 
\begin{equation}
	e^{-\pi\alpha' k^2 G_0}=\left(1+{i\over 2}\theta\bar\theta{\partial\over\partial y}\right)e^{-\pi\alpha' k^2 f\left(\frac{2y}{\sqrt\epsilon}\right)}\,.
	\label{G0}
\end{equation}
The exponential factor on the right hand side of the above equation has been seen in the previous sections and for fixed $k^{2}$ it damps the integrand at all points in the domain except for those close to the boundary $y \lesssim \sqrt{\epsilon}$. We thus repeat our procedure of integrating (\ref{intosusy}) a distance $\Lambda$ into the bulk, where $\Lambda \rightarrow 0$ as $\epsilon \rightarrow 0$ but we arrange for $\frac{\Lambda}{\sqrt{\epsilon}}$ to diverge. This means that to leading order in $\epsilon$ we can replace the components of the classical super-field $\mathbf{X}_{c}$ by their boundary values.

\bigskip
\noindent The integral of the first term in (\ref{intosusy}) cancels against the boundary term present in the interaction term. To see this consider
\begin{equation}
-2i\int dx\,d^2\theta\,e^{ik\cdot {\bf X}_c}\bar D {\bf X}_c^{[\mu} D{\bf X}_c^{\nu]}\int_0^\Lambda dy \,\left(1+{i\over 2}\theta\bar\theta{\partial\over\partial y}\right)e^{-\pi\alpha' k^2 f\left(\frac{2y}{\sqrt\epsilon}\right)}
\label{clas}\,.
\end{equation}
Both parts of the $y$ integral are known from previous work. The monotonicity of $f\left(s\right)$ allows us to bound the first term $\left|\int_0^\Lambda dy \exp (-\pi\alpha' k^2 f(2y/\sqrt\epsilon))\right|<\Lambda$ which vanishes as the cut-off is removed. The second term is a total derivative and in the limit as $\epsilon \rightarrow 0$ evaluates to $-\frac{i}{2}\theta\bar{\theta}$. We must still integrate over $\theta$ which means that we must determine the $\theta$- and $\bar{\theta}$-independent parts of the slowly varying terms on the boundary. The result is 
\begin{equation}
	\int dx~e^{ik\cdot { X}_c}\bar\Psi^{[\mu}_c\Psi^{\nu]}_c
	\label{bsusy}
\end{equation}
which is as claimed.

\bigskip
\noindent The remaining term in (\ref{intosusy}) can be written as
\begin{align}
	&-2i\int dx\,d^2\theta\,e^{ik\cdot {\bf X}_c}\left(\bar D {\bf X}_c^{[\mu} ik^{\nu]}+ D{\bf X}_c^{[\nu}ik^{\mu]}
\right)\nonumber \\
&\qquad \times {(\theta-\bar\theta)\over\pi\alpha' k^2}
\int_0^\Lambda dy ~{\partial f\over\partial y}\,\left(1 + \frac{i}{2}\theta\bar{\theta}\frac{\partial}{\partial y}\right)e^{-\pi\alpha' k^2 f\left(\frac{2y}{\sqrt{\epsilon}}\right)}.
\label{intoosusy}
\end{align} 
The second term in the rounded brackets of the $y$-integral cannot contribute due to its $\theta$ dependence and the first term is again a total derivative which tends to unity as the regulator is removed. We again expand the slowly varying fields on the boundary in powers of anti-commuting variables and seek terms with a single factor of $\theta$ or $\bar{\theta}$. A little algebra leads to
\begin{equation}
	{1\over\pi\alpha' k^2}\int dx~e^{ik\cdot { X}_c}\left(ik\cdot\left(\Psi_c+\bar\Psi_c\right)(\Psi_c+\bar\Psi_c)^{[\mu}+{\frac{\partial X_c^{[\mu}}{\partial x}}\right)ik^{\nu]}.
\end{equation}
We have preserved the global supersymmetry with our regularisation and it is straightforward to verify that this result is indeed invariant under (\ref{deltasusy}). We can now use the boundary conditions (\ref{bconSusy}) to relate the boundary values of the worldsheet variables to the variables on the one dimensional worldlines to obtain the $\epsilon \rightarrow 0$ limit of (\ref{intosusy}) as
\begin{equation}
	-2e^{-S_{\textrm{spin}}[{\bf X}_c]-S_L}\int_{B} dx~e^{ik\cdot w}
\left({\frac{d w^{[\mu}}{d x}}+\sqrt h \,ik\cdot\psi\psi^{[\mu}\right)\frac{ik^{\nu]}}{k^2}.
\label{result}
\end{equation}
which we recognise contains the supersymmetric Wilson loop. It is only in $S_{\textrm{spin}}\left[\mathbf{X}_{c}\right]$ that the string length scale $\sqrt{\alpha^{\prime}}$ appears and only in $S_{L}$ that the conformal scale and its super-partner are present. The classical action can be removed by taking the tensionless limit $\alpha^{\prime}k^{2} \rightarrow 0$\footnote{As was discussed at the end of section \ref{GenExp} the tensionless limit corresponds to taking $\alpha^{\prime}$ large as measured with respect to the length scale, $l$, of the closed loop B -- that is $l / \sqrt{\alpha^{\prime}} \rightarrow 0$.}. The result does not contain any further dependence on the metric which we have treated as constant, absorbing the conformal scale into the cut-off $\epsilon$. This has occurred despite there being no mass-shell restrictions on $k^{2}$. Since there is no $\epsilon$-dependence in (\ref{result}) we conclude that the result is independent of this constant scale. Spatial variations in this scale 
contribute at higher order in $\epsilon$ so vanish as the cut-off is removed. So the conformal scale and its super-partner decouple from the calculation (if we assume that the metric on the world-line is independent of that on the world-sheet) and are present only in $S_{L}$; they can be removed completely if we assume further internal degrees of freedom to take us to a critical string-theory.

\bigskip
\noindent Similarly to the bosonic case the interaction contains terms which involve points inserted on different world-sheets and other terms with multiple insertions on the same world-sheet. For the former we can use (\ref{result}) to average over two distinct world-sheets to determine the leading order behaviour in the tensionless limit:
\begin{align}
&\int{\mathscr{D}{\bf X}_{i}\over Z_0}{\mathscr{D}{\bf X}_{j}\over Z_0}
\, e^{-S_{spin}[{\bf X}_{i}]-S_{spin}[{\bf X}_{j}]}\,
S_{int}[{\bf X}_{i},{\bf X}_{j}] =\nonumber \\ 
%\backsimeq q^2\int {d^dk\over (2\pi)^d} \,\left<V_{i}^{\mu\nu}(k)
%V_{j}^{\mu\nu}(-k)\right>\nonumber\\
%\backsimeq 
q^2&\int {d^dk\over (2\pi)^d}
\int_{BB^{\prime}} dx\,dx'~{e^{ik\cdot(w-w')}\over k^2}\left({dw\over dx}+\sqrt h\, 
\psi\cdot i k\, \psi\right)\cdot\left({dw'\over dx'}+\sqrt {h'} 
\,\psi'\cdot i k\, \psi'\right)
\label{reso}
\end{align}
This result is the order $q^{2}$ contribution to the expectation value of two super-Wilson loops parameterised by $x$ and $x^{\prime}$ in spinor QED. This demonstrates our result holds at leading order in the case that the worldsheets are distinct. Following the bosonic theory we shall consider extending this to arbitrary order and also treat the case that multiple vertices are on the same worldsheet.

\subsection{Generalisation to arbitrary order}
\label{arbord}
When some insertions approach one another on the same world-sheet we may find divergences that change our result in a similar way to that we found for the bosonic theory. In this section we demonstrate that no such divergences arise and the calculation reduces to the result we seek. It is because our interaction and regularisation procedure preserves the residual supersymmetry that such divergences are forbidden from arising since their possible forms are not themselves supersymmetric. 

\bigskip
\noindent We follow the same steps as in the bosonic case by considering a general term at order $N$ in the expansion of the interaction which has $2N$ vertex insertions on a single world-sheet:
\begin{equation}
	\int\mathscr{D}{\bf X}\, e^{-S_{\textrm{spin}}}\prod _{i = 1}^{2N}\int d^2\theta_{i}\left(\int \,d^2z_{i}~\bar D {\bf X}_{i}^{[\mu_i} D{\bf X}_{i}^{\nu_i]}-\int_{y_{i}=0}\!\!dx_{i}~\theta_{i}\bar{\theta}_{i}{\bar\Psi}^{[\mu_i}_{i} {\Psi}^{\nu_i]}_{i}\right)e^{ik_i\cdot {\bf X}_{i}}
\,.
\end{equation}
The functional integral over $\mathbf{X}$ will lead to the ubiquitous factor $\linebreak \exp{\left(-\pi\alpha^{\prime}\sum_{ij}k_{i}\cdot k_{j}G_{ij}\right)}$ where we continue to denote the Green function $\linebreak G_{ij} \equiv G\left(z_{i}, \theta_{i}; z_{j}, \theta_{j}\right).$ When all of the points are separated by a distance much greater than $\sqrt{\epsilon}$ the exponential factors $\exp{\left(-\pi\alpha^{\prime}k^{2}G_{0}\right)}$ which involve (\ref{G0}) suppress the integrand unless the points are close to the boundary. In the latter case we follow section \ref{boundn} by integrating each point a distance $\Lambda$ into the bulk, focussing on contractions that take place separately within each vertex. At leading order in the cut-off the components of the super-fields and the slowly varying Green functions between the separated points will be replaced by their boundary values.

\bigskip
\noindent Using (\ref{intosusy}) take the contribution involving $r$ copies of the second term which arises from a single contraction of the quantum fields and integrate these a distance $\Lambda$ into the bulk:
\begin{align}
	\prod_{j = 1}^{r}\int\!\int_{0}^{\Lambda}& d^{2}\theta_{j} dy_{j}~2\pi\alpha^{\prime} \left(\bar{D}_{j}\mathbf{X}_{cj}^{\left[\mu_{j}\right.}ik_{j}^{\left.\nu_{j}\right]} + D_{j}\mathbf{X}_{cj}^{\left[\nu_{j}\right.}ik_{j}^{\left.\mu_{j}\right]}\right)\nonumber \\
	& \times \frac{i}{2}\left(\theta_{j} - \bar{\theta}_{j}\right)\frac{\partial f}{\partial y_{j}} \left(1 + \frac{\theta_{j}\bar{\theta}_{j}}{2}\frac{\partial}{\partial y_{j}}\right)e^{-\pi\alpha^{\prime}k_{j}^{2}f\left(\frac{2y_j}{\sqrt{\epsilon}}\right)}  e^{i k_{j} \cdot \mathbf{X}_{cj}} \nonumber \\
	& \times \prod_{i = r + 1}^{2N}\int\!\int_{0}^{\Lambda}d^{2}\theta_{i}dy_{i}~\bar{D}_{i}\mathbf{X}_{ci}D_{i}\mathbf{X}_{ci}\left(1 + \frac{\theta_{i}\,\bar{\theta}_{i}}{2}\frac{\partial}{\partial y_{i}}\right)e^{-\pi\alpha^{\prime}k_{i}^{2}f\left(\frac{2y_{i}}{\sqrt{\epsilon}}\right)} e^{i k_{i} \cdot \mathbf{X}_{ci}}.
	\label{Nboundsusy}
\end{align}
To extract the $\epsilon$-dependence of this expression it is useful to generalise the scaling carried out in the bosonic case by setting
\begin{equation}
	y \rightarrow \epsilon^{\frac{1}{2}}y; ~~~~ \theta\rightarrow \epsilon^{\frac{1}{4}}\theta; ~~~~ \bar{\theta} \rightarrow \epsilon^{\frac{1}{4}}\bar{\theta}
	\label{scalesusy}
\end{equation}
for all variables in (\ref{Nboundsusy}). Under these changes of variables and simplifying the anti-commuting variables a little we get
\begin{align}
	\epsilon^{-\frac{r}{4}} \prod_{j = 1}^{r}&\int\!\int_{0}^{\frac{\Lambda}{\sqrt{\epsilon}}} d^{2}\theta_{j} dy_{j}~2\pi\alpha^{\prime} \left(\bar{D}_{j}\mathbf{X}_{cj}^{\left[\mu_{j}\right.}ik_{j}^{\left.\nu_{j}\right]} + D_{j}\mathbf{X}_{cj}^{\left[\nu_{j}\right.}ik_{j}^{\left.\mu_{j}\right]}\right)\frac{i}{2}\left(\theta_{j} - \bar{\theta}_{j}\right)\frac{\partial f}{\partial y_{j}}e^{-\pi\alpha^{\prime}k_{j}^{2}f\left(2y_j\right)}  \nonumber \\
	&\times e^{i k_{j} \cdot \mathbf{X}_{cj}}\prod_{i = r + 1}^{2N}\int\!\int_{0}^{\frac{\Lambda}{\sqrt{\epsilon}}}d^{2}\theta_{i}dy_{i}~\bar{D}_{i}\mathbf{X}_{ci}D_{i}\mathbf{X}_{ci}\left(1 + \frac{\theta_{i}\,\bar{\theta}_{i}}{2}\frac{\partial}{\partial y_{i}}\right)e^{-\pi\alpha^{\prime}k_{i}^{2}f\left(2y_{i}\right)} e^{i k_{i} \cdot \mathbf{X}_{ci}}.
	\label{2int}
\end{align}
The first $r$ integrals with respect to $y_{j}$ evaluate to 
\begin{equation}
	\frac{1}{\pi\alpha^{\prime}k_{j}^{2}}\left(1 - \left(\frac{\Lambda}{\sqrt{\epsilon}}\right)^{-\frac{1}{2}\alpha^{\prime}k_{j}^{2}}\right)
\end{equation}
and the $2N - r$ remaining integrals with respect to $y_{i}$ contain two terms. As described above the first can be bounded by $\frac{\Lambda}{\sqrt{\epsilon}}$ and the second is equal to $-\frac{i}{2}\theta_{i}\bar{\theta_{i}}$. The latter contribution gives the result we seek as can be seen by carrying out the integrals over the Grassmann variables:
\begin{align}
	\epsilon^{-\frac{r}{4}}\prod_{j = 1}^{r}\int d^{2}\theta_{j}~2\pi\alpha^{\prime} \left(\bar{D}_{j}\mathbf{X}_{cj}^{\left[\mu_{j}\right.}ik_{j}^{\left.\nu_{j}\right]} + D_{j}\mathbf{X}_{cj}^{\left[\nu_{j}\right.}ik_{j}^{\left.\mu_{j}\right]}\right)\frac{i\left(\theta_{j} - \bar{\theta}_{j}\right)}{2\pi\alpha^{\prime} k_{j}^{2}}\left(1 - \left(\frac{\Lambda}{\sqrt{\epsilon}}\right)^{-\frac{1}{2}\alpha^{\prime}k_{j}^{2}}\right)& \nonumber \\
	\times e^{i k_{j} \cdot \mathbf{X}_{cj}}\prod_{i = r + 1}^{2N} \int d^{2}\theta_{i}~ \frac{i}{2}\theta_{i}\bar{\theta}_{i}\bar{D}_{i}\mathbf{X}_{ci}D_{i}\mathbf{X}_{ci} e^{i k_{i} \cdot \mathbf{X}_{ci}}&.
\end{align} 
The integrals with respect to $\theta_{j}$ require us to find the $\bar{\theta}_{j}$ and $\theta_{j}$ terms in the super-fields. Under the scaling (\ref{scalesusy}) such terms pick up a factor $\epsilon^{\frac{1}{4}}$. The $r$ such terms cancel the leading factor of $\epsilon^{-\frac{r}{4}}$ so the Grassmann integration selects a single term which is independent of the cut-off $\epsilon$. The remaining integrals with respect to $\theta_{i}$ require the $\theta_{i}$- and $\bar{\theta}_{i}$-independent parts of the super-fields which do not change under scaling. Following the same algebra as at first order, summing over $r$, integrating around the boundary and enforcing the boundary conditions the result is 
\begin{equation}
	\sum_{r = 0}^{2N} \prod_{j = 1}^{r}\int dx_{j}~e^{ik\cdot w_{j}}
\left({\frac{d w_{j}^{[\mu_{j}}}{d x_{j}}}+\sqrt{h_{j}} \,ik_{j}\cdot\psi_{j}\psi_{j}^{[\mu_{j}}\right)\frac{ik_{j}^{\nu_{j}]}}{k_{j}^2} \prod_{i = r + 1}^{2N}\int dx_{i}~e^{ik_{i}\cdot X_{c i}}\bar\Psi_{ci}^{[\mu_{i}}\Psi_{ci}^{\nu_{i}]}
\end{equation}
With the exception of the $r = 2N$ case the contributions in this sum cancel terms arising out of the boundary term in the interaction (\ref{susyInt}) which conspire to ensure supersymmetry is maintained. This leaves the contribution occurring from $2N$ contractions between fields which corresponds to $2N$ points inserted on the boundary of a single Wilson loop:
\begin{equation}
	q^{2N}\prod_{j = 1}^{2N}\int dx_{j}~e^{ik\cdot w_{j}}
\left({\frac{d w_{j}^{[\mu_{j}}}{d x_{j}}}+\sqrt{h_{j}} \,ik_{j}\cdot\psi_{j}\psi_{j}^{[\mu_{j}}\right)\frac{ik_{j}^{\nu_{j}]}}{k_{j}^2}.
\end{equation}
As in the bosonic case this result is independent of the string tension $\alpha^{\prime}$. It remains to enforce the contractions of the space-time indices and impose pairwise $k_{j + 1} = - k_{j}$ as defined in the interaction to produce
%\begin{equation}
%	\prod_{j = 1}^{N}\int_{B} dx_{j} dx^{\prime}_{j}~e^{ik_{j}\cdot \left(w_{j} - w^{\prime}_{j}\right)}
%\left({\frac{d w^{\left[\mu_{j}\right.}_{j}}{d x_{j}}}+\sqrt{h_{j}} \,ik_{j}\cdot\psi_{j}\psi^{\left[\mu_{j}\right.}_{j}\right)\left({\frac{d w^{\prime\left[\mu_{j}\right.}_{j}}{d x^{\prime}_{j}}}+\sqrt{h^{\prime}_{j}} \,ik_{j}\cdot\psi^{\prime}_{j}\psi^{\prime\left[\mu_{j}\right.}_{j}\right)\frac{k^{\left.\nu_{j}\right]}_{j} k^{\left.\nu_{j}\right]}_{j}}{\left(k_{j}^2\right)^{2}}
%	\label{resultSusy}
%\end{equation}
\begin{equation}
	q^{2N}\prod_{j = 1}^{N}\int_{B} dx_{j} dx^{\prime}_{j}~\frac{e^{ik_{j}\cdot \left(w_{j} - w^{\prime}_{j}\right)}}{k^{2}}
\left({\frac{d w_{j}}{d x_{j}}}+\sqrt{\vphantom{h^{\prime}}h_{j}} \,ik_{j}\cdot\psi_{j}\psi_{j}\right)\cdot \left({\frac{d w^{\prime}_{j}}{d x^{\prime}_{j}}}-\sqrt{h^{\prime}_{j}} \,ik_{j}\cdot\psi^{\prime}_{j}\psi^{\prime}_{j}\right)
	\label{resultSusy}
\end{equation}
showing how pairs of points on the boundary interact.

\bigskip 
\noindent Returning to the integrals over the $y_{i}$ in (\ref{2int}) we bounded the second term by $\frac{\Lambda}{\sqrt{\epsilon}}$ and for the $\theta_{i}$ integrals we now seek the $\theta_{i}\bar{\theta}_{i}$ term from the boundary super-fields. This term scales as $\sqrt{\epsilon}$ which leaves a contribution of order $\Lambda^{2N - r}$. The integration over the $\theta_{j}$ variables remains the same as above so the Grassmann integration selects a single term which vanishes as $\epsilon \rightarrow 0$ because $\Lambda$ vanishes in this limit too. Other contributions from this configuration of points are subleading in $\epsilon$. This completes our treatment of the case where all $2N$ points are close to the boundary (and a distance greater than $\sqrt{\epsilon}$ apart from one another) and demonstrates the result (\ref{resultSusy}) we sought. We now consider what happens when these points are close to one another in the bulk or the boundary to show that in contrast to the bosonic case no divergences 
appear.

\bigskip
\noindent Suppose that of the $2N$ points a number $n + 1$ are within $\Lambda$ of one another (but that this set is separated by more than $\Lambda$ from any other points on the same worldsheet). Now it is the contractions between different vertices which are rapidly varying. Following the procedure taken for the bosonic case (see section \ref{bulkn}) Wick's theorem allows us to replace this by a sum of terms involving various contractions between this set of points and normal ordered terms which have not been contracted with other operators outside of this set. The leading order contribution comes from expanding the normal ordered terms about the position of the final point $z_{n + 1}$. We then integrate the first $n$ points in a region of size $\Lambda$ about this reference point which remains to be integrated about the worldsheet. The $\epsilon$-dependence can be extracted by counting derivatives of rapidly varying fields. Now
\begin{align}
G^{\epsilon}\left(z_r, \theta_{r}; z_s, \theta_{s}\right)&=-f\left(\frac{\sqrt{\vphantom{^{R}}z_{rs}\bar z_{rs}}}{\sqrt\epsilon}\right)+
f\left(\frac{\sqrt{z_{rs}^R\bar z_{rs}^R}}{\sqrt\epsilon}\right)\nonumber\\
&=
-f\left(\frac{\sqrt{\vphantom{^{R}}z_{rs}\bar z_{rs}}} {\sqrt\epsilon}\right)+
{1\over4\pi}\log\left({(2iy_{n+1}-\theta_r\bar\theta_s)(-2iy_{n+1}-\bar\theta_r\theta_s)\over\epsilon}\right)\nonumber \\
&+\mathcal{O}\left(\frac{\Lambda}{y_{n+1}}\right)
	\label{Gsusy}
\end{align}
and it is the first term of this which varies rapidly as the points $z_{r}$ and $z_{s}$ move apart. Wick contractions between fields evaluated at the point $z_{r}$ and $z_{s}$ produce various derivatives of this Green function. In parallel to the bosonic string the leading order contribution comes from contractions which have all $2\left(n + 1\right)$ possible derivatives acting on the first term in (\ref{Gsusy}). This can be seen by scaling the relative coordinates (but not $z_{n + 1}$ or $\bar z_{n + 1}$) and the $\theta_r$, $\bar\theta_r$:
\begin{equation}
z_r-z_s\rightarrow \epsilon^{\frac{1}{2}} \left(z_r-z_s\right);\quad\quad\theta_r\rightarrow \epsilon^{\frac{1}{4}} \theta_r;\quad\quad\bar\theta_r\rightarrow \epsilon^{\frac{1}{4}} 
\bar\theta_r,
\end{equation}
so that
\begin{align}
&f\left({\sqrt{z_{rs}\bar z_{rs}}\over\sqrt\epsilon}\right))\rightarrow f\left((\sqrt{z_{rs}\bar z_{rs}}\right); \nonumber \\
&{1\over4\pi}\log\left({(2iy_{n+1}-\theta_r\bar\theta_s)(-2iy_{n+1}-\bar\theta_r\theta_s)\over\epsilon}\right)\rightarrow \frac{1}{4\pi}\log{\left(\frac{4y_{n+1}^{2}}{\epsilon}\right)} + \mathcal{O}\left(\epsilon^{\frac{1}{2}}\right)
	\label{Gslow}
\end{align}
under which the super-derivatives and integration measures transform as 
\begin{equation}
	 D\rightarrow \epsilon^{-\frac{1}{4}} D,\quad \bar D\rightarrow \epsilon^{-\frac{1}{4}} \bar D, 
\end{equation}
\begin{equation}
 d^2 z_r d^2\theta_r\rightarrow \epsilon^{\frac{1}{2}}d^2 z_r d^2\theta_r ,\quad d^2 z_{n+1} d^2\theta_{n+1}\rightarrow \epsilon^{-\frac{1}{2}}d^2 z_{n+1} d^2\theta_{n+1}
\end{equation}
so the integral with respect to $d^{2}\theta_{n + 1}\prod_{r} d^2 z_r d^2\theta_r $ of the term containing $2(n+1)$ derivatives, $D$ and $\bar D$, acting on $f(\sqrt{z_{rs}\bar z_{rs}/\epsilon})$  scales into 
$1/\epsilon$ multiplied by an integral independent of $\epsilon$. This depends on the momenta $k_r$ in a potentially complicated way but because of the way the contractions were carried out to form the $2\left(n + 1\right)$ derivatives it is possible to integrate by parts to enforce the $\mathbf{X}$ dependence to sit only in the exponent:
\begin{align}
	\frac{1}{\epsilon} \int d^{2}z_{n + 1} d^{2}\theta_{n + 1} \prod_{r = 1}^{n} d^{2}z_{r}d^{2}\theta_{r}&F^{\mu_{1}\ldots \nu_{n+1}} \left(z_{1}, \theta_{1},\ldots,z_{n + 1}, \theta_{n+1}\right) :e^{i \sum_{r = 1}^{n + 1} k_{r} \cdot \mathbf{X}_{n + 1}}: \nonumber \\
	 &\times \exp{\bigg(-\pi\alpha^{\prime}\sum_{r, s = 1}^{n + 1} k_{r}\cdot k_{s} G^{\epsilon} \left(z_{r}, \theta_{r}; z_{s}\theta_{s}\right)\bigg)}.
\end{align}
At leading order in $\epsilon$ after carrying out the integral over the relative coordinates and the $\theta_{r}$, $\bar{\theta}_{r}$ we are left with
\begin{equation}
	{1\over\epsilon}\tilde F^{\mu_1\ldots\nu_{n+1}}(k_1,..,k_{n+1})\,
\int d^2z_{n+1} :e^{iK\cdot { X}(z_1)}:\left({\epsilon\over y_{n+1}^2}\right)^{\alpha'K^2/4}
	\label{nonsusy}
\end{equation}
where we have defined
\begin{align}
	\tilde F^{\mu_1\ldots\nu_{n+1}}\left(k_1,..,k_{n+1}\right)=
\int d^2\theta_{n+1}&\left(\prod_{r=1}^{n} d^2z_r\,d^2\theta_r\right)\,F^{\mu_1  \ldots \nu_{n+1}}\left(z_1,\theta_{1},\ldots,z_{n+1}, \theta_{n+1}\right) \nonumber \\
&\times \exp{\bigg(\pi\alpha'\sum_{r, s = 1}^{n + 1} k_r\cdot k_s f\left(\sqrt {z_{rs}\bar z_{rs}}\right)\bigg)}
\end{align}
and $K=\sum_{r=1}^{n+1} k_r$. Since this is not invariant under the residual supersymmetry (\ref{deltasusy}) it must vanish so there can be no $1/\epsilon$ divergence present.

\bigskip
\noindent Subleading terms of order $\epsilon^{-3/4}$ could in principle appear if we were to have $2n + 1$ derivatives acting on $G_{\epsilon}$ or from an expansion of the super-field components. However the required factors of $\epsilon^{1/4}$ are paired with fermionic fields $\Psi$ and $\bar{\Psi}$. They cannot be present since the final result must be bosonic. The first non-trivial divergence which could potentially occur is of order $\epsilon^{-1/2}$ and can arise in a number of ways. The second order expansion in $\theta$ and $\bar{\theta}$ of the exponentiated super-field contains $\epsilon^{1/2} \theta\bar{\theta} k\cdot\Psi k\cdot \Psi$; taking two derivatives off the rapidly varying part of $G^{\epsilon}$ reduces the power of $\epsilon$ picked up under scaling by $1/2$ and leaves two super-derivatives of the super-field or derivatives of the slowly varying part of $G^{\epsilon}$; taking only one derivative off $G^{\epsilon}$ in combination with expanding one super-derivative of the super-field to 
first order gives a similar expression and an expansion of the components of the super-field about the point $z_{n + 1}$ gives $\epsilon^{1/2} \left(z - z_{n+1}\right)\cdot \partial X$. The latter two of these vanish again by rotational symmetry whilst contributions from the slowly varying part of $G^{\epsilon}$ would have the same $X$-dependence as (\ref{nonsusy}). From two super-derivatives or an expansion of the super-field components the contribution at this order has an $X$-dependence proportional to 
\begin{equation}
	{c^{\rho\sigma}\over\sqrt\epsilon}\int d^2z_1:\bar\Psi^\rho\Psi^\sigma e^{iK\cdot { X}(z_1)}:\left({\epsilon\over y_{n+1}^{2}}\right)^{\alpha'K^2/4},
	\label{div}
\end{equation}
Under the residual supersymmetry this too changes, although if the coefficient $c^{\rho\sigma} = K^{\rho}K^{\sigma}$ its variation takes the same form as the variation of the boundary term $\epsilon^{-1/2}\int dx \,\exp(ik\cdot w)$. Were this boundary term to be generated as the insertions approach one another close to the boundary then it would be possible for this divergence to be present. However a term proportional to $k \cdot \bar{\Psi} k\cdot \Psi$ can only be generated from expanding the super-fields in the exponential for the $\theta\bar{\theta}$ contribution. The coefficient of this term would be 
\begin{equation}
	\int d^2\theta_{n+1}\left(\prod_{r=1}^{n} d^2z_r\,d^2\theta_r\right)\,F^{\mu_1\ldots\nu_{n+1}}\left(z_1,\theta_{1}\ldots,z_{n+1}, \theta_{n_1}\right)e^{\pi\alpha'\sum k_r\cdot k_s f(\sqrt {z_{rs}\bar z_{rs}})}\,\bar\theta_r\theta_s 
	\label{poss}
\end{equation}
independent of the choice of $r$ and $s$. We can demonstrate that this vanishes by virtue of its $\theta$ dependence. $F^{\mu_{1}\ldots\nu_{n+1}}$ arose out of $2\left(n + 1\right)$ derivatives acting on $f\left(\sqrt{z_{rs}\bar{z}_{rs}}\right)$ and (\ref{poss}) requires it to contain a total of $n$ $\theta$s and $n$ $\bar{\theta}$s to be non-zero\footnote{The requirement is of course more stringent than this in that the $\theta$s and $\bar{\theta}$s must have the correct indices but the general argument does not rely on this detail.}. Schematically, $f\left(\sqrt{z_{rs}\bar{z}_{rs}}\right)$ has a dependence on anti-commuting variables of the form
\begin{equation}
	f\left(\sqrt{z_{rs}\bar{z}_{rs}}\right) = f\left(\left|z_{r} - z_{s}\right|^{2}\right) - \theta_{r}\theta_{s} g_{1}\left(z_{r} - z_{s}\right) - \bar{\theta}_{r}\bar{\theta}_{s} g_{2}\left(z_{r} - z_{s}\right) - \theta_{r}\bar{\theta}_{r}\theta_{s}\bar{\theta}_{s} h\left(z_{r} - z_{s}\right)
\end{equation}
%\begin{equation}
%	f\left(\sqrt{z_{rs}\bar{z}_{rs}}\right) = f\left(\left|z_{r} - z_{s}\right|^{2}\right) - \theta_{r}\theta_{s}\left(\bar{z}_{r}-\bar{z}_{s}\right)\left.f^{\prime}\right|_{\left|z_{r} - z_{s}\right|^{2}} - \bar{\theta}_{r}\bar{\theta}_{s}\left(z_{r}-z_{s}\right)\left.f^{\prime}\right|_{\left|z_{r} - z_{s}\right|^{2}} - \theta_{r}\bar{\theta}_{r}\theta_{s}\bar{\theta}_{s}\left. f^{\prime\prime}\right|_{\left|z_{r} - z_{s}\right|^{2}}
%\end{equation}
where the functions $g_{1}$, $g_{2}$ and $h$ depend on the relative separation of the points and involve derivatives of $f\left(s\right)$. Now suppose that the $n + 1$ derivatives $D$ and $n + 1$ derivatives $\bar{D}$ contained in $F$ produce $p$ copies of $D_{r}f$, $q$ of $\bar{D}
_{r}f$, $r$ $D_{s}D_{r}f$, $s$ $\bar{D}_{s}\bar{D}_{r}f$ and $t$ lots of $\bar{D}_{s}D_{r}f$ with $p + 2r + t = n + 1 = q + 2s + t$. It follows that the schematic $\theta$ and $\bar{\theta}$ dependence of each of these terms is respectively $\theta + \theta \bar{\theta}\bar{\theta}$, $\bar{\theta} + \theta \theta \bar{\theta}$, $1 + \theta\theta + \bar{\theta}\bar{\theta} + \theta\bar{\theta}\theta\bar{\theta}$, $1 + \theta\theta + \bar{\theta}\bar{\theta} + \theta\bar{\theta}\theta\bar{\theta}$ and $\theta\bar{\theta}$. Counting modulo 2 we thus have a total of $p + t = n + 1$ factors of various $\theta$s and $q + t = n + 1$ factors of $\bar{\theta}$s. So $F^{\mu_{1}\ldots\nu_{n+1}}$ cannot contain the correct number of $\theta$s and $\bar{\theta}$s for (\ref{poss}) to produce a non-zero result. 

\bigskip
\noindent The next possible divergence is of order $\epsilon^{-1/4}$ but it too vanishes because its field content would have to be fermionic. The next order in $\epsilon$ consists of finite terms, but these are suppressed by the overall factor of $\left({\epsilon/ y_{n+1}^2}\right)^{\alpha'K^2/4}$ which comes from the slowly varying part of $G^{\epsilon}$. As $K^{2} \geq 0$ in Euclidean signature such terms vanish for all $K^{2}$ except those close to zero in terms of $\epsilon$. Since $K$ is eventually to be integrated over we also need to consider the contribution of these small values. Following the discussion in section \ref{bulkn} we recall that for $\alpha'$ large and $\epsilon$ small this factor behaves effectively as
\begin{equation}
	\frac{\delta\left(K^{2}\right)}{\left(\frac{1}{2} \alpha^{\prime} \ln{\frac{y_{1}}{\epsilon}}\right)^{\frac{D}{2}}}
\end{equation}
and so is also suppressed in the tensionless limit. We conclude that there are no terms associated with a set of points meeting one another in the bulk of the worldsheet that survive in the tensionless limit as the cut-off is removed. In the bosonic case we were unable to deduce whether our arguments extended to the case that the points are also close to the boundary but in the current case we can use the supersymmetry to show that there are no further contributions from this case.

\bigskip
\noindent Close to the boundary the second term in $G^{\epsilon}$ also varies rapidly. To consider its variation too we must also scale $y_{n + 1}$ along with the other variables. This means that the integration measure $d^2z_{n + 1}d^{2}\theta_{n + 1}$ is unchanged by the scaling and the leading order divergence is $\mathcal{O}\left(\epsilon^{-1/2}\right)$. There is also no suppression by the second term in $G^{\epsilon}$ but the $X$-dependence remains the same as that when the points are far from the boundary. A term proportional to 
\begin{equation}
	\frac{1}{\sqrt{\epsilon}}\int dx~e^{i K\cdot X} 
\end{equation} 
has been seen before; it is not supersymmetric and so this divergence is not present. This time the presence of the boundary breaks the symmetry of the integration domain so a term of order $\epsilon^{-1/4}$ is not forbidden -- it could arise out of an expansion of the super-field in the exponent or by taking one of the derivatives off $G^{\epsilon}$ and onto a super-field. The possible $\mathbf{X}$-dependence has the form
\begin{equation}
	\frac{c^{\rho}}{\epsilon^{\frac{1}{4}}}\int dx~ \left(\Psi + \bar{\Psi}\right)^{\rho} e^{i K \cdot X}
\end{equation}
which changes under the residual supersymmetry unless $c^{\rho} \propto K^{\rho}$ in which case the change is a total derivative. It is fermionic, however, and by applying the same counting of $\theta$s and $\bar{\theta}$s as before (since such a term can only arise from an expansion of the super-field in the exponent) it is straightforward to show that such a term cannot arise out of an integral over the Grassmann variables since it requires $n$ $\theta$s or $n$ $\bar{\theta}$s.

\bigskip
\noindent The final order in $\epsilon$ to consider provides finite terms. These may arise out of two super-derivatives of the super-field or various expansions of the super-field about the point $z_{n + 1}$ in tandem with super-derivatives. But there is only one potential term which remains invariant under the residual supersymmetry which is the electromagnetic coupling
\begin{equation}
	\int dx~ e^{iK\cdot X} \left( dX^\mu/dx+iK\cdot (\Psi+\bar\Psi)
(\Psi+\bar\Psi)^\mu\right)
\label{posso}
\end{equation}
where $\mu$ here must be equal to $\mu_{q}$ of the vertex $V^{\mu_{q}\nu_{q}}$ which was used to generate this finite piece. However the supersymmetric generalisation of Gauss' law, (\ref{genGauss}), can be used to show that this cannot be formed. Indeed contracting $k_{q}$ with the integral of the $q$-th vertex
\begin{align}
	k_{q}^{\mu_{q}}&\int \,d^2\theta_{q}\left(\int d^2z_q\,\bar D_q {\bf X}_q^{[\mu_q} D_q{\bf X}_q^{\nu_q]}-\int_{y_{q} = 0}\!\!dx_{q}~\theta_{q}\bar{\theta}_{q}{\bar\Psi}_q^{[\mu_q} {\Psi}_q^{\nu_q]}\right)e^{ik_q\cdot {\bf X}_q}\nonumber \\
	=&\int_{y_{q} = 0} dx_q~ \left({dX_q^{\nu_{q}}\over dx_q} + ik_{q}\cdot\left(\Psi_{q} + \bar{\Psi}_{q}\right) \left(\Psi_{q} + \bar{\Psi}_{q}\right)^{\nu_{q}}\right)e^{ik_q\cdot X_q}
\label{gausso}
\end{align}
which is a boundary term that does not contain the quantum variables $\mathbf{\bar{X}}_q$. This means that it cannot take place in any contractions with other terms in the set so factors out of the normal ordered expansion of the other vertices. Therefore this boundary integral of the $q$-th field would have to factor out of the contraction of (\ref{posso}) with $k_{q}$. This is not possible because (\ref{posso}) contains only one field integrated around the boundary so the contraction could not produce an integral involving the $k_{q}$ dependence and the field $X\left(z_{q}\right)$ multiplied into an integral involving the remaining momenta with a field content expanded about the point $z_{n + 1}$.

\bigskip
\noindent This completes the argument that supersymmetry prevents divergent or finite corrections from appearing when insertions approach one another on the same worldsheet and proves our claim that (\ref{reso}) exponentiates. This leads directly to (\ref{mainf}) and allows the replacement of an integral over the gauge field by an integral over fluctuating spinning strings interacting upon contact. Since the result we have found is independent of the cut-off $\epsilon$ we have also shown that the scale of the worldsheet metric decouples from the calculation, appearing only in $S_{L}$. Spatial variations of this scale could only contribute at higher order in $\epsilon$ and so vanish as we remove the regulator by taking the limit $\epsilon \rightarrow 0$. The final step to return to the world-line formulation of spinor QED is to integrate over the worldsheet metric and boundaries weighted by the world-line action for
\begin{align}
	\int\left( \prod_{j}^n {\mathscr{D}(g,{\bf X},w,\psi,h,\chi)_j\over Z_0}\right)\, e^{-S_s-S_{BdVH}} =\nonumber \\
\int \left(\prod_{j}^n {\mathscr{D}(w,\psi,h,\chi)_j}\right){{\mathscr{D}}A\over N}\,e^{-S_{gf}-S_{BdVH}}\,\prod_j
W_s[A].
\label{sexpWL2}
\end{align}
Summing over $n$ then re-expresses the partition function of QED in terms of the partition function of spinning strings with contact interactions. To also express the generating functional (\ref{intooD}) requires the world-line Green function which in analogous fashion to the bosonic theory requires the inclusion of open strings. The calculation proceeds in the same way but for the differing boundary conditions on each end of the spinning string and again it is only the Dirichlet end of the string which contributes to the interaction. It is also possible to introduce a background gauge field to source photon amplitudes on the world-line, as described for scalar QED.

\section{Conclusion}
We have investigated how strings with contact interactions can be used to model Abelian gauge fields. We were able to construct $\delta$-functions on the world-sheet that decoupled from the Liouville degree of freedom because their contribution was negligible except close to the world-sheet boundary where they generated the electromagnetic coupling. Although the purely bosonic theory proved to be problematic the world-sheet supersymmetry present in the spinning string provided the structure needed to eliminate unwanted divergences and also generate the super-Wilson loops needed to couple spinor matter to electromagnetism in the world-line approach. The string world-sheets correspond to the trajectories of lines of electric flux joined to charged particles. 

\bigskip
\noindent It proved necessary to take the tensionless limit   to remove dependence on the classical string action %$S_{\textrm{spin}}\left[\mathbf{X_{c}}\right]$ 
so the string length-scale is large compared to the size of the Wilson loops. The strings themselves, therefore, can be very large and it may be possible to distinguish this theory from conventional QED where the interactions are mediated by point-like particles by direct observation of these extended objects. Additionally it may be possible to detect string-like corrections to QED at large distances, (although we have not calculated these). Since the scale of the world-sheet metric decouples from our calculations we could argue that the super-Liouville degrees of freedom only lead to an overall multiplicative factor that 
cancels out of physical amplitudes. Alternatively we might modify the model to include sufficient internal degrees of freedom to ensure a critical string-theory. This decoupling allows us to apply our string theory in four-dimensional space-time dimension.

\bigskip
\noindent QED is of course an extremely successful theory, having been tested to high accuracy in experiments, but nonetheless it is an effective theory arising out of the Standard Model,
so our string model must also be just an effective theory. Understanding how it relates to the more fundamental non-Abelian case will require some development of the model.

\bigskip
\noindent Both authors are grateful  to STFC: PM for support under the Consolidated Grants ST/J000426/1 and ST/L000407/1, and JPE for a studentship. This research is also supported in part by the the Marie Curie network GATIS 
(gatis.desy.eu) of the European Union's Seventh Framework Programme FP7/2007-2013/ under REA Grant Agreement No 317089.

\bibliographystyle{JHEP}
\bibliography{bibWeyl}

%\end{document}

\section*{Appendix A -- Mixed Boundary Conditions and the Green functions}
\label{AppA}
Recall that out of the worldline formalism of the field theories appear Green function factors $\left(-\mathcal{D}^{2} + m^{2}\right)^{-1}\left(b, a\right)$ which in our work are represented as curves running between positions $a^{\mu}$ and $b^{\mu}$. We associate a string worldsheet to these curves but impose Dirichlet boundary conditions at one end of the string -- fixing it to follow the curve between $a^{\mu}$ and $b^{\mu}$ -- and Neumann boundary conditions at the other. In the main text it was shown that Dirichlet boundary conditions ensure that the general damping to integrands caused by the coincident Green function is not present near the boundary, since here $\exp{\left(- \pi\alpha^{\prime}k^{2} G\left(\xi, \xi\right)\right)} \sim \mathcal{O}\left(1\right)$. Neumann boundary conditions do not impose this and so contributions arising from points close to this end of the string will be exponentially damped. So we expect to receive contributions to our integrals only from a strip close to the 
Dirichlet end of the string. There is one distinguished point which may spoil this argument which is the at the point where these two boundaries coincide. 

\bigskip
\noindent In order to simplify the effect of these mixed boundary conditions it is favourable to instead work on a worldsheet domain which consists of the upper-right quadrant of the complex plane via the simple conformal mapping from the upper half plane $z \rightarrow \sqrt{z}$. The positive real axis in this plane corresponds to the end of the string with Dirichlet boundary conditions and the positive imaginary axis corresponds to the end of the string on which Neumann boundary conditions are imposed. Again we shall expand about $\phi = \textrm{const}$ and will 
specialise to $\phi = 0$ to determine the leading order behaviour. The only real change to the calculations we have presented in previous sections is that the Green function on the worldsheet must be modified to respect the mixed boundary conditions. The method of images in the upper-right quadrant gives the Green function as
\begin{equation}
	G\left(w, z^{\prime}\right) = \ln{\left|z - z^{\prime}\right|^{2}} - \ln{\left|z - \bar{z^{\prime}}\right|^{2}} + \ln{\left|z + \bar{z^{\prime}}\right|^{2}} - \ln{\left|z + z^{\prime}\right|^{2}}
\end{equation}
The coincident limit of this function requires regularisation as in the previous case so we shall apply the heat-kernel representation. It is straightforward to verify that in terms of $z = x + iy$ the coincident limit can be written as
\begin{align}
	G_{\epsilon}\left(z, z\right) &= \int_{\epsilon}^{\infty} \frac{d\tau}{4\pi\tau} \left[1 - \exp{\left(-\frac{y^{2}}{\tau}\right)} + \exp{\left(-\frac{x^{2}}{\tau}\right)} - \exp{\left(-\frac{x^2 + y^{2}}{\tau}\right)}\right] \nonumber \\
	&= f\left(\frac{y}{\sqrt{\epsilon}}\right) - f\left(\frac{x}{\sqrt{\epsilon}}\right) + f\left(\frac{\sqrt{x^{2} + y^{2}}}{\sqrt{\epsilon}}\right).
\end{align}
At a distance much greater than $\sqrt{\epsilon}$ from both boundaries the coincident Green function is of order $\ln{\frac{y^{2}}{\epsilon}}$. When approaching the positive imaginary axis it increases to $2\ln{\frac{y^{2}}{\epsilon}}$. Close to the positive real axis (corresponding to the Dirichlet end of the string) $G_{\epsilon}\left(z, z\right)$ is of order $\frac{y^{2}}{\epsilon}$, except at the corner where the axes meet; here it varies from $2\frac{y^{2}}{\epsilon}$ to $\frac{y^{2}}{\epsilon}$ when moving along the positive real axis and from $2\frac{y^{2}}{\epsilon}$ to $2\ln{\frac{y^{2}}{\epsilon}}$ when moving along the positive imaginary axis, both over a distance of order $\sqrt{\epsilon}$. So $G_{\epsilon}\left(z, z\right)$ is of order $\ln{\frac{y^{2}}{\epsilon}}$ everywhere on the worldsheet, except in a small strip close to 
the positive real axis where it is of order $\frac{y^{2}}{\epsilon}$.

\bigskip

\noindent This demonstrates more concretely that indeed all integrands of relevance will be heavily damped except for a small strip close to the Dirichlet boundary of the string. We shall not repeat the entire calculation for an arbitrary number of vertex operators inserted onto the worldsheet since it suffices to consider the behaviour of a single insertion, in much the same way as the calculation that preceded the careful treatment of Section \ref{bound1}. We shall therefore consider the expectation value
\begin{equation}
	\int d^{2}z~ e^{i k \cdot x^{\prime}} \left<V_{k}\left(z\right)\right>
	\label{vertMixed}
\end{equation}
integrated over $\Re(z) > 0,~\Im(z) > 0$, which contains two terms. A non-vanishing contribution arises out of a single contraction between the pieces of the vertex operator which leads an integral
\begin{equation}
	2\pi\alpha^{\prime}k^{\left[\mu\right.}\int_{0}^{\infty}\int_{0}^{\infty}dxdy~\epsilon^{ab}\partial_{a}G\left(z, z\right) \partial_{b}X_{C}^{\left.\nu\right]}e^{-\pi\alpha^{\prime}k^{2}G\left(z, z\right)} e^{i k\cdot \left(x^{\prime} - X_{c}\right)}.
	\label{wickMixed}
\end{equation}
The integrand is damped by the exponent involving the Green function, except close to $y = 0$ so we integrate a distance $\Lambda$ into the bulk and replace the slowly varying terms involving the field $X^{\mu}$ and its derivatives with their values on the Dirichlet boundary. We shall first consider the term 
\begin{align}
	2\pi\alpha^{\prime}k^{\mu}&\int_{0}^{\infty} dx~ \partial_{x}X_{c}^{\nu}e^{ik\cdot \left(x^{\prime} - X_{c}\right)}  \int_{0}^{\infty} dy~ \partial_{y}G\left(z, z\right) e^{-\pi\alpha^{\prime}k^{2}G} \nonumber \\
	=\frac{k^{\mu}}{k^{2}}&\int_{0}^{\infty} dx~ \partial_{x}X_{c}^{\nu}e^{ik\cdot \left(x^{\prime} - X_{c}\right)}\left[e^{-\pi\alpha^{\prime}k^{2}G\left(z,z\right)}\right]_{0}^{\infty},
\end{align}
where the fields $X^{\mu}$ and derivatives are evaluated on the boundary $y = 0$. The only contribution is from the lower bound of the integration,
\begin{equation}
	\int_{0}^{\infty} dx \frac{k^{\left[\mu\right.}}{k^{2}} \partial_{x}X_{c}^{\left.\nu\right]} e^{i k\cdot \left(x^{\prime} - X_{c}\right)} = \int_{B_{0}} dw^{\left[\nu\right.} k^{\left.\mu\right]} ~\frac{e^{i k\cdot \left(x^{\prime} - w\right)}}{k^{2}},
\end{equation}
which is the result we sought. It can be represented diagrammatically as the interaction of points on the worldline $B_{0}$ with a background massless vector field (see Fig \ref{fig0open}) and is independent of $\alpha^{\prime},~ \epsilon$ and the scale of the worldsheet metric. The second term which arises out of (\ref{wickMixed}) does not contribute. It is equal to
\begin{align}
&2\pi\alpha^{\prime}k^{\mu}\int_{0}^{\infty}dy ~\partial_{y}X_{c}^{\nu} e^{i k \cdot \left(x^{\prime} - X_{c}\right)} \int_{0}^{\infty} dx ~\partial_{x}G\left(z, z\right) e^{-\pi\alpha^{\prime} k^{2} G\left(z,z\right)}\nonumber \\
=&\frac{k^{\mu}}{k^{2}}\int_{0}^{\infty}dy~\partial_{y}X_{c}^{\nu}e^{ik \cdot \left(x^{\prime} - X_{c}\right)}\left( e^{-\pi\alpha^{\prime}k^{2}f\left(\frac{y}{\sqrt{\epsilon}}\right)} - e^{-2\pi\alpha^{\prime}k^{2}f\left(\frac{y}{\sqrt{\epsilon}}\right)}\right) 
\end{align}
where once again the fields take on their values at the boundary $y = 0$. The two terms in rounded brackets vary rapidly over the domain of integration but we have met their like in the previous sections and it has already been demonstrated that they vanish as the regulator is removed. 

\begin{figure}
	\centering
	\def\svgwidth{0.4 \columnwidth}
	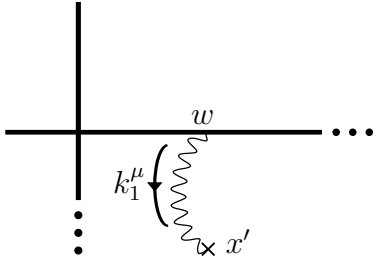
\caption{A single vertex operator corresponds to an interaction with a background field. The only contribution comes from points on the Dirichlet boundary of the string -- this boundary is the worldline of the particle representing the Green function in the worldline approach.}
\label{fig0open}
\end{figure}

\bigskip

\noindent It remains to show that the other form of expression arising out of (\ref{vertMixed}) does not contribute to the expectation value. Since the analysis follows the exact same form as in the main text we shall focus here only on the distinguished corner $x = y = 0$. To do so we consider the region of the domain $0 \leq x \leq a$, $ 0 \leq y \leq a$ where $a$ is chosen to enforce $x^{2} + y^{2} \leq 2a^{2} << \epsilon$. Within this region we have
\begin{equation}
	G\left(z, z\right) \approx \frac{2y^{2}}{\epsilon}.
\end{equation}
The term in question is $\epsilon^{ab}\partial_{a}X^{\mu}\partial_{b}X^{\nu} e^{-\pi\alpha^{\prime} k^{2}G\left(z, z\right)}$ so that the rapidly varying part of the integrand is
\begin{align}
	&\int_{0}^{a\sqrt{\epsilon}}dx\int_{0}^{a\sqrt{\epsilon}}dy~e^{-2\pi\alpha^{\prime}k^{2}\frac{y^{2}}{\epsilon} } \nonumber \\
	=& \epsilon\int_{0}^{a}\int_{0}^{a} dy ~e^{-\pi\alpha^{\prime}k^{2}y^{2}}.
\end{align}
This integral is bounded by $a^{2}\epsilon$ which vanishes as $\epsilon \rightarrow 0$ because $\frac{a}{\sqrt{\epsilon}} \rightarrow 0$ with $\epsilon$. By applying the exact same analysis as previous sections the remaining regions can be shown to also offer a contribution which vanishes as the regulator is removed. 

\bigskip

\noindent It is clear from this result how the calculation would proceed in the general case involving multiple vertex operators. The Green function supplies a similar damping for an arbitrary placement of the points on the worldsheet; it is only when all of the points are within a strip of size $\sqrt{\epsilon}$ of the Dirichlet boundary of the string that a finite contribution can be expected as the regulator is removed, or when the points are arranged in clusters in the bulk. In the latter case the boundary has no effect so the results of the main text apply. In the former we would see a copy of the above calculation for each point and the surviving terms are those involving contractions only amongst the constituents of each vertex operator, rather than between operators inserted at different points. A repeat of the previous 
calculations leads to the result at order $N$
\begin{equation}
	4^{N}\prod_{j = 1}^{N} \int_{B_{0}} dw \cdot dw^{\prime}\frac{e^{i k_{j}\cdot \left(w- w^{\prime}\right)}}{k_{j}^{2}}	
\end{equation}
corresponding to the interaction between pairs of points on the boundary mediated by a massless vector boson; those pairs of points which interact arise from the two vertex operators with equal and opposite momenta. This expression can be represented diagrammatically as in Fig. \ref{figNopen}.

\begin{figure}
	\centering
	\def\svgwidth{0.5 \columnwidth}
	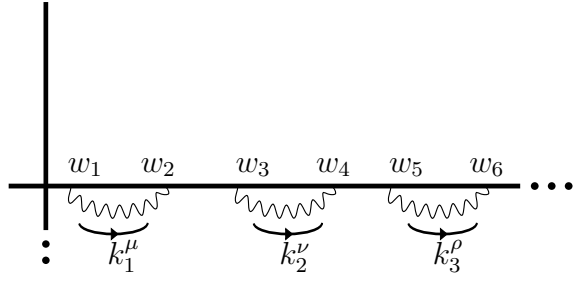
\caption{With multiple vertex operators inserted on the worldsheet there is an interaction between pairs of points on the Dirichlet boundary. These pairs share equal and opposite momentum and the picture corresponds to exchange of massless vectors between points on the worldline representing the Green function in the worldline theory.}
\label{figNopen}
\end{figure}

\section*{Appendix B -- Integrating over the world-line metric}
\label{AppB}
We show that gauge fixing the world-line reparametrisation invariance by choosing the intrinsic metric to be constant, $h=T^2$, allows us to replace ${\mathscr{D}}h$ by $dT$ in the case of open world-lines and $dT/T$ for closed ones. The details in this appendix follow the approach taken by Polyakov\cite{poly}, elaborated upon in his book\cite{polyB}. We present the calculations in the bosonic case because we extend the principle to the fermionic case in the next appendix, for which there is less information apparent in the literature. Consider open world-lines first: following Polyakov we construct the volume element $ {\mathscr{D}}{h}$ by first defining a reparametrisation invariant inner product on variations of $h$:
$$
(\delta_1h,\delta_2h)_h=\int_0^1 h^{-2}\delta_1h\,\delta_2h
\sqrt{h}\,d\xi
$$
The powers of $h$ in this expression may be easily checked by attaching the appropriate indices to the intrinsic metric $h\rightarrow h_{ab}$. If the function space of metrics is itself parametrised by co-ordinates $\zeta_r$ then
\be {\mathscr{D}}{h}
=\sqrt {{\rm Det} \left({\partial h\over\partial \zeta_r},{\partial h\over\partial \zeta_s} \right)_h}\,\prod_r\,d\zeta_r \label{voldem}
\ee
We use this to factorise ${\mathscr{D}}h$ into a piece corresponding to reparametrisations and a piece corresponding to `physical' changes of the metric. An infinitesimal change of co-ordinate $\xi$:
$$
\tilde\xi=\xi+V(\xi)$$
is parametrised by the one-dimensional contravariant vector field $V(\xi)$ which vanishes at the ends of the interval so that $\tilde\xi$
also lies between $0$ and $1$. The induced change in $h$ is
$$
\delta_V h(\xi)=\tilde h(\xi)-h(\xi)=-\left\{V(\xi){d\over d\xi}+2{dV(\xi)\over d\xi}\right\}\,h(\xi)
\equiv -2 DV\,.
$$
We can decompose an arbitrary variation of $h$ into a piece of this form (for some $V$) and a piece orthogonal to all such changes $\delta_\perp h$. This last piece represents a `physical' change in $h$, i.e. one that cannot be written as a reparametrisation. It has to satisfy, for all $V$
$$
0=(\delta_\perp h,\delta_V h)_h \Leftrightarrow 
{d\over d\xi}{\delta_\perp h\over h}=0
$$
so $\delta_\perp h=h\,\delta c$ with $\delta c$ independent of $\xi$, i.e. a global scaling, and $(\delta_\perp h,\delta_\perp h)_h=(\delta c)^2
\int \sqrt{h}\,d\xi$. Defining a reparametrisation invariant inner product on contravariant vectors
$$
(\delta_1V,\delta_2V)_V=\int_0^1 h\,\delta_1V\,\delta_2V
\sqrt{h}\,d\xi
$$
enables us to write 
$$
(\delta_{V_1} h,\delta_{V_2} h)_h=4(D{V_1},D{V_2})_g=4({V_1},D^\dagger D{V_2})_V
$$
so that the volume element ${\mathscr{D}}{h}$ factorises as
\be
{\mathscr{D}}h=dc\,{\mathscr{D}} V\,{\sqrt{\left(\int_0^1 \sqrt{h}\,d\xi\right)\left(\vphantom{\int_{0}^{1}}\,
{\rm Det} (D^\dagger D)\right)}}\label{factvol}
\ee
with ${\mathscr{D}} V$ being constructed using the inner product on vectors. Expanding around $h=T^2$
the eigenvalue equation for $D^\dagger D$, $D^\dagger DV=\lambda V$ 
$$
-{1\over g}{d\over d\xi}\left({1\over {\sqrt g}}{d\over d\xi}\left( {\sqrt g}\,V\right)\right)=\lambda V\,,
$$
becomes
\be
-{1\over T^2}{d^2\over d\xi^2}\,V=\lambda \,V\,.\label{redofff}
\ee
The eigenfunctions vanish at the end-points, and so are $\sin (n\pi\xi)$ and the eigenvalues are $\lambda=(n\pi/T)^2$ with $n=1,2,,$. Hence
$$
\sqrt{{\rm Det} (D^\dagger D)}=\prod_{n=1..\infty} {n\pi\over T}\propto
T^{-\zeta(0)}=\sqrt T
$$
where we have effectively used $\zeta$-function regularisation. Also
\be
dc={2dT\over T},\quad \int_0^1 \sqrt{h}\,d\xi=T,\quad \label{redof}
\ee
so up to a numerical constant
$$
{\mathscr{D}}h=dT\,{\mathscr{D}}V\,.
$$
Now $\int{\mathscr{D}}V$ is the volume of the reparametrisation symmetry and so can be absorbed into the normalisation allowing us to replace ${\mathscr{D}}h$ by $dT$ for open world-lines.

\bigskip
\noindent For closed world-lines we again have (\ref{redofff}) and (\ref{redof}), but now the boundary condition on $V$ is that it should be periodic, so the eigenfunctions are  $\sin (2n\pi\xi)$ and $\cos (2n\pi\xi)$ $n=1,2,,$ with twofold degenerate eigenvalues $\lambda=(2n\pi/T)^2$ with $n=1,2,,$. So now
$$
\sqrt{{\rm Det} (D^\dagger D)}=\left(\prod_{n=1..\infty} {2n\pi\over T}\right)^2\propto
T^{-2\zeta(0)}=T
$$
The constant eigenfunction, which has vanishing eigenvalue, is excluded since it does not change the gauge-fixed value of the metric, so ${\mathscr{D}}V$ in (\ref{factvol}) has to be modified to ${\mathscr{D}}V'$ which excludes constant $V$. So
\be
{\mathscr{D}}{h}={dT\over T}{\mathscr{D}}V'\,{\sqrt T}\,T
\ee
The inner product of two constant vectors is $(\delta_1v,\delta_2v)_V=T^3\delta_1v\,\delta_2v$ so
\be
{\mathscr{D}} V={\mathscr{D}} V' dv \,T^{3/2}
\ee
Thus, when we divide by the volume of the reparametrisation symmetry there is a factor remaining. Putting this all together means that we can replace  ${\mathscr{D}}h$ by $dT/T$ for closed world-lines.

\section*{Appendix C -- Spin-half}
\label{AppSH}
Usually the action of Brink, di Vecchia and Howe\cite{BdVH}
is treated within canonical quantisation but here we need to develop the functional integral approach. We extend the material presented in the previous appendix for the bosonic point particle. If we consider first the case of closed world-lines
\be
Z=\int {\mathscr{D}} (h,w,\chi,\psi)\,e^{-S_0-S_F-S_A}\label{ssyi}
\ee
so that $w^\mu(\xi)$ is periodic, then it is natural to take the anti-commuting degrees of freedom to be anti-periodic.
Expanding around $h=T^2$, as in the previous appendix, we can again replace ${\mathscr{D}}h$ by $dT/T$. An arbitrary value of $\chi$ can be transformed to zero by a supersymmetry transformation. Expanding around this we can write  an arbitrary change in $\chi$ as a supersymmetry transformation, so
\be
{\mathscr{D}} \chi={\rm Det}^{-1/2}\left( -\left({1\over \sqrt h}{d\, \over d\xi}\right)^2\right)
{\mathscr{D}}\alpha\label{FP1}
\ee
for appropriate $\alpha$. The form of the operator on the right hand side follows from the reparametrisation invariant inner products on variations of $\chi$ and $\alpha$ used to construct the volume elements:
\be
(\delta_1\chi,\delta_2\chi)_h=\int_0^1 h^{-1}\delta_1\chi\,\delta_2\chi
\sqrt{h}\,d\xi,\quad (\delta_1\alpha,\delta_2\alpha)_h=\int_0^1 \delta_1\alpha\,\delta_2\alpha
\sqrt{h}\,d\xi
\ee
With anti-periodic boundary conditions the functional determinant is independent of $\sqrt h=T$:
\be
\sqrt{{\rm Det}\bigg( -\bigg({1\over T}{d\, \over d\xi}\bigg)^2\bigg)}
=\left(\prod_{n=0}^\infty {(2n+1)\pi\over T}\right)^2
=\left({\prod_{n=1}^\infty {n\pi\over T}{\Big /} \prod_{n=1}^\infty {2n\pi\over T}}\right)^2\propto
\frac{T^{-2\zeta(0)}}{(T/2)^{-2\zeta(0)}}
\ee
so integrating over $\alpha$ and using $t=T\xi$ puts (\ref{ssyi}) into the gauge-fixed form
\be
Z=\int_0^\infty {dT\over T}\int {\mathscr{D}}(w,\psi)e^{-{1\over 2}\int_0^T dt\,(\dot{w}^2
+\psi\cdot\dot\psi+{i}F_{\mu\nu}\psi^\mu\psi^\nu
-i\dot w\cdot A)}\label{gfform}
\ee
The $\psi^\mu$ play the role of $\gamma$-matrices; specifically
if we introduce an anti-commuting source $\eta^\mu$ then
\be
\int {\mathscr{D}}\psi \,e^{-\int_0^T dt\,({1\over 2}\psi\cdot\dot\psi-\sqrt 2\eta\cdot\psi)}
={\mathscr{T}}\,{\rm tr}\left (e^{\int_0^T dt\,\eta\cdot\gamma}\right )
\nonumber
\ee
\be
=\sum_{n=0}^\infty \int_0^Tdt_1\int^{t_1}_0dt_2..\int_0^{t_{n-1}} dt_{n}\,{\rm tr}\left (\eta(t_1)\cdot\gamma...\eta(t_{n})\cdot\gamma\right)\label{gamrep}
\ee
where ${\mathscr{T}}$ denotes time-ordering.
To show this consider differentiating from the left with respect to $\eta^\mu(t')$. The integrand of the n-th term becomes 
\be
\sum_{r=1}^{n}(-)^{r+1}\,{\rm tr}\left (\eta(t_1)\cdot\gamma...\eta(t_{r-1})\cdot\gamma\,\,\gamma^\mu\,\,
\eta(t_{r+1})\cdot\gamma...\eta(t_n)\cdot\gamma)\right)
\ee
with $t_1>t_2>..>t_{r-1}>t'>t_{r+1}>..>t_n$.
The $\gamma^\mu$ can be moved to the right using the anti-commutation relations $\{\gamma^\lambda,\gamma^\rho\}=2\delta^{\lambda\rho}$ to give a sum of terms:
\begin{align}
&\sum_{r=1}^{n}\Big((-)^{n-1}{\rm tr}\left (\eta(t_1)\cdot\gamma...\eta(t_{r-1})\cdot\gamma\,
\eta(t_{r+1})\cdot\gamma...\eta(t_n)\cdot\gamma\,\gamma^\mu\right)\nonumber\\
&+\sum_{p=r+1}^n 2\,(-)^{p}\,
{\rm tr}\left (\eta(t_1)\cdot\gamma...\eta(t_{r-1})\cdot\gamma\,
\eta(t_{r+1})\cdot\gamma...\right.\nonumber \\
&\left. \qquad\qquad ...\eta(t_{p-1})\cdot\gamma\,\eta^\mu(t_p)\,\eta(t_{p+1})\cdot\gamma...\eta(t_n)\cdot\gamma\right)\Big)
\end{align}
Moving $\eta(t_p)$ to the extreme left removes the $(-)^{p}$, giving
\begin{align}
&\sum_{r=1}^{n}\Big(
(-)^{n-1}{\rm tr}\left (\eta(t_1)\cdot\gamma...\eta(t_{r-1})\cdot\gamma\,
\eta(t_{r+1})\cdot\gamma...\eta(t_n)\cdot\gamma\,\gamma^\mu\right)\nonumber\\
&+
\sum_{p=r+1}^n 2\,\eta^\mu(t_p)\,
{\rm tr}\left (\eta(t_1)\cdot\gamma...\eta(t_{r-1})\cdot\gamma\,
\eta(t_{r+1})\cdot\gamma...\right. \nonumber \\
&\qquad \qquad \left. ...\eta(t_{p-1})\cdot\gamma\,\eta(t_{p+1})\cdot\gamma...\eta(t_n)\cdot\gamma\right)\Big)
\label{exppa}
\end{align}
Alternatively we can move $\gamma^\mu$ to the left to give
\begin{align}
&\sum_{r=1}^{n}\Big(
{\rm tr}\left (\gamma^\mu\eta(t_1)\cdot\gamma...\eta(t_{r-1})\cdot\gamma\,
\eta(t_{r+1})\cdot\gamma...\eta(t_n)\cdot\gamma\,\right)\nonumber\\
&+\sum_{p=1}^{r-1} -2\,\eta^\mu(t_p)\,
{\rm tr}\left (\eta(t_1)\cdot\gamma...\eta(t_{r-1})\cdot\gamma\,
\eta(t_{r+1})\cdot\gamma... \right. \nonumber \\
&\qquad \qquad \left. ...\eta(t_{p-1})\cdot\gamma\,\eta(t_{p+1})\cdot\gamma...\eta(t_n)\cdot\gamma\right)\Big)
\label{exppb}
\end{align}
Taking the average of (\ref{exppa}) and (\ref{exppb}) the first terms cancel because of cyclicity of the trace and the fact that $n$ should be even for the trace to be non-zero. Performing the integrations then gives
\be
{\delta\over\delta\eta(t')}
{\mathscr{T}}\,{\rm tr}\left (e^{\int_0^T dt\,\eta\cdot\gamma}\right )=\left(
\int_0^T{\epsilon(t'-t)}\eta(t)\, dt\,\right)
{\mathscr{T}}\,{\rm tr}\left (e^{\int_0^T dt\,\eta\cdot\gamma}\right )
\ee
(with $\epsilon (t)=t/|t|$), implying 
\be
{\mathscr{T}}\,{\rm tr}\left (e^{\int_0^T dt\,\eta\cdot\gamma}\right )=4e^{{1\over 2}\int_0^T dt\,dt'\,\eta(t)\cdot \eta(t')\epsilon(t-t')}
\ee
which compares with 
\be
\int {\mathscr{D}}\psi \,e^{-\int_0^T dt\,({1\over 2}\psi\cdot\dot\psi-\sqrt 2\eta\cdot\psi)}
=Z_0\,e^{\int_0^T dt\,dt'\,\eta(t)\cdot \eta(t')\,G(t,t')}
\ee
where $Z_0$ is a power of the determinant of $d/dt$, but since this operates on anti-periodic functions $Z_0$ will be independent of $T$ and so just a numerical constant. $G(t,t')=\epsilon(t-t')/2$ is the Green function for this operator.

Using this representation of the $\gamma$-matrices we can write the gauge-fixed expression (\ref{gfform}) as
\bq
&&
\int_0^\infty {dT\over T}\int {\mathscr{D}}(w,\psi)e^{-{1\over 2}\int_0^T dt\,(\dot{w}^2
+\psi\cdot\dot\psi
-i\dot w\cdot A)}
{\mathscr{T}}\,{\rm tr}\left (e^{-\int_0^T dt\,
{i\over 4}F_{\mu\nu}\gamma^\mu\gamma^\nu}\right )\nonumber\\
&&
=
\int_0^\infty {dT\over T} \,{\rm Tr}\left(e^{-{T\over 2}\left({\cal D}^2
+{i\over 2}F_{\mu\nu}\gamma^\mu\gamma^\nu\right )}\right)
\nonumber\\
&&
=
\int_0^\infty {dT\over T}\,{\rm Tr}\left(e^{-T\left({(\gamma\cdot\cal D)}^2\right)}
\right)\nonumber
\eq
which gives our result for closed world-lines
\be\int {\mathscr{D}} (h,w,\chi,\psi)\,e^{-S_0-S_F-S_A}
=-{\rm ln\,Det}\left(\left(\gamma\cdot\cal D\right)^2\,\right)
\ee

We now turn to the case of open world-lines. For this we will need a version of (\ref{gamrep}) involving matrices rather than their traces. In the representation:
\be
\gamma^4=\left( \begin{array}{cc}
0 &   1\\
1& 0  \end{array} \right)
\quad
\gamma^j=\left( \begin{array}{cc}
0 &   -i\sigma^j\\
i\sigma^j& 0  \end{array} \right),
\ee
the combination $(\gamma^4+i\gamma^3)(\gamma^4-i\gamma^3)
(\gamma^2-i\gamma^1)(\gamma^2+i\gamma^1)$ is represented by
$(4,0,0,0)\otimes (4,0,0,0)^t$. So we can extract the $11$ component of the matrix $e^{\int dt \eta\cdot \gamma}$ by inserting this product into the trace, which can be accomplished by differentating with respect to the source at either end of the interval using the time-ordering to obtain the correct ordering of $\gamma$-matrices:
\be
\left({\partial\over\partial \eta^4}+i{\partial\over\partial \eta^3}\right)_0
\left({\partial\over\partial \eta^4}-i{\partial\over\partial \eta^3}\right)_T
\left({\partial\over\partial \eta^2}-i{\partial\over\partial \eta^1}\right)_0
\left({\partial\over\partial \eta^2}+i{\partial\over\partial \eta^1}\right)_T
{\mathscr{T}}\,{\rm tr}\left (e^{\int_0^T dt\,\eta\cdot\gamma}\right )
\nonumber
\ee
Applying the derivatives to the integral representation (\ref{gamrep}) generates an insertion of 
\be
(\psi^4+i\psi^3)_0(\psi^4-i\psi^3)_T
(\psi^2-i\psi^1)_0(\psi^2+i\psi^1)_T
\ee
which, given the rules of integration over anti-commuting Grassman numbers, acts as a set of $\delta$-functions which change the boundary conditions on the functional integral.
So 
\be
{\mathscr{T}}\,\left (e^{\int_0^T dt\,\eta\cdot\gamma}\right )_{11}
=\int {\mathscr{D}}\psi \,e^{-\int_0^T dt\,({1\over 2}\psi\cdot\dot\psi-\eta\cdot\psi)}\Big |_{\psi_2=i\psi_1,\psi_4=-i\psi_3\,{\rm at\,t=0}}^{\psi_2=-i\psi_1,\psi_4=i\psi_3\,{\rm at\,t=T}}
\ee
The other matrix components can be obtained from this by appropriate insertions of $\gamma^\mu$ at the ends of the interval, we denote this as
\be
{\mathscr{T}}\,\left (e^{\int_0^T dt\,\eta\cdot\gamma}\right )_{ab}
=\int {\mathscr{D}}\psi \,e^{-\int_0^T dt\,({1\over 2}\psi\cdot\dot\psi-\eta\cdot\psi)}\Big |_{ab}
\ee

We return to the evaluation of 
\be
Z_{ab}=\int {\mathscr{D}} (h,w,\chi,\psi)\,e^{-S_0-S_F-S_A}\Big|_{ab}
\ee 
attaching spinor indices $ab$ corresponding to boundary conditions on the $\psi$-integration.
For open world-lines, we can replace ${\mathscr{D}}h$ by 
$dT$ on gauge-fixing the reparametrisation invariance by setting $h=T^2$ as in Appendix A. The integral over $\chi$ can similarly be done by using the supersymmetry. Given that $w^\mu$ is fixed at the ends of the world-line the supersymmetry parameter $\delta\alpha$ must vanish there. This means that an arbitrary change in $\chi$ cannot always be written as a supersymmetry transformation, instead we put
\be
\delta \chi=\delta_\perp \chi+2{d\delta\alpha\over d\xi}
\ee
and require that these two pieces be orthogonal, which requires 
that $\delta_\perp \chi/\sqrt h$ be constant. Expanding about $h = T^{2}$ we also expand $\chi$ about the $\xi$-independent value $\chi_0$. The inner product on variations of $\chi_0$ is 
\be
(\delta_1\chi_0,\delta_2\chi_0)_h=\int_0^1h^{-1}
\delta_1\chi_0\,\delta_2\chi_0\, \sqrt h d\xi=\delta_1\,\chi_0\delta_2\chi_0\,/T
\ee
and the boundary conditions require eigenfunctions which vanish at either end of the interval. This, together with the usual rules for integrating over Grassman numbers, implies that
\be
{\mathscr{D}} \chi=\sqrt T\, d\chi_0\,{\rm Det}^{-1/2}\left( -\left({1\over T}{d\, \over d\xi}\right)^2\right)
{\mathscr{D}}\alpha= d\chi_0\,
{\mathscr{D}}\alpha={1\over T}\,d\left({\chi_0\over  T}\right)\,
{\mathscr{D}}\alpha
\,.
\ee
As before the integral over $\alpha$ gives a constant since the integrand is invariant under supersymmetry. So $Z_{ab}$ becomes
\be
\int_0^\infty {dT\over T}\int {\mathscr{D}}(w,\psi)\,d\left({\chi_0\over T}\right)\,e^{-{1\over 2}\int_0^T dt\,(\dot{w}^2
+\psi\cdot\dot\psi+{\chi_0\over T}\dot w\cdot\psi+{i}F_{\mu\nu}\psi^\mu\psi^\nu
+2i\dot w\cdot A)}\Big|_{ab}
\ee
Scaling $\chi_0$ and factoring the $\psi$-dependence we obtain
\be
\int\,d\chi_0 \int_0^\infty {dT\over T}\int {\mathscr{D}}w\,e^{-{1\over 2}\int_0^T dt\,(\dot{w}^2
+2i\dot w\cdot A)}\left\{\int {\mathscr{D}}\psi\,
e^{-{1\over 2}\int_0^T dt\,(\psi\cdot\dot\psi+{\chi_0}\dot w\cdot\psi+{i}F_{\mu\nu}\psi^\mu\psi^\nu)}
\Big|_{ab}\right\}
\ee
Integrating out the $\psi$ replaces them by $\gamma$-matrices:
\begin{align}
&\int\,d\chi_0 
\int_0^\infty {dT\over T}\int {\mathscr{D}}w\,e^{-{1\over 2}\int_0^T dt\,(\dot{w}^2
+2i\dot w\cdot A)}{\mathscr{T}}\left\{
e^{-{1\over 2}\int_0^T dt\,({\chi_0}\dot w\cdot\gamma+{i}F_{\mu\nu}\gamma^\mu\gamma^\nu/2)}
\right\}\Big|_{ab}\nonumber \\
\end{align}
We identify this as the path-integral representation of the matrix element:
\be
\int\,d\chi_0 \int_0^\infty {dT\over T}\,
\langle w_f,a|e^{-T\hat H}
|w_i,b\rangle
=\int\,d\chi_0 \,
\langle w_f,a|\,{\rm ln}\left(\hat H\right )
|w_i,b\rangle
\ee
where the Hamiltonian is 
\be
\hat H=\left((\hat p+A)^2 
+\chi_0\gamma\cdot (\hat p+A)/\sqrt 2
+{i}F_{\mu\nu}\gamma^\mu\gamma^\nu/2\right)/2
\ee
In the Schr\"odinger representation the Hamiltonian becomes the operator
\be
\hat H_s=\left(-{\cal D}^2 
-\chi_0i\gamma\cdot{\cal D}
-{i}F_{\mu\nu}\gamma^\mu\gamma^\nu\right)/2
=\left(-(\gamma\cdot{\cal D})^2-\chi_0i\gamma\cdot{\cal D}\right)/2
\ee
so we arrive at 
\be
Z_{ab}=\int\,d\chi_0 \,
\langle w_f,a|\,{\rm ln}\left(-(\gamma\cdot{\cal D})^2-\chi_0i\gamma\cdot{\cal D}
\right ) | w_i, b\rangle
=\langle w_f,a|\,\left(\gamma\cdot{\cal D}\right)^{-1}
|w_i,b\rangle
\ee
i.e. the Dirac propagator.

\section*{Appendix D -- Gauge-fixing Maxwell theory}
\label{AppC}
In the text we have adopted a gauge-fixing procedure that imposes the gauge condition using a delta-function. We have to compute the integral
\be
Z[J]=\int {\mathscr{D}} A \, e^{\int d^4x\left( \left(A_\mu \partial ^2 A_\mu -(\partial\cdot A)^2\right)/(2q^2) +J\cdot A+{\cal A}\cdot\partial^2 A/q^2\right)}\,.
\ee
with $J=-i\sum_j\int dy_j \delta(y_j-x)$ and $\cal A$ on-shell: $\partial^2{\cal A}=\partial\cdot{\cal A}=0$.
Gauge invariance prevents the operator in the kinetic term from being inverted. Inserting a delta-function that imposes the gauge condition $\partial \cdot A=0$ (we absorb the associated Faddeev-Popov determinant into the normalisation as it is independent of $A$) gives
 \be
Z[J]=\int {\mathscr{D}} (A,\lambda) \, e^{\int d^4x\left(\left( A_\mu \partial ^2 A_\mu -(\partial\cdot A)^2\right)/(2q^2) +J\cdot A+{\cal A}\cdot\partial^2 A/q^2+i \lambda \partial\cdot A\right)}
\ee
Shifting the integration variable $\lambda$ by $-\frac{i\partial\cdot A}{2q^{2}}$ modifies the differential operator to one that is invertible
\bq
Z[J]&=&
\int {\mathscr{D}} (A,\lambda) \, e^{\int d^4x\left( A_\mu \partial ^2 A_\mu/(2q^2)  +J\cdot A+{\cal A}\cdot\partial^2 A/q^2+i \lambda \partial\cdot A\right)}\nonumber\\
&=&\int {\mathscr{D}} \lambda \, e^{{q^2\over 2}\int d^4x\left( J
\partial^{-2}J+2{\cal A}\cdot J/q^2+{\cal A}\partial^2{\cal A}/q^4+ 2i\lambda\,\partial^{-2}\partial\cdot J+2i\lambda\partial\cdot{\cal A}/q^2-\lambda^2
\right)}\nonumber\\
&=&e^{{q^2\over 2}\int d^4x\left( J
\partial^{-2}J-\partial\cdot J\,(\partial^{-2})^2\partial\cdot J+2{\cal A}\cdot J/q^2
\right)}\nonumber\\
&=&e^{{q^2\over 2}\sum_{jk}\int dy_j\cdot\Delta\cdot dy_k-i\sum_{j}\int dy_j\cdot{\cal A}}
\eq
\section*{Appendix E -- Contraction Algebra}
\label{AppD}
In this appendix we calculate some of the important contractions used in the main work. From (\ref{Z0}) in the main text it is straightforward to calculate simple correlation functions. For example, for the product of derivatives of fields we may put $k=0$ in the generating function (since the $k$-dependence arose there because of the exponential factors)
\begin{align}
	\left<\partial_{1}X_{\mu}\left(\xi\right)\partial_{1}X^{\prime}_{\nu}\left(\xi^{\prime}\right)\right> &= \left.\frac{\delta}{\delta j^{\mu}_{1}\left(\xi\right)} \frac{\delta}{\delta j^{\nu}_{1}\left(\xi^{\prime}\right)} \mathcal{Z}\left(j, k = 0\right)\right|_{j = 0} \nonumber \\
	&= \left.\frac{\delta}{\delta j^{\mu}_{1}\left(\xi\right)}\left[\left(\int d^{2}\tilde{\xi}~4\pi\alpha^{\prime}\eta_{\alpha \nu}j^{\nu a}\left(\tilde{\xi}\right) \partial_{a}\partial^{\prime}_{1}G\left(\tilde{\xi}, \xi^{\prime}\right) \right.\right.\right. \nonumber \\
	&\qquad\qquad\qquad \left.\left.\left.+~\vphantom{\int d^{2}\tilde{\xi}}\partial^{\prime}_{1}X^{\nu}_{0}\left(\xi^{\prime}\right) \right)\mathcal{Z}\left(j, k=0\right)\right]\right|_{j=0} \\
	&= 4\pi\alpha^{\prime}\partial_{1}\partial_{1}^{\prime}G\left(\xi, \xi^{\prime}\right) + \partial_{1}X^{\mu}_{0}\left(\xi\right) \partial^{\prime}_{1}X^{\nu}_{0}\left(\xi^{\prime}\right)
\end{align}
where the latter term is the contribution due to the presence of the boundary. Furthermore, to justify the expectation value of the exponential factor we may set $j = 0$ and consider
\begin{align}
	\left<\exp{\left[i k\cdot \left(X\left(\xi\right) - X\left(\xi^{\prime}\right)\right)\right]}\right> &= \mathcal{Z}\left(j=0, k\right) \nonumber \\
	&= \exp{-\pi\alpha^{\prime} k^{2} \Psi\left(\xi, \xi^{\prime}\right)} \cdot \exp{ik\cdot\left(X\left(\xi\right) - X\left(\xi^{\prime}\right)\right)}.
\end{align}
The above results provide a further justification for the forms of (\ref{wicks}) - (\ref{exp}) and highlight the r\^{o}le played by the boundary.

\bigskip

\noindent We also present in this appendix the calculation of contractions between fields. This can be determined by use of $\mathcal{Z}\left(j, k\right)$ but for the operator product expansion used it is easier to see it by making use of Wick's theorem and the previous results. Specifically we are interested in the product of two fields of the form
\begin{equation}
	A^{\mu} e^{i k \cdot B}
\end{equation}
considered inside a correlation function. We note, however, that these fields are not considered normal ordered by themselves. By expanding the exponential and applying Wick's theorem we arrive at
\begin{equation*}
	A^{\mu} e^{i k \cdot B} = \sum_{n = 0}^{\infty} \frac{\left(ik\right)^{n}}{n!} A^{\mu} B^{n} ~~~~~~~~~~~~~~~~~~~~~~~~~~~~~~~~~~~~~~~~~~~~~~~~~~~~~~~~~~~~~~~~~~~~~~~~~~~~~~~~~~~~
\end{equation*}
\begin{equation*}
\bcontraction{\hphantom{A^{\mu} e^{i k \cdot B} } = \sum_{n=0}^{\infty} \sum_{\lambda =0}^{\lambda < \frac{n-1}{2}} ik_{\nu}}    {\!A^{\mu}}    {}    {B}
\bcontraction{\hphantom{A^{\mu} e^{i k \cdot B} } = \sum_{n=0}^{\infty} \sum_{\lambda =0}^{\lambda < \frac{n-1}{2}} ik_{\nu} A^{\mu} B^{\nu} \cdot \left(-k_{\nu}k_{\rho}\right) }    {\left(B^{\nu}\right.}    {\!\!}    {B}
	\hphantom{A^{\mu} e^{i k \cdot B} } = \sum_{n=0}^{\infty} \sum_{\lambda =0}^{\lambda < \frac{n-1}{2}} ik_{\nu} A^{\mu} B^{\nu} \cdot \left(-k_{\nu}k_{\rho}\right)^{\lambda} \left(B^{\nu} B^{\rho}\right)^{\lambda} ~\frac{\left(ik\right)^{n-2\lambda -1}}{(n-1)!}\cdot \begin{pmatrix}n\!-\!1 \\ 2\lambda\end{pmatrix} \frac{\left(2\lambda\right)!}{2^\lambda \lambda!}:B^{n-2\lambda -1}:
\end{equation*}
\begin{equation*}
\bcontraction{\hphantom{A^{\mu} e^{i k \cdot B}} + \sum_{n=0}^{\infty} \sum_{\lambda =0}^{\lambda < \frac{n}{2}} \left(-k_{\nu}k_{\rho}\right) }     {\left(B^{\nu}\right.}    {\!\!}    {B}
	\hphantom{A^{\mu} e^{i k \cdot B}} + \sum_{n=0}^{\infty} \sum_{\lambda =0}^{\lambda < \frac{n}{2}} \left(-k_{\nu}k_{\rho}\right)^{\lambda} \left(B^{\nu} B^{\rho}\right)^{\lambda} \frac{\left(ik\right)^{n-2\lambda}}{n!}\cdot \frac{\left(2\lambda\right)!}{2^{\lambda} \lambda !}\begin{pmatrix}n \\ 2\lambda \end{pmatrix} :A^{\mu} B^{n - 2\lambda}: ~~~~~~~~~~~~~~~~~~~~~~~~
\end{equation*}
\begin{equation*}
\bcontraction{\hphantom{A^{\mu} e^{i k \cdot B} } = \sum_{n=0}^{\infty} \sum_{\lambda =0}^{\lambda < \frac{n-1}{2}} ik_{\nu}}     {A^{\mu}}    {}    {B}
\bcontraction{\hphantom{A^{\mu} e^{i k \cdot B} } = \sum_{n=0}^{\infty} \sum_{\lambda =0}^{\lambda < \frac{n-1}{2}} ik_{\nu} A^{\mu} B^{\nu} \cdot \frac{\left(-k_{\nu}k_{\rho}\right)^{\lambda}}{2^{\lambda} \lambda!}}    {\left(B^{\nu}\right.}    {\!\!}    {B}
	\hphantom{A^{\mu} e^{i k \cdot B} } = \sum_{n=0}^{\infty} \sum_{\lambda =0}^{\lambda < \frac{n-1}{2}} ik_{\nu} A^{\mu} B^{\nu} \cdot \frac{\left(-k_{\nu}k_{\rho}\right)^{\lambda}}{2^{\lambda} \lambda!}\cdot \left(B^{\nu} B^{\rho}\right)^{\lambda} ~\frac{\left(ik\right)^{n-2\lambda -1}}{\left(n\!-\!2\lambda\!-\!1\right)!}~:B^{n\!-\!2\lambda\! -\!1}:  ~~~~~~~~~~~~~~~~~~~
\end{equation*}
\begin{equation*}
\bcontraction{\hphantom{A^{\mu} e^{i k \cdot B}} + \sum_{n=0}^{\infty} \sum_{\lambda =0}^{\lambda < \frac{n}{2}}  \frac{\left(-k_{\nu}k_{\rho}\right)^{\lambda}}{2^{\lambda} \lambda !}}    {\left(B^{\nu}\right.}    {\!\!}    {B} 
	\hphantom{A^{\mu} e^{i k \cdot B}} + \sum_{n=0}^{\infty} \sum_{\lambda =0}^{\lambda < \frac{n}{2}}  \frac{\left(-k_{\nu}k_{\rho}\right)^{\lambda}}{2^{\lambda} \lambda !} \left(B^{\nu} B^{\rho}\right)^{\lambda} \frac{\left(ik\right)^{n-2\lambda}}{\left(n\!-\!2\lambda\right)!} :A^{\mu}B^{n - 2\lambda}: ~~~~~~~~~~~~~~~~~~~~~~~~~~~~~~~~~~~~~~~~
\end{equation*}
\begin{equation}
\bcontraction{\hphantom{A^{\mu} e^{i k \cdot B}} = ik_{\nu}}    {A^{\mu}}    {}    {B}
\bcontraction{\hphantom{A^{\mu} e^{i k \cdot B}} = ik_{\nu} A^{\mu} B^{\nu} \exp{\left(-\frac{1}{2}k_{\nu}k_{\rho}\right.}\!\!\!\!}        {\left(B^{\nu}\right.}    {\!\!}    {B} 
\bcontraction{\hphantom{A^{\mu} e^{i k \cdot B}} = ik_{\nu} A^{\mu} B^{\nu}  \exp{\left(-\frac{1}{2}k_{\nu}k_{\rho}B^{\nu}B^{\rho}\right)}:e^{i k \cdot B}: +~ \exp{\left(-\frac{1}{2}k_{\nu}k_{\rho}\right.}\!\!\!\!}        {\left(B^{\nu}\right.}    {\!\!}    {B} 
	\hphantom{A^{\mu} e^{i k \cdot B}} = ik_{\nu} A^{\mu} B^{\nu}  \exp{\left(-\frac{1}{2}k_{\nu}k_{\rho}B^{\nu}B^{\rho}\right)} :e^{i k \cdot B}: +~ \exp{\left(-\frac{1}{2}k_{\nu}k_{\rho}B^{\nu}B^{\rho}\right)}:A^{\mu}e^{ik\cdot B}:\:
\end{equation}
In the above equations the combinatoric factors come from the number of ways of choosing the fields to take place in contraction and from the ordering of these contractions amongst themselves. The form of the last line shows that we may consider the product by working out contractions of term with itself and the cross terms with others in the product. This result is used extensively in section \ref{calc0} to determine the product of fields arising in the perturbative expansion; in the main work the field $B$ is represented by the field $X$ and the prefactor $A$ by a derivative $\partial X$. We may, for example, immediately extract the product 
\begin{equation}
	\partial_{1}X^{\mu} e^{i k\cdot X} = 4\pi\alpha^{\prime}ik^{\mu} \partial_{1}G \exp{\left(-\pi\alpha^{\prime}k^{2} \psi\right)}:e^{i k \cdot X}: + \exp{\left(-\pi\alpha^{\prime}k^{2} \psi\right)}:\partial_{1}X^{\mu} e^{ik\cdot X}:
\end{equation}
One may use the general iterative nature of Wick's theorem along with the results above to calculate more complicated products, as has been done in the main text. We note here that the lack of normal ordering of each term in the product means that contractions between fields at coincident points are generated. Typically this leads to Green functions at coincident points which diverge. This behaviour, coupled with the boundary terms remaining in the normal ordered fields leftover after contractions conspires to provide the results discussed in the paper.

\end{document}

%% file: first.pdf_tex
%% Creator: Inkscape inkscape 0.48.5, www.inkscape.org
%% Accompanies image file 'first.pdf' (pdf, eps, ps)
\begingroup%
  \makeatletter%
  \providecommand\color[2][]{%
    \errmessage{(Inkscape) Color is used for the text in Inkscape, but the package 'color.sty' is not loaded}%
    \renewcommand\color[2][]{}%
  }%
  \providecommand\transparent[1]{%
    \errmessage{(Inkscape) Transparency is used (non-zero) for the text in Inkscape, but the package 'transparent.sty' is not loaded}%
    \renewcommand\transparent[1]{}%
  }%
  \providecommand\rotatebox[2]{#2}%
  \ifx\svgwidth\undefined%
    \setlength{\unitlength}{200bp}%
    \ifx\svgscale\undefined%
      \relax%
    \else%
      \setlength{\unitlength}{\unitlength * \real{\svgscale}}%
    \fi%
  \else%
    \setlength{\unitlength}{\svgwidth}%
  \fi%
  \global\let\svgwidth\undefined%
  \global\let\svgscale\undefined%
  \makeatother%
  \begin{picture}(1,0.56)%
    \put(0,0){\includegraphics[width=\unitlength]{first.pdf}}%
    \put(0.21802557,0.38005127){\color[rgb]{0,0,0}\makebox(0,0)[lb]{\smash{$\Lambda$}}}%
    \put(0.8,0.52005127){\color[rgb]{0,0,0}\makebox(0,0)[lb]{\smash{$\xi^{1}$}}}%
    \put(0.22,0.48005127){\color[rgb]{0,0,0}\makebox(0,0)[lb]{\smash{$\xi^{2}$}}}%
  \end{picture}%
\endgroup%

%% file: firstBound1.pdf_tex
%% Creator: Inkscape inkscape 0.48.5, www.inkscape.org
%% Accompanies image file 'firstBound1.pdf' (pdf, eps, ps)
\begingroup%
  \makeatletter%
  \providecommand\color[2][]{%
    \errmessage{(Inkscape) Color is used for the text in Inkscape, but the package 'color.sty' is not loaded}%
    \renewcommand\color[2][]{}%
  }%
  \providecommand\transparent[1]{%
    \errmessage{(Inkscape) Transparency is used (non-zero) for the text in Inkscape, but the package 'transparent.sty' is not loaded}%
    \renewcommand\transparent[1]{}%
  }%
  \providecommand\rotatebox[2]{#2}%
  \ifx\svgwidth\undefined%
    \setlength{\unitlength}{200bp}%
    \ifx\svgscale\undefined%
      \relax%
    \else%
      \setlength{\unitlength}{\unitlength * \real{\svgscale}}%
    \fi%
  \else%
    \setlength{\unitlength}{\svgwidth}%
  \fi%
  \global\let\svgwidth\undefined%
  \global\let\svgscale\undefined%
  \makeatother%
  \begin{picture}(1,0.56)%
    \put(0,0){\includegraphics[width=\unitlength]{firstBound1.pdf}}%
    \put(0.21802557,0.38005127){\color[rgb]{0,0,0}\makebox(0,0)[lb]{\smash{$\Lambda$}}}%
    \put(0.8,0.52005127){\color[rgb]{0,0,0}\makebox(0,0)[lb]{\smash{$\xi^{1}$}}}%
    \put(0.36,0.30005127){\color[rgb]{0,0,0}\makebox(0,0)[lb]{\smash{$\xi^{2}$}}}%
  \end{picture}%
\endgroup%

%% file: firstBound2.pdf_tex
%% Creator: Inkscape inkscape 0.48.5, www.inkscape.org
%% Accompanies image file 'firstBound2.pdf' (pdf, eps, ps)
\begingroup%
  \makeatletter%
  \providecommand\color[2][]{%
    \errmessage{(Inkscape) Color is used for the text in Inkscape, but the package 'color.sty' is not loaded}%
    \renewcommand\color[2][]{}%
  }%
  \providecommand\transparent[1]{%
    \errmessage{(Inkscape) Transparency is used (non-zero) for the text in Inkscape, but the package 'transparent.sty' is not loaded}%
    \renewcommand\transparent[1]{}%
  }%
  \providecommand\rotatebox[2]{#2}%
  \ifx\svgwidth\undefined%
    \setlength{\unitlength}{200bp}%
    \ifx\svgscale\undefined%
      \relax%
    \else%
      \setlength{\unitlength}{\unitlength * \real{\svgscale}}%
    \fi%
  \else%
    \setlength{\unitlength}{\svgwidth}%
  \fi%
  \global\let\svgwidth\undefined%
  \global\let\svgscale\undefined%
  \makeatother%
  \begin{picture}(1,0.56)%
    \put(0,0){\includegraphics[width=\unitlength]{firstBound2.pdf}}%
    \put(0.21802557,0.38005127){\color[rgb]{0,0,0}\makebox(0,0)[lb]{\smash{$\Lambda$}}}%
    \put(0.8,0.30005127){\color[rgb]{0,0,0}\makebox(0,0)[lb]{\smash{$\xi^{1}$}}}%
    \put(0.22,0.30005127){\color[rgb]{0,0,0}\makebox(0,0)[lb]{\smash{$\xi^{2}$}}}%
  \end{picture}%
\endgroup%

%% file: firstBulk.pdf_tex
%% Creator: Inkscape inkscape 0.48.5, www.inkscape.org
%% Accompanies image file 'firstBulk.pdf' (pdf, eps, ps)
\begingroup%
  \makeatletter%
  \providecommand\color[2][]{%
    \errmessage{(Inkscape) Color is used for the text in Inkscape, but the package 'color.sty' is not loaded}%
    \renewcommand\color[2][]{}%
  }%
  \providecommand\transparent[1]{%
    \errmessage{(Inkscape) Transparency is used (non-zero) for the text in Inkscape, but the package 'transparent.sty' is not loaded}%
    \renewcommand\transparent[1]{}%
  }%
  \providecommand\rotatebox[2]{#2}%
  \ifx\svgwidth\undefined%
    \setlength{\unitlength}{200bp}%
    \ifx\svgscale\undefined%
      \relax%
    \else%
      \setlength{\unitlength}{\unitlength * \real{\svgscale}}%
    \fi%
  \else%
    \setlength{\unitlength}{\svgwidth}%
  \fi%
  \global\let\svgwidth\undefined%
  \global\let\svgscale\undefined%
  \makeatother%
  \begin{picture}(1,0.56)%
    \put(0,0){\includegraphics[width=\unitlength]{firstBulk.pdf}}%
    \put(0.21802557,0.38005127){\color[rgb]{0,0,0}\makebox(0,0)[lb]{\smash{$\Lambda$}}}%
    \put(0.48,0.42005127){\color[rgb]{0,0,0}\makebox(0,0)[lb]{\smash{$\xi^{1}$}}}%
    \put(0.56,0.52005127){\color[rgb]{0,0,0}\makebox(0,0)[lb]{\smash{$\xi^{2}$}}}%
    \put(0.428,0.48405127){\color[rgb]{0,0,0}\makebox(0,0)[lb]{\smash{$\Lambda$}}}%
  \end{picture}%
\endgroup%

%% file: stripRegion.pdf_tex
%% Creator: Inkscape inkscape 0.48.4, www.inkscape.org
%% Accompanies image file 'stripRegion.pdf' (pdf, eps, ps)
\begingroup%
  \makeatletter%
  \providecommand\color[2][]{%
    \errmessage{(Inkscape) Color is used for the text in Inkscape, but the package 'color.sty' is not loaded}%
    \renewcommand\color[2][]{}%
  }%
  \providecommand\transparent[1]{%
    \errmessage{(Inkscape) Transparency is used (non-zero) for the text in Inkscape, but the package 'transparent.sty' is not loaded}%
    \renewcommand\transparent[1]{}%
  }%
  \providecommand\rotatebox[2]{#2}%
  \ifx\svgwidth\undefined%
    \setlength{\unitlength}{124.46400146bp}%
    \ifx\svgscale\undefined%
      \relax%
    \else%
      \setlength{\unitlength}{\unitlength * \real{\svgscale}}%
    \fi%
  \else%
    \setlength{\unitlength}{\svgwidth}%
  \fi%
  \global\let\svgwidth\undefined%
  \global\let\svgscale\undefined%
  \makeatother%
  \begin{picture}(1,0.61479625)%
    \put(0,0){\includegraphics[width=\unitlength]{stripRegion.pdf}}%
    \put(0.43424604,0.19584138){\color[rgb]{0,0,0}\makebox(0,0)[lb]{\smash{a}}}%
    \put(0.43343446,0.35490704){\color[rgb]{0,0,0}\makebox(0,0)[lb]{\smash{b}}}%
    \put(0.43424604,0.56932098){\color[rgb]{0,0,0}\makebox(0,0)[lb]{\smash{$\frac{\Lambda}{\sqrt{\epsilon
}}$}}}%
    \put(0.82347246,0.00755746){\color[rgb]{0,0,0}\makebox(0,0)[lb]{\smash{$x$}}}%
    \put(0.90527715,0.21905145){\color[rgb]{0,0,0}\makebox(0,0)[lb]{\smash{$y$}}}%
    \put(0.08073017,0.02297827){\color[rgb]{0,0,0}\makebox(0,0)[lb]{\smash{$\partial D$}}}%
  \end{picture}%
\endgroup%

%% file: O1.pdf_tex
%% Creator: Inkscape 0.48.1, www.inkscape.org
%% Accompanies image file 'O1.pdf' (pdf, eps, ps)

\begingroup
  \makeatletter
  \providecommand\color[2][]{%
    \errmessage{(Inkscape) Color is used for the text in Inkscape, but the package 'color.sty' is not loaded}
    \renewcommand\color[2][]{}%
  }
  \providecommand\transparent[1]{%
    \errmessage{(Inkscape) Transparency is used (non-zero) for the text in Inkscape, but the package 'transparent.sty' is not loaded}
    \renewcommand\transparent[1]{}%
  }
  \providecommand\rotatebox[2]{#2}
  \ifx\svgwidth\undefined
    \setlength{\unitlength}{37.2015625pt}
  \else
    \setlength{\unitlength}{\svgwidth}
  \fi
  \global\let\svgwidth\undefined
  \makeatother
  \begin{picture}(1,0.87979196)%
    \put(0,0){\includegraphics[width=\unitlength]{O1.pdf}}%
    \put(0.49863497,0.47673012){\color[rgb]{0,0,0}\makebox(0,0)[lb]{\smash{$k^{\mu}$}}}%
    \put(0.02553656,0.45522598){\color[rgb]{0,0,0}\makebox(0,0)[lb]{\smash{$w^{1}$}}}%
    \put(0.76844398,0.27153499){\color[rgb]{0,0,0}\makebox(0,0)[lb]{\smash{$w^{2}$}}}%
  \end{picture}%
\endgroup

%% file: bulkRegion.pdf_tex
%% Creator: Inkscape inkscape 0.48.4, www.inkscape.org
%% Accompanies image file 'bulkRegion.pdf' (pdf, eps, ps)
\begingroup%
  \makeatletter%
  \providecommand\color[2][]{%
    \errmessage{(Inkscape) Color is used for the text in Inkscape, but the package 'color.sty' is not loaded}%
    \renewcommand\color[2][]{}%
  }%
  \providecommand\transparent[1]{%
    \errmessage{(Inkscape) Transparency is used (non-zero) for the text in Inkscape, but the package 'transparent.sty' is not loaded}%
    \renewcommand\transparent[1]{}%
  }%
  \providecommand\rotatebox[2]{#2}%
  \ifx\svgwidth\undefined%
    \setlength{\unitlength}{78.05564575bp}%
    \ifx\svgscale\undefined%
      \relax%
    \else%
      \setlength{\unitlength}{\unitlength * \real{\svgscale}}%
    \fi%
  \else%
    \setlength{\unitlength}{\svgwidth}%
  \fi%
  \global\let\svgwidth\undefined%
  \global\let\svgscale\undefined%
  \makeatother%
  \begin{picture}(1,1.00039777)%
    \put(0,0){\includegraphics[width=\unitlength]{bulkRegion.pdf}}%
    \put(0.5512454,0.50016323){\color[rgb]{0,0,0}\makebox(0,0)[lb]{\smash{a}}}%
    \put(0.65373639,0.7051452){\color[rgb]{0,0,0}\makebox(0,0)[lb]{\smash{b}}}%
    \put(0.37701073,0.58215602){\color[rgb]{0,0,0}\makebox(0,0)[lb]{\smash{$r$}}}%
    \put(0.51024901,0.37717405){\color[rgb]{0,0,0}\makebox(0,0)[lb]{\smash{$\theta$}}}%
    \put(0.75622734,0.96137271){\color[rgb]{0,0,0}\makebox(0,0)[lb]{\smash{$\frac{\Lambda}{\sqrt{\epsilon}}$}}}%
  \end{picture}%
\endgroup%

%% file: Nbound.pdf_tex
%% Creator: Inkscape inkscape 0.48.4, www.inkscape.org
%% Accompanies image file 'Nbound.pdf' (pdf, eps, ps)
\begingroup%
  \makeatletter%
  \providecommand\color[2][]{%
    \errmessage{(Inkscape) Color is used for the text in Inkscape, but the package 'color.sty' is not loaded}%
    \renewcommand\color[2][]{}%
  }%
  \providecommand\transparent[1]{%
    \errmessage{(Inkscape) Transparency is used (non-zero) for the text in Inkscape, but the package 'transparent.sty' is not loaded}%
    \renewcommand\transparent[1]{}%
  }%
  \providecommand\rotatebox[2]{#2}%
  \ifx\svgwidth\undefined%
    \setlength{\unitlength}{200bp}%
    \ifx\svgscale\undefined%
      \relax%
    \else%
      \setlength{\unitlength}{\unitlength * \real{\svgscale}}%
    \fi%
  \else%
    \setlength{\unitlength}{\svgwidth}%
  \fi%
  \global\let\svgwidth\undefined%
  \global\let\svgscale\undefined%
  \makeatother%
  \begin{picture}(1,0.56)%
    \put(0,0){\includegraphics[width=\unitlength]{Nbound.pdf}}%
    \put(0.21802557,0.38005127){\color[rgb]{0,0,0}\makebox(0,0)[lb]{\smash{$\Lambda$}}}%
  \end{picture}%
\endgroup%

%% file: ON.pdf_tex
%% Creator: Inkscape 0.48.1, www.inkscape.org
%% Accompanies image file 'ON.pdf' (pdf, eps, ps)

\begingroup
  \makeatletter
  \providecommand\color[2][]{%
    \errmessage{(Inkscape) Color is used for the text in Inkscape, but the package 'color.sty' is not loaded}
    \renewcommand\color[2][]{}%
  }
  \providecommand\transparent[1]{%
    \errmessage{(Inkscape) Transparency is used (non-zero) for the text in Inkscape, but the package 'transparent.sty' is not loaded}
    \renewcommand\transparent[1]{}%
  }
  \providecommand\rotatebox[2]{#2}
  \ifx\svgwidth\undefined
    \setlength{\unitlength}{34.3pt}
  \else
    \setlength{\unitlength}{\svgwidth}
  \fi
  \global\let\svgwidth\undefined
  \makeatother
  \begin{picture}(1,0.95421678)%
    \put(0,0){\includegraphics[width=\unitlength]{ON.pdf}}%
    \put(0.58673469,0.84358944){\color[rgb]{0,0,0}\makebox(0,0)[lb]{\smash{$k_{1}^{\alpha}$}}}%
    \put(0.19170591,0.75839397){\color[rgb]{0,0,0}\makebox(0,0)[lb]{\smash{$w_{1}$}}}%
    \put(0.84329446,0.44708798){\color[rgb]{0,0,0}\makebox(0,0)[lb]{\smash{$w_{i+1}$}}}%
    \put(0.35349854,0.07390978){\color[rgb]{0,0,0}\makebox(0,0)[lb]{\smash{$k_{N}^{\chi}$}}}%
    \put(0.07361516,0.40044075){\color[rgb]{0,0,0}\makebox(0,0)[lb]{\smash{$w_{2N-1}$}}}%
    \put(0.72667638,0.14388063){\color[rgb]{0,0,0}\makebox(0,0)[lb]{\smash{$w_{2N}$}}}%
    \put(0.54008746,0.51705883){\color[rgb]{0,0,0}\makebox(0,0)[lb]{\smash{$k_{i}^{\mu}$}}}%
    \put(0.07361516,0.54038244){\color[rgb]{0,0,0}\makebox(0,0)[lb]{\smash{$w_{i}$}}}%
    \put(0.84329446,0.68032413){\color[rgb]{0,0,0}\makebox(0,0)[lb]{\smash{$w_{2}$}}}%
  \end{picture}%
\endgroup

%% file: Nbulk.pdf_tex
%% Creator: Inkscape inkscape 0.48.4, www.inkscape.org
%% Accompanies image file 'Nbulk.pdf' (pdf, eps, ps)
\begingroup%
  \makeatletter%
  \providecommand\color[2][]{%
    \errmessage{(Inkscape) Color is used for the text in Inkscape, but the package 'color.sty' is not loaded}%
    \renewcommand\color[2][]{}%
  }%
  \providecommand\transparent[1]{%
    \errmessage{(Inkscape) Transparency is used (non-zero) for the text in Inkscape, but the package 'transparent.sty' is not loaded}%
    \renewcommand\transparent[1]{}%
  }%
  \providecommand\rotatebox[2]{#2}%
  \ifx\svgwidth\undefined%
    \setlength{\unitlength}{200bp}%
    \ifx\svgscale\undefined%
      \relax%
    \else%
      \setlength{\unitlength}{\unitlength * \real{\svgscale}}%
    \fi%
  \else%
    \setlength{\unitlength}{\svgwidth}%
  \fi%
  \global\let\svgwidth\undefined%
  \global\let\svgscale\undefined%
  \makeatother%
  \begin{picture}(1,0.56)%
    \put(0,0){\includegraphics[width=\unitlength]{Nbulk.pdf}}%
    \put(0.21802557,0.38005127){\color[rgb]{0,0,0}\makebox(0,0)[lb]{\smash{$\Lambda$}}}%
    \put(0.55292896,0.51500049){\color[rgb]{0,0,0}\makebox(0,0)[lb]{\smash{$\Lambda$}}}%
  \end{picture}%
\endgroup%

%% file: O0.pdf_tex
%% Creator: Inkscape 0.48.1, www.inkscape.org
%% Accompanies image file 'O0.pdf' (pdf, eps, ps)

\begingroup
  \makeatletter
  \providecommand\color[2][]{%
    \errmessage{(Inkscape) Color is used for the text in Inkscape, but the package 'color.sty' is not loaded}
    \renewcommand\color[2][]{}%
  }
  \providecommand\transparent[1]{%
    \errmessage{(Inkscape) Transparency is used (non-zero) for the text in Inkscape, but the package 'transparent.sty' is not loaded}
    \renewcommand\transparent[1]{}%
  }
  \providecommand\rotatebox[2]{#2}
  \ifx\svgwidth\undefined
    \setlength{\unitlength}{42.4140625pt}
  \else
    \setlength{\unitlength}{\svgwidth}
  \fi
  \global\let\svgwidth\undefined
  \makeatother
  \begin{picture}(1,0.58424887)%
    \put(0,0){\includegraphics[width=\unitlength]{O0.pdf}}%
    \put(0.23459201,0.16658685){\color[rgb]{0,0,0}\makebox(0,0)[lb]{\smash{$k_{1}^{\mu}$}}}%
    \put(0.40434703,0.31748023){\color[rgb]{0,0,0}\makebox(0,0)[lb]{\smash{$w$}}}%
    \put(0.4797937,0.03455513){\color[rgb]{0,0,0}\makebox(0,0)[lb]{\smash{$x^{\prime}$}}}%
  \end{picture}%
\endgroup

%% file: ONO.pdf_tex
%% Creator: Inkscape 0.48.1, www.inkscape.org
%% Accompanies image file 'ONO.pdf' (pdf, eps, ps)

\begingroup
  \makeatletter
  \providecommand\color[2][]{%
    \errmessage{(Inkscape) Color is used for the text in Inkscape, but the package 'color.sty' is not loaded}
    \renewcommand\color[2][]{}%
  }
  \providecommand\transparent[1]{%
    \errmessage{(Inkscape) Transparency is used (non-zero) for the text in Inkscape, but the package 'transparent.sty' is not loaded}
    \renewcommand\transparent[1]{}%
  }
  \providecommand\rotatebox[2]{#2}
  \ifx\svgwidth\undefined
    \setlength{\unitlength}{62.9390625pt}
  \else
    \setlength{\unitlength}{\svgwidth}
  \fi
  \global\let\svgwidth\undefined
  \makeatother
  \begin{picture}(1,0.47194116)%
    \put(0,0){\includegraphics[width=\unitlength]{ONO.pdf}}%
    \put(0.19120314,0.01199076){\color[rgb]{0,0,0}\makebox(0,0)[lb]{\smash{$k_{1}^{\mu}$}}}%
    \put(0.12194335,0.17459415){\color[rgb]{0,0,0}\makebox(0,0)[lb]{\smash{$w_{1}$}}}%
    \put(0.24905042,0.17459415){\color[rgb]{0,0,0}\makebox(0,0)[lb]{\smash{$w_{2}$}}}%
    \put(0.41428962,0.17459415){\color[rgb]{0,0,0}\makebox(0,0)[lb]{\smash{$w_{3}$}}}%
    \put(0.5541074,0.17459415){\color[rgb]{0,0,0}\makebox(0,0)[lb]{\smash{$w_{4}$}}}%
    \put(0.68121447,0.17459415){\color[rgb]{0,0,0}\makebox(0,0)[lb]{\smash{$w_{5}$}}}%
    \put(0.82103225,0.17459415){\color[rgb]{0,0,0}\makebox(0,0)[lb]{\smash{$w_{6}$}}}%
    \put(0.49055386,0.01199076){\color[rgb]{0,0,0}\makebox(0,0)[lb]{\smash{$k_{2}^{\nu}$}}}%
    \put(0.75747871,0.01199076){\color[rgb]{0,0,0}\makebox(0,0)[lb]{\smash{$k_{3}^{\rho}$}}}%
  \end{picture}%
\endgroup